\documentclass{article} %
\PassOptionsToPackage{varqu}{inconsolata}
\usepackage{colm2024_conference}

\usepackage{booktabs}
\usepackage{arydshln}

\usepackage{graphicx}
\usepackage{enumitem}
\usepackage{wrapfig}
\usepackage{algorithm}
\usepackage{algpseudocode}

\usepackage{microtype}
\usepackage{amsmath}
\usepackage{amssymb}
\usepackage{colortbl}
\usepackage[utf8]{inputenc}
\usepackage[T1]{fontenc}

\definecolor{lightgray}{rgb}{0.9,0.9,0.9}
\usepackage{caption}
\usepackage{subcaption}
\usepackage{xcolor}
\usepackage{setspace}
\usepackage{url}
\usepackage{multirow}
\usepackage{tabularx}
\usepackage{pgfplots}
\pgfplotsset{compat=1.18}
\usepackage{tikz}
\usetikzlibrary{er,positioning,bayesnet}
\usepackage{makecell}
\usepackage{tipa}
\usepackage{siunitx}
\usepackage{nicefrac}
\usepackage{tocloft}
\usepackage{listings}
\usepackage[raster,skins]{tcolorbox}
\usepackage{xltabular}
\usepackage{adjustbox}
\usepackage{xurl}
\usepackage{xspace}
\usepackage{pifont}

\usepackage{amsmath,amsfonts,bm}

\def\eqref#1{equation~\ref{#1}}

\def\1{\bm{1}}

\def\rc{{\textnormal{c}}}

\DeclareMathAlphabet{\mathsfit}{\encodingdefault}{\sfdefault}{m}{sl}
\SetMathAlphabet{\mathsfit}{bold}{\encodingdefault}{\sfdefault}{bx}{n}

\lstdefinestyle{domainex}{%
  basicstyle=\ttfamily\fontsize{5.2pt}{6.2pt}\selectfont,
  breaklines=true,
  breakatwhitespace=false,
  columns=fullflexible,
  frame=none,
  aboveskip=0pt,
  belowskip=0pt,
  xleftmargin=0pt,
  xrightmargin=0pt,
  keepspaces=true,
  showstringspaces=false,
}

\newcommand{\yes}{\textcolor{green!60!black}{\ding{51}}} % ✔
\newcommand{\no}{\textcolor{red}{\ding{55}}}   % ✘
\newcommand{\NA}{\textemdash}
\newcommand{\benchname}{\textsc{SecRespond}}

\colorlet{colorscale}{cyan!40!red}
\colorlet{colorctrl}{cyan!40!black}
\definecolor{colordecouple}{HTML}{d2e2f0}
\definecolor{colorunify}{HTML}{dad8e8}
\definecolor{lightblue}{RGB}{220,235,250}
\tcbset{
  takeawaysbox/.style={
    title=Takeaways,
    colback=lightblue!80,
    colframe=black,
    fonttitle=\bfseries\small,
    coltitle=white,
    colbacktitle=black,
    enhanced,
    attach boxed title to top left={xshift=2.5mm,yshift=-2.5mm},
    boxed title style={rounded corners, size=small, colframe=black, colback=black},
    width=\linewidth,
    arc=3.5mm
  }
}

\title{SecRespond: Benchmarking AI Agents\\ for Real-World Post-Compromise Incident Response}

\author{
Lehan Wang$^{1,3}$, Boli Chen$^{1}$, Ruixue Ding$^{1}$, Pengjun Xie$^{1}$, Jinwei Huang$^{2}$,\\ Zhendong Liu$^{2}$, Shuo Wang$^{2}$, Tao Lei$^{2}$, Xin Ouyang$^{2}$, Xiaomeng Li$^{3}$\\$^{1}$ \includegraphics[height=0.4cm]{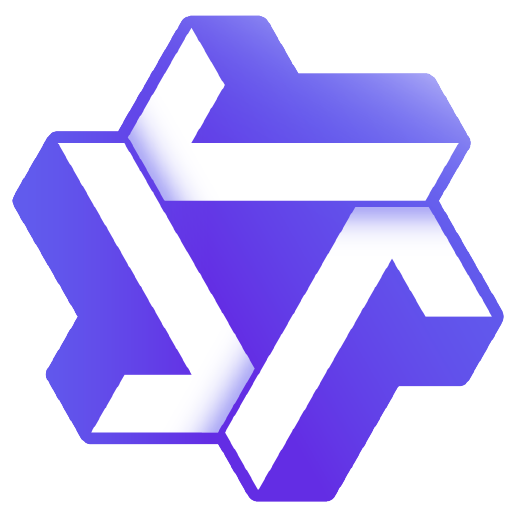} Tongyi Lab, Alibaba Group \quad $^{2}$ \includegraphics[height=0.4cm]{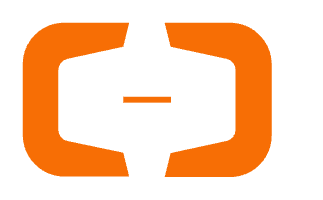} Alibaba Cloud Computing, Alibaba Group
\\ $^{3}$ \includegraphics[height=0.45cm]{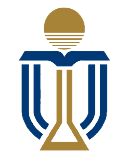} The Hong Kong University of Science and Technology
}

\newcommand{\fix}{\marginpar{FIX}}

\begin{document}

\maketitle

% \newpage
% \makeatletter
% \def\addcontentsline#1#2#3{%
%   \addtocontents{#1}{\protect\contentsline{#2}{#3}{\thepage}{\@currentHref}}}
% \makeatother
% \setcounter{tocdepth}{3}
% \setcounter{secnumdepth}{3}
% \tableofcontents
% \newpage

\lstset{
    basicstyle=\ttfamily\small,
    frame=single,
    breaklines=true,
    columns=fullflexible,
    keepspaces=true,
    backgroundcolor=\color{gray!8},
    literate={—}{{\textemdash}}1 {…}{{\textellipsis}}1,
}

\begin{abstract}
Large Language Model (LLM) agents are increasingly adopted in real-world security operations with access to host artifacts and command-line interfaces (CLIs), making it critical to thoroughly assess their security capabilities. However, existing cybersecurity benchmarks focus on pre-compromise settings where agents are placed in a clean and idealized environment before an attack occurs.
This leaves the post-compromise setting underexplored. To address this gap, we introduce \benchname{}, the first benchmark for evaluating LLM agents on the post-compromise incident-response workflow. Given a forensic disk snapshot of a compromised host together with the alerts, vulnerability scans, and baseline checks reported by a host security product, agents are required to produce forensic reports on intrusions, baseline risks, and vulnerability risks, together with a remediation plan. We instantiate this task across 10 cyber ranges, each constructed from a distinct compromised cloud host, spanning 4 entry-point types, 21 ATT\&CK techniques, and 5 operating systems. We evaluate 23 frontier LLMs on the OpenCode agent harness. Experimental results show that although current agents can reliably uncover the problems exposed by alerts, they struggle to proactively investigate the disk for silent intrusions and to produce comprehensive, verified remediation plans, with no model achieving complete detection and remediation on any single range. This reveals a fundamental bottleneck in building agents for real-world incident response.
The benchmark is publicly available at \url{https://github.com/Alibaba-NLP/qqr/tree/main/data/secrespond}.
\end{abstract}

\section{Introduction}
%% Agent's potential application on cybersecurity
Large Language Model (LLM) agents, equipped with iterative reasoning and autonomous tool-use abilities~\citep{openai2026_gpt54,qwen2026_qwen37,zai2026_glm51}, demonstrate remarkable potential in tackling complex, professional real-world tasks~\citep{merrill2026terminal,ye2026claw,li2026skillsbench}, among which cybersecurity is a significant deployment target. With access to host artifacts and command-line interfaces (CLIs), LLM agents are increasingly expected to assist in real-world security operations from vulnerability discovery to incident response~\citep{wang2025cybergym,deason2025cybersoceval}. Therefore, it is critical to assess how reliably current agents can fulfill such pipelines and how they can be involved in human workflows.

%% Existing benchmarks on cybersecurity and their limitation
Recently, several cybersecurity benchmarks have been developed to cover various offensive-security subtasks.
\cite{zhang2025cybench} and~\cite{shao2024nyu} assess the offensive capabilities of LLM agents with CTF challenges, while~\cite{zhu2025cve} and~\cite{wang2025cybergym} evaluate agents' vulnerability discovery and exploitation abilities using real-world web applications or software projects. Recent efforts such as CyberModelArena~\citep{wiz_cyber_model_arena_2025} have begun to assess harness-model combinations across the full attack lifecycle. In parallel, defensive benchmarks evaluate LLMs on security knowledge, vulnerability discovery, and patching. \cite{jing2024secbench} collects multiple-choice question-answer pairs to assess LLMs' defensive knowledge, while \cite{autopatchbench_meta_2025} evaluates LLMs on vulnerability patching.

However, these existing security benchmarks focus on \textbf{\textit{pre-compromise}} settings, where the agent operates in a clean and idealized environment to find, exploit, or patch weaknesses before an attack occurs~\citep{zhang2025cybench,wang2025cybergym}.
Even on the defensive side, existing works simplify the problem to reasoning over isolated alerts~\citep{deason2025cybersoceval}, system logs~\citep{wu2025excytin}, or incident reports~\citep{lin2025ircopilot} rather than confronting the compromised system with real artifacts left by the attack, such as persistence mechanisms, deliberately erased traces, and noisy concurrent host activity.
This leaves the \textbf{\textit{post-compromise}} workflow underexplored, which includes investigating a successful intrusion on the compromised host, reconstructing the attack chain and remediating the system. As post-compromise response covers the full incident-response cycle, evaluating LLM agents on this pipeline measures whether they can actively discover problems within the compromised disk and respond effectively to live alerts and real intrusion artifacts, thereby serving as the prerequisite to facilitating agent assistance for security operations in production.

To address this gap, we introduce \benchname{}, the first benchmark for evaluating LLM agents on the post-compromise incident-response workflow. The agent is given a forensic disk snapshot of a compromised host with the alerts, vulnerability scans, and baseline checks reported by a host security product. Given these inputs, the agent is required to investigate the host, reconstruct what happened, and produce one progress file and four reports covering the detected intrusion, vulnerabilities and baseline risks, along with a remediation plan to resolve them.
\benchname{} instantiates the above task in 10 cyber range scenarios, each built from a frozen snapshot of a distinct, fully instantiated cloud-host environment compromised through an end-to-end attack over real network protocols. Together, these ranges cover the diversity and complexity of real-world intrusion scenarios across 4 entry-point types, 21 ATT\&CK techniques, and 5 operating systems.
Moreover, we propose a hierarchical rubric that decomposes each range into detailed checkpoints that validate whether each problem is comprehensively discovered and effectively handled, and grade them with the LLM-as-a-Judge method.
We further design a five-dimensional Capability (CAP) taxonomy including intrusion entity, persistence mechanism, baseline risk, vulnerability risk, and investigation \& response quality, and map each checkpoint to its corresponding capability items.
In this way, checkpoint results from heterogeneous ranges become comparable and can be aggregated through the capability taxonomy to benchmark model capabilities.
In total, \benchname{} defines 52 capability items and 280 checkpoints, reflecting fine-grained model performance in the full incident response workflow from investigation to remediation.

We conduct extensive experiments on a representative agent harness, OpenCode~\citep{opencode2026}, paired with 23 different large language models (LLMs), spanning multiple model families and successive releases within each family.
Our experiments reveal that current agents can surface the problems exposed by alerts but struggle to investigate the forensic disk proactively and to complete the remediation required for incident response. Even the strongest model, Claude Opus 4.7, achieves only 72.4\% averaged over detection and planning, leaving malicious artifacts untouched and remediation incomplete, especially in the ranges where the attack chain grows longer and broader.
All the models score higher on detection than on planning, with the gap widening to 34.7\% for GPT-5.5, since agents tend to apply the obvious first fix but rarely complete the remaining remediation.
Furthermore, we observe a significant imbalance across capability dimensions. Average detection reaches 75.4\% on intrusion entity but only 58.8\% on persistence mechanism, whereas the average planning score peaks at 55.3\% on baseline risk and reaches only 31.8\% on investigation \& response quality.
This exposes a fundamental bottleneck in versatile incident-response models.

%% Generalize contribution
Our contributions are as follows:
\begin{itemize}
    \item \textbf{A benchmark for post-compromise incident response.} We release \benchname{}, the first benchmark targeting the full post-compromise response workflow, which comprises 10 cyber ranges built on fully instantiated cloud-host environments, covering 4 entry-point types, 21 ATT\&CK techniques, and 5 operating systems.
    \item \textbf{A hierarchical capability evaluation framework.} We propose a hierarchical LLM-as-a-Judge framework where each task is decomposed into fine-grained checkpoints manually designed by security experts. The resulting 280 checkpoints are graded along both detection and planning axes and mapped to 52 capability items in our five-dimensional Capability (CAP) taxonomy, supporting per-capability diagnosis and cross-range analysis.
    \item \textbf{An empirical study of frontier agents on real-world post-compromise tasks.} We evaluate 23 LLMs on a representative agent harness, OpenCode, spanning multiple model families and successive releases within each family, with each task independently judged by three strong proprietary LLMs to mitigate individual bias.
    \item \textbf{Key findings on incident response capability of agent models.} Our study shows that current agents reliably discover alerted problems but struggle with proactive investigation and complete remediation. We also observe that agent models' incident response capabilities are significantly imbalanced across different dimensions.
\end{itemize}

\section{Related Work}

\subsection{Agent Benchmarks}
To comprehensively evaluate the capabilities of LLM agents from multiple aspects, many benchmarks have been developed. To validate agent models' coding abilities, SWE-Bench~\citep{jimenez2024swe} tests the model on resolving real GitHub issues, SWE-Bench-Pro~\citep{deng2025swe} extends it to more realistic long-horizon problems that require examining multiple files and modifying substantial code, and Terminal-Bench~\citep{merrill2026terminal} targets interactive shell execution in a terminal environment. To evaluate agents in GUI environments, WebArena~\citep{zhou2024webarena}  places agents in realistic web applications to conduct web navigation, while OSWorld~\citep{xie2024osworld} extends to real desktop environments where agents are supposed to perceive screenshots and operate applications. To benchmark agents in multi-turn conversational interaction, $\tau$-bench~\citep{yao2024tau} prompts agents to serve a simulated user in customer-service dialogues,
$\tau^2$-bench~\citep{barres2506tau2} establishes a dual-control environment in which the user can also modify the shared state, and Vitabench~\citep{he2025vitabench} scales the setting to versatile real-world applications including food delivery, in-store consumption, and online travel services. Recently, agentic harnesses such as OpenClaw and OpenCode have encouraged evaluation of agents operating within these scaffolds.
CocoaBench~\citep{team2026cocoabench} evaluates agents' compositional capability in vision, search, and coding on long-horizon tasks, ClawBench~\citep{zhang2026clawbench} constructs daily tasks on live websites, and Claw-Eval~\citep{ye2026claw} and Claw-Eval-Live~\citep{li2026claw} cover end-to-end workflows across service orchestration, multimodal interaction, and professional dialogue.
Despite these broad evaluations, existing agent benchmarks target software engineering, web, and general productivity tasks, largely overlooking security operations.

\subsection{Benchmarking LLMs in Cybersecurity} 
Evaluating cybersecurity capabilities of LLMs and LLM agents has drawn growing attention, typically from both offensive and defensive aspects.
On the offensive side, early work~\citep{li2024wmdp} assesses intrinsic security knowledge of LLMs with curated multiple-choice questions. These benchmarks focus on knowledge probing rather than automatic operational abilities. Subsequently, Cybench~\citep{zhang2025cybench} and NYU CTF Bench~\citep{shao2024nyu} place LLM agents inside Docker containers to evaluate their offensive abilities through interactive CTF challenge solving, and CyberGym~\citep{wang2025cybergym} extends the scenario to vulnerability discovery and exploitation in real-world software projects, where LLM agents uncover zero-day bugs that human maintainers have missed. Recent work, CyberModelArena~\citep{wiz_cyber_model_arena_2025}, unifies the full attack lifecycle into a single benchmark of five categories and evaluates different combinations of agent designs and underlying models to analyze their independent and joint offensive capabilities.
On the defensive side, SecBench~\citep{jing2024secbench} measures LLMs' defensive knowledge in network, endpoint, application, and cloud security, while CyberSOCEval~\citep{deason2025cybersoceval} evaluates LLMs from the perspective of Security Operations Center (SOC) analysts on understanding malware analysis and threat intelligence reports. AutoPatchBench~\citep{autopatchbench_meta_2025} evaluates AI-assisted patch generation for repairing vulnerabilities.
ExCyTIn-Bench~\citep{wu2025excytin} evaluates LLM agents on cyber threat investigation by reasoning over alerts and security logs.
However, as shown in Table~\ref{tab:comparison}, none of the existing work reproduces actual end-to-end host intrusion that leaves forensic artifacts, including persistence mechanisms, partially cleaned traces, and noisy concurrent legitimate activity. As a result, there remains a gap in evaluating whether LLM agents can investigate a post-compromise disk snapshot and produce detection and remediation analysis.

\begin{table*}[t]
\centering
\caption{Comparison with existing cybersecurity benchmarks. We compare \benchname{} with representative cybersecurity benchmarks along the following aspects: \textbf{Scope} denotes the security task; \textbf{Post-Compromise} indicates whether the benchmark targets the real-world scenario after a host has already been attacked; \textbf{Real Filesystem} denotes whether the task is grounded in a host disk rather than synthesized logs or text; \textbf{Multi-Step Analysis} indicates whether solving the task requires multiple investigative steps; \textbf{Cross-File Analysis} summarizes whether the evidence should be correlated across multiple files; \textbf{Rubric-Based Evaluation} indicates whether the evaluation metrics use fine-grained rubrics; \textbf{CLI-Compatible} denotes whether the benchmark task can be executed through a command-line interface.}
\label{tab:comparison}
\resizebox{1.0\textwidth}{!}{
\setlength{\tabcolsep}{6pt}
\begin{tabular}{l ccccccc}
\toprule
\textbf{Benchmark} & \textbf{Scope} & \textbf{\makecell{Post-Compromise}} & \textbf{\makecell{Real\\Filesystem}} & \textbf{\makecell{Multi-Step\\Analysis}} & \textbf{\makecell{Cross-File\\Analysis}} & \textbf{\makecell{Rubric-Based\\Evaluation}} & \textbf{\makecell{CLI-\\Compatible}} \\
\midrule
SecBench~\citep{jing2024secbench} & Knowledge & \no & \no & \no & \no & \no & \no \\
NYU CTF Bench~\citep{shao2024nyu} & CTF & \no & \no & \yes & \yes & \no & \no \\
CyberSecEval~\citep{wan2024cyberseceval} & Offensive & \no & \no & \yes & \no & \yes & \no \\
CyBench~\citep{zhang2025cybench} & CTF & \no & \no & \yes & \yes & \no & \no \\
CVE-Bench~\citep{zhu2025cve} & CVE & \no & \no & \yes & \no & \no & \no \\
AutoPatchBench~\citep{autopatchbench_meta_2025} & Patch & \no & \yes & \yes & \yes & \no & \no  \\
CyberGym~\citep{wang2025cybergym} & Vuln & \no & \yes & \yes & \yes & \no & \no \\
CyberSOCEval~\citep{deason2025cybersoceval} & SOC & \no & \no & \no & \no & \no & \no \\
ExCyTIn-Bench~\citep{wu2025excytin} & Threat Detection & \yes & \no & \yes & \yes & \no & \no \\
CyberModelArena~\citep{wiz_cyber_model_arena_2025} & Offensive & \no & \yes & \yes & \yes & \no & \yes \\
\midrule
\rowcolor{blue!12} \textbf{\benchname{} (Ours)} & Forensic & \yes & \yes & \yes & \yes & \yes & \yes \\
\bottomrule
\end{tabular}
}
\end{table*}

\subsection{Evaluation Methodology}
Cybersecurity benchmarks predominantly rely on deterministic verification with a fixed ground truth, such as multiple-choice accuracy~\citep{li2024wmdp}, CTF flag matching~\citep{zhang2025cybench}, or verifying the agent's execution output~\citep{wang2025cybergym,autopatchbench_meta_2025}.
Such protocols cannot be directly applied to our benchmark, where the task output is open-ended investigation reports. 
Therefore, we adopt the LLM-as-a-Judge method~\citep{gu2026survey}, which decouples the target into verifiable units, such as the rubric items for trajectory evidence in Claw-Eval~\citep{ye2026claw}, thus making it well-suited for open-ended tasks.
To keep the judgment reliable and interpretable, we design a hierarchical LLM-as-a-Judge framework that decomposes each task into fine-grained checkpoints, which are further mapped to capability items in our five-dimensional capability taxonomy, enabling fine-grained analysis within each range and capability-level comparison across ranges.

\section{\benchname{} Benchmark}

\begin{figure}[t]
\centering
\includegraphics[width=1.0\textwidth]{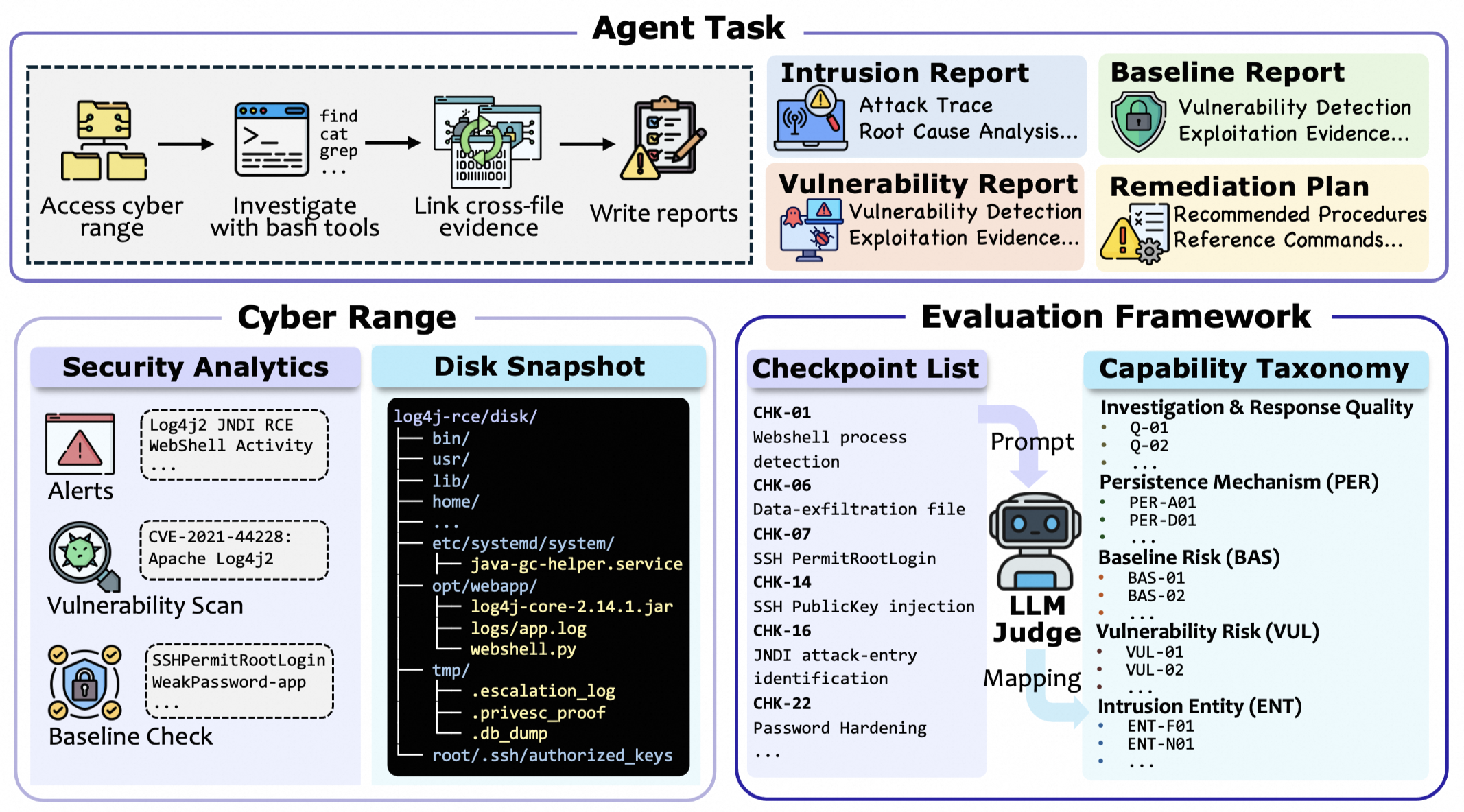}
\caption{Overview of \benchname{}. In our benchmark task, the agent is expected to investigate a cyber range and produce incident-response reports. Each cyber range includes a forensic disk snapshot of a real, fully instantiated cloud host compromised through an end-to-end attack over real network protocols, together with security analytics comprising alerts and scan findings from a security product. To evaluate the agent's incident-response capability, we organize the required skills into a capability taxonomy and map each capability to a list of checkpoints for each range. These checkpoints serve as evaluation rubrics for the LLM-as-a-Judge framework.}
\label{fig:overview}
\end{figure}

\benchname{} evaluates LLM agents' ability to investigate a post-compromise host snapshot and produce comprehensive forensic reports on intrusion analysis, baseline assessment, and vulnerability verification, together with a structured remediation plan. In this section, we first describe the benchmark task and explain how we construct cyber ranges on real, fully instantiated cloud hosts across diverse attack types and operating systems (\S\ref{sec:env}).
Then, we define a capability taxonomy that specifies what the agent should be able to accomplish in the incident-response process (\S\ref{sec:cap}). Finally, we introduce the evaluation framework, which decomposes each range into fine-grained checkpoints that link concrete forensic evidence to capability items, enabling hierarchical, cross-range evaluation (\S\ref{sec:eval}).

\subsection{Task Construction}
\label{sec:env}
\subsubsection{Overview of Tasks}
\label{sec:task}
To comprehensively evaluate the agents' incident response abilities, we construct 10 cyber ranges covering multiple attack scenarios. A cyber range includes a frozen, reproducible, read-only forensic disk snapshot of a host that has been compromised by a complete, end-to-end attack delivered over real network protocols, along with a set of security analytics including real-time alerts and scan findings from a security product. The agent is expected to investigate the exposed alerts and the snapshot, to reconstruct the intrusion from the traces the attack leaves behind. 
These ranges cover different operating systems, entry points, ATT\&CK techniques, and attack stages, which are selected to mimic the distribution in actual incidents based on a correlation analysis of alerts, vulnerabilities, and baselines from 372 real compromised cloud hosts.
In detail, the ranges are summarized in Table~\ref{tab:ranges}. 

The forensic disk snapshot typically covers all the directories of the compromised host, and contains the naturally generated intrusion artifacts left by the attack which the agent can read as the primary forensic evidence, including logs, configuration changes, and residual files.
Beyond the disk snapshot, the agent also receives three types of security analytics from the host security product on the victim machine: (i) real-time alerts from the anomaly detection engine monitoring suspicious processes, network connections, and login events, (ii) vulnerability findings from scans of installed packages and applications, and (iii) baseline check results from configuration checks. 
Given these inputs, the agent is required to produce one progress file and four reports: an \textbf{\textit{intrusion report}} reconstructing the attack chain covering entry point, lateral movement, persistence mechanisms, and final impact, a \textbf{\textit{vulnerability report}} verifying which exploited weaknesses are present on the host, a \textbf{\textit{baseline report}} assessing host configuration against security baselines, and a \textbf{\textit{remediation plan}} with recommended steps and reference commands for executing the fixes. The overview of \benchname{} is demonstrated in Figure~\ref{fig:overview}.

\begin{table*}[t]
\centering
\caption{Overview of the cyber ranges. We construct 10 cyber ranges in total, spanning 4 entry points and 5 operating systems. Each range reproduces a complete multi-stage attack chain from initial access through privilege escalation, persistence, and impact, covering 21 ATT\&CK techniques.}
\label{tab:ranges}
\renewcommand{\arraystretch}{1.15}
\resizebox{1.0\textwidth}{!}{
\begin{tabular}{@{}
>{\raggedright\arraybackslash}m{1.8cm}
>{\raggedright\arraybackslash}m{1.8cm}
>{\raggedright\arraybackslash}m{1.8cm}
>{\raggedright\arraybackslash}m{5.0cm}
>{\raggedright\arraybackslash}m{7.5cm}
@{}}
\toprule
\textbf{Range} & \textbf{Entry type} & \textbf{OS} & \textbf{ATT\&CK techniques} & \textbf{Attack chain} \\
\midrule
SSH-Miner & Baseline Weak & CentOS 7 &
T1110, T1021.004, T1059.004, T1552, T1053.003, T1546.004, T1543.002, T1548.003, T1496 &
SSH brute force $\to$ mining and persistence (crontab/bashrc/systemd) \\
\midrule
Shiro-Fastjson & Known CVE & CentOS 8 &
T1190, T1059.004, T1505.003, T1548.003, T1098.004, T1053.003, T1543.002, T1496 &
Shiro default key $+$ Fastjson $\to$ webshell $\to$ privilege escalation $\to$ mining \\
\midrule
Log4j-RCE & Known CVE & CentOS 8 &
T1190, T1059.006, T1505.003, T1548.003, T1098.004, T1053.003, T1546.004, T1543.002, T1556, T1070, T1496 &
Log4j RCE $\to$ webshell $+$ persistence \\
\midrule
Docker-Escape & Baseline Weak & Ubuntu 20.04 &
T1190, T1611, T1059.004, T1098.004, T1053.003, T1546.004, T1543.002, T1071.001, T1496 &
Exposed Docker API $\to$ container escape $\to$ host takeover \\
\midrule
Redis-RCE & Baseline Weak & Ubuntu 22.04 &
T1190, T1021.004, T1059.004, T1552, T1098.004, T1053.003, T1546.004, T1496 &
Unauthenticated Redis $\to$ SSH-key write $\to$ mining \\
\midrule
Jenkins-RCE & Business code & Ubuntu 22.04 &
T1190, T1059.004, T1548.001, T1548.003, T1574.006, T1053.003, T1071.001, T1496 &
Jenkins Script Console $\to$ command execution $\to$ mining \\
\midrule
Next.js-RCE & Known CVE & Ubuntu 22.04 &
T1190, T1059.007, T1505.003, T1548.001, T1574.006, T1098.004, T1053.003, T1546.004, T1543.002, T1496 &
Next.js CVE $\to$ SUID privilege escalation $\to$ \texttt{LD\_PRELOAD} rootkit \\
\midrule
NPM-Worm & Supply chain & Ubuntu 22.04 &
T1195.002, T1059.007, T1059.006, T1552, T1098.004, T1053.003, T1543.002, T1071.001, T1496 &
Malicious npm package $\to$ worm propagation $+$ credential theft \\
\midrule
ASP.NET-\newline ViewState & Known CVE & Windows Server &
T1190, T1505.003, T1552, T1071.001 &
ASP.NET ViewState $\to$ RCE $\to$ MSSQL backdoor $+$ WMI persistence \\
\midrule
RDP-\newline Service-Abuse & Baseline Weak & Windows Server &
T1110, T1552, T1071.001 &
Exposed RDP $\to$ password spraying $\to$ weak service DACL abuse $\to$ service restart for SYSTEM payload $+$ disguised service \& scheduled task $+$ credential dump and C2 \\
\bottomrule
\end{tabular}
}
\end{table*}

\subsubsection{Design Principles}
\label{sec:design}
As described above, each task is centered on a cyber range that the agent should investigate.
To reproduce the operational properties of real incidents, we build each range as three decoupled layers. The \textbf{\textit{blueprint}} layer specifies the evaluation purpose of the scenario, including the targeted stack, the attack chain annotated with MITRE ATT\&CK technique IDs, and the capability items the scenario aims to cover.
The \textbf{\textit{instance}} layer realizes the blueprint on a victim host under three constraints: (i) \textit{real vulnerabilities} in the form of disclosed CVEs or configuration flaws rather than synthetic bugs, (ii) \textit{real attacks} delivered through network protocols rather than directly placing malicious files, and (iii) \textit{real intrusion traces} produced naturally by the attack. The \textbf{\textit{checklist}} layer converts the resulting artifacts into checkpoints, each with a pass/fail criterion, an evidence source, and a capability mapping.

\subsubsection{Construction Pipeline.}

\begin{figure}[t]
\centering
\includegraphics[width=1.0\textwidth]{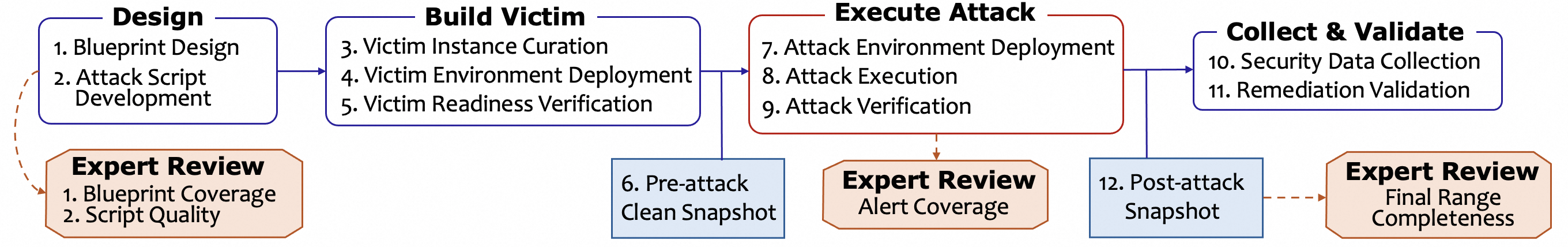}
\caption{Construction Pipeline of Cyber Range. The construction pipeline comprises 12 stages, with human experts involved in key steps to audit and validate the results.}
\label{fig:flow}
\end{figure}

The above three layers are realized by a 12-stage construction pipeline shown in Figure~\ref{fig:flow}: blueprint design, attack script development, victim instance curation, victim environment deployment, victim readiness verification, pre-attack clean snapshot, attack environment deployment, attack execution, attack verification, security data collection, remediation validation, and post-attack snapshot.
Among these stages, human experts are involved in auditing the plausibility of the attack chain and capability coverage of the blueprint, reviewing the code quality and attack authenticity of the attack scripts, verifying that alert coverage meets the standard during data collection, and checking the completeness of the final range. To distinguish alert-driven shallow analysis from deep forensic capability in each range, we include 30--40\% of attack actions that can trigger security-product alerts, while the remaining 60--70\% leave silent file modification and configuration artifacts. The alerts provide anchor points to begin reasoning, while the hidden evidence on disk forces investigation and discovery. These analytics are collected through the security platform API after the attack and provided as static files.

\subsection{Capability Taxonomy}
\label{sec:cap}
To benchmark agent models' incident response capability, we organize the required capabilities to investigate and remediate a compromised host into a five-dimensional Capability (CAP) taxonomy. 
\begin{itemize}
    \item \textbf{Intrusion Entity (ENT)} denotes the ability to locate and handle the artifacts left by an intrusion, spanning malicious processes, malicious or tampered files, and malicious network entities.
    \item \textbf{Persistence Mechanism (PER)} indicates the ability to recover the mechanisms an attacker installs to survive reboot and cleanup, including scheduled tasks, services, shell and environment initialization, account and privilege backdoors, web-component tampering, database backdoors, and kernel- or loader-level techniques.
    \item \textbf{Baseline Risk (BAS)} is the ability to examine host configuration against security baselines, covering SSH hardening, database and middleware access control, credential safety, privilege and service-permission auditing, container and orchestration configuration, web-service configuration, and cloud-resource configuration.
    \item \textbf{Vulnerability Risk (VUL)} represents the ability to confirm the weaknesses on the host, spanning Java component vulnerabilities, web-application code vulnerabilities, container-escape vulnerabilities, and front-end framework vulnerabilities.
    \item \textbf{Investigation \& Response Quality (Q)} reflects the ability to produce a thorough, well-evidenced, and trustworthy investigation, covering entry-point localization, attack-chain reconstruction, attacker-information extraction, honesty, investigation comprehensiveness, cross-language and cross-service tracing, remediation completeness, and business-impact assessment.
\end{itemize}
In total, the taxonomy comprises 52 items across five dimensions as shown in Table~\ref{tab:cap-taxonomy}, including 12 in ENT, 21 in PER, 7 in BAS, 4 in VUL, and 8 in Q. These capability items represent the benchmark target and form the foundation to build compromised ranges.

\begin{table}[t]
\centering
\caption{The \benchname{} evaluation taxonomy with 52 capability items across five CAP dimensions.}
\label{tab:cap-taxonomy}
\renewcommand{\cellalign}{tl}
\resizebox{0.9\textwidth}{!}{
\begin{tabular}{@{}
>{\raggedright\arraybackslash}m{3cm}
>{\raggedright\arraybackslash}m{12.5cm}
@{}}
\toprule
\textbf{Dimension} & \textbf{Capability Items} \\
\midrule
\makecell[l]{Intrusion\\Entity} & \textbf{ENT-F01}~Webshell file; \textbf{ENT-F02}~Linux malware file; \textbf{ENT-F03}~Malicious SO / kernel module; \textbf{ENT-F04}~Residual data file; \textbf{ENT-F05}~Tampered-file restoration; \textbf{ENT-F06}~Windows malware file; \textbf{ENT-N01}~Attacker-IP inbound block; \textbf{ENT-N02}~Malicious-IP outbound block; \textbf{ENT-N03}~Malicious-domain block; \textbf{ENT-P01}~Mining process; \textbf{ENT-P02}~C2 beacon process; \textbf{ENT-P03}~Windows malicious process \\
\midrule
\makecell[l]{Persistence\\Mechanism} & \textbf{PER-A01}~authorized\_keys tampering; \textbf{PER-A02}~sudoers implant; \textbf{PER-A03}~Rogue local account; \textbf{PER-D01}~MySQL backdoor; \textbf{PER-D02}~MSSQL backdoor; \textbf{PER-E01}~profile.d / udev rule; \textbf{PER-E02}~WMI event subscription; \textbf{PER-H01}~ld.so.preload hijack; \textbf{PER-H02}~DLL hijacking; \textbf{PER-I01}~Shell init (bashrc/profile); \textbf{PER-I02}~/etc/environment injection; \textbf{PER-M01}~SUID/SGID backdoor; \textbf{PER-S01}~Cron task; \textbf{PER-S02}~systemd timer; \textbf{PER-S03}~at job; \textbf{PER-S04}~Windows scheduled task; \textbf{PER-V01}~systemd service; \textbf{PER-V02}~init.d / rc.local; \textbf{PER-V03}~Windows service; \textbf{PER-W01}~Nginx config tampering; \textbf{PER-W02}~App-container config tampering \\
\midrule
\makecell[l]{Baseline\\Risk} & \textbf{BAS-01}~SSH hardening; \textbf{BAS-02}~DB / middleware access control; \textbf{BAS-03}~Credential safety; \textbf{BAS-04}~Privilege / service audit; \textbf{BAS-05}~Container / orchestration config; \textbf{BAS-06}~Web-service config; \textbf{BAS-07}~Cloud-resource config \\
\midrule
\makecell[l]{Vulnerability\\Risk} & \textbf{VUL-01}~Java component vuln; \textbf{VUL-02}~Web-app code vuln; \textbf{VUL-03}~Container-escape vuln; \textbf{VUL-04}~Node.js / front-end framework vuln \\
\midrule
Investigation \& Response Quality & \textbf{Q-01}~Entry-point localization; \textbf{Q-02}~Attack-chain reconstruction; \textbf{Q-03}~Attacker-info extraction; \textbf{Q-04}~Honesty \& confidence calibration; \textbf{Q-05}~Investigation thoroughness; \textbf{Q-06}~Cross-language/service tracing; \textbf{Q-07}~Remediation-verification completeness; \textbf{Q-08}~Business-impact assessment \\
\bottomrule
\end{tabular}
}
\end{table}

\subsection{Checkpoint Decomposition}
\label{sec:eval}
To evaluate the capabilities above, we decompose them into fine-grained checkpoints within each range to measure whether the corresponding capability is fulfilled in the specific scenario.
A checkpoint (CHK) is the smallest scoring unit, specifying the evidence required to discover a finding in a specific range. Each checkpoint matches one or more items in the CAP taxonomy, and multiple checkpoints may be assigned to the same CAP. The capability taxonomy is operationalized through 280 checkpoints in total, as detailed in Table~\ref{tab:cap-coverage}.
We first define \textbf{CHK-score}, which grades each CHK along two axes, detection and planning. 
Detection measures whether the agent correctly uncovers problems on the compromised host, including intrusion entities, baseline risks, and vulnerability risks.
Planning validates whether the agent proposes a correct and complete remediation plan that resolves the existing problems.
The detection axis has a total score of 3, comprising discovery of the problem, sufficiency of the supporting evidence, and correctness of attribution to the root cause. The planning axis has a maximum score of 2, comprising correctness of the proposed remediation step and completeness of the plan to fully resolve the attack artifacts. The normalized score of a single checkpoint on axis $a$ is defined as the CHK-score, which represents the agent model's performance at the checkpoint level:
\begin{equation}
\text{CHK-score}_c^{a} = \frac{s_c^{a}}{M^{a}} \in [0, 1],
\quad a \in \{\text{det}, \text{plan}\}
\label{eq:chk-score}
\end{equation}
where $M^{\text{det}} = 3$ and $M^{\text{plan}} = 2$ are the maximum scores of the two axes, and $s_c^{a} \in \{0, 1, \dots, M^{a}\}$ is the score assigned to checkpoint $c$ on axis $a$.

Since each checkpoint is anchored to a capability item, checkpoint results from different ranges become comparable. We aggregate the checkpoints mapped to the same capability item into a capability-level score, which we define as the \textbf{CAP-score}. The CAP-score on a given axis is the ratio of the achieved scores to the maximum scores over all checkpoints mapped to the corresponding capability, computed as:
\begin{equation}
\text{CAP-score}^{a} =
\frac{1}{\left| \mathcal{C}^{a}_{\text{CAP}} \right|}
\sum_{c \in \mathcal{C}^{a}_{\text{CAP}}} \text{CHK-score}_c^{a}
\times 100\%
\label{eq:cap-score}
\end{equation}
where $\mathcal{C}^{a}_{\text{CAP}}$ is the set of checkpoints mapped to the capability item.

\section{Experiments}
\subsection{Experimental Settings}
We evaluate 23 large language models spanning 8 model series and their successive versions, with the full list reported in Table~\ref{tab:main-noskill}. 
We include multiple versions of the same series to track how incident response capability evolves across releases. 
All models are driven by the same agent harness, OpenCode~\citep{opencode2026}, and receive identical task inputs (\S\ref{sec:task}), instruction prompt (see Appendix~\ref{sec:task-prompt}), and tool set to ensure a fair comparison. Each model is accessed using default parameters with reasoning enabled.

\subsection{Evaluation Metrics}
Following the LLM-as-a-Judge paradigm, we score each generated report against the range 
checklist defined in \S\ref{sec:eval}. To reduce single-judge bias, we adopt three strong proprietary models, Claude Opus 4.7~\citep{anthropic_claude_opus_4_7_2026}, Gemini 3.1 Pro~\citep{gemini_team_gemini_3_1_pro_2026}, and GPT-5.4 Pro~\citep{openai_gpt_5_4_2026}, as independent judges. Each judge runs in the same OpenCode harness, receives the instruction prompt (see Appendix~\ref{sec:eval-prompt}) and the checkpoint list with criteria, and uses a bash tool set to read the reports produced by the assessed model.
Each judge then assigns a score for every checkpoint, and the $\text{CHK-score}_c^{a}$ (Eq.~\ref{eq:chk-score}) for each checkpoint is averaged over the three judges. We then aggregate the checkpoint-level CHK-scores to the range level. For each range $r$ and axis $a \in \{\text{det}, \text{plan}\}$, the range-level $\text{CHK-score}_{r}^{a}$ is the mean value of the achieved $\text{CHK-score}_c^{a}$ over the range's checkpoints:
\begin{equation}
  \text{CHK-score}_{r}^{a} =
  \frac{\displaystyle\sum_{c \in \mathcal{C}_{r}^a} \text{CHK-score}_c^{a}}
  {\left| \mathcal{C}_{r}^a \right|}
  \times 100\%,
  \label{eq:range-score}
\end{equation}
where $\mathcal{C}_{r}^a$ is the set of checkpoints in range $r$ for axis $a$. We report the range-level CHK-score (Eq.~\ref{eq:range-score}) for range-specific analysis, and the CAP-score (Eq.~\ref{eq:cap-score}) for range-aggregated analysis.

\begin{table*}[t]
\centering
\caption{The range-level $\text{CHK-score}_{r}^{a}$ of Detection and Planning for each range. ``Det'' denotes $\text{CHK-score}_{r}^{det}$, the percentage of achieved detection scores over all its checkpoints, averaged across three LLM judges; ``Plan'' denotes the corresponding $\text{CHK-score}_{r}^{plan}$ on the planning dimension.}
\label{tab:main-noskill}
\resizebox{1.0\textwidth}{!}{
\setlength{\tabcolsep}{4pt}
\begin{tabular}{lcccccccccccccccccccccc}
\toprule
\multirow{2}{*}{\textbf{Model}} & \multicolumn{2}{c}{\textbf{SSH-Miner}} & \multicolumn{2}{c}{\textbf{Redis-RCE}} & \multicolumn{2}{c}{\textbf{Docker-Escape}} & \multicolumn{2}{c}{\textbf{Jenkins-RCE}} & \multicolumn{2}{c}{\textbf{Shiro-Fastjson}} & \multicolumn{2}{c}{\textbf{Log4j-RCE}} & \multicolumn{2}{c}{\textbf{Next.js-RCE}} & \multicolumn{2}{c}{\textbf{NPM-Worm}} & \multicolumn{2}{c}{\textbf{\makecell{ASP.NET-\\ViewState}}} & \multicolumn{2}{c}{\textbf{\makecell{RDP-Ser-\\vice-Abuse}}} & \multicolumn{2}{c}{\textbf{Overall}} \\
\cmidrule(lr){2-3} \cmidrule(lr){4-5} \cmidrule(lr){6-7} \cmidrule(lr){8-9} \cmidrule(lr){10-11} \cmidrule(lr){12-13} \cmidrule(lr){14-15} \cmidrule(lr){16-17} \cmidrule(lr){18-19} \cmidrule(lr){20-21} \cmidrule(lr){22-23}
 & Det & Plan & Det & Plan & Det & Plan & Det & Plan & Det & Plan & Det & Plan & Det & Plan & Det & Plan & Det & Plan & Det & Plan & Det & Plan \\
\midrule
% Claude Opus 4.8 \\
Claude Opus 4.7$^\dagger$ & \textbf{77.8} & \textbf{83.3} & 82.4 & 69.6 & 90.8 & 75.5 & \textbf{78.9} & \textbf{70.2} & 71.6 & 52.5 & 88.4 & \underline{78.7} & 77.8 & 60.1 & \NA & \NA & 83.3 & 37.9 & 59.9 & \underline{63.9} & \textbf{79.0} & \textbf{65.7} \\
Claude Opus 4.6 & 75.3 & 48.0 & 79.6 & 60.8 & \textbf{92.1} & 62.3 & 66.7 & 60.6 & 66.7 & 43.2 & 88.4 & 68.5 & 76.5 & 64.9 & \underline{81.5} & 60.9 & 77.8 & \textbf{49.3} & 77.8 & 61.1 & \underline{78.2} & 58.0 \\
Claude Opus 4.5 & 72.5 & 61.7 & 70.4 & 55.9 & 79.2 & 57.0 & 71.9 & 62.3 & 69.6 & 55.2 & 85.9 & 71.3 & 66.7 & 60.8 & 69.8 & \underline{65.4} & 56.1 & 27.3 & 53.1 & 47.2 & 69.5 & 56.4 \\
Claude Sonnet 4.6 & 75.8 & 44.1 & 68.5 & 61.8 & 83.3 & 63.2 & 67.2 & 59.7 & 76.8 & 60.9 & 87.9 & \textbf{81.5} & 71.6 & \textbf{70.2} & 68.7 & 53.8 & 58.1 & 29.5 & \textbf{89.5} & \textbf{66.7} & 74.7 & 59.1 \\
Claude Sonnet 4.5 & \textbf{77.8} & \underline{76.5} & 86.1 & \underline{75.6} & 84.2 & 78.1 & 62.6 & 55.3 & 73.8 & \textbf{71.9} & 78.3 & 39.8 & 65.7 & \underline{69.1} & 41.0 & 49.4 & 73.8 & \underline{41.7} & 53.7 & 54.7 & 69.7 & \underline{61.2} \\
GPT-5.5 & 67.6 & 27.4 & 78.2 & 49.0 & 68.0 & 35.1 & 73.7 & 33.4 & 63.9 & 27.0 & \textbf{96.5} & 33.3 & 58.5 & 40.4 & 55.8 & 41.0 & 81.3 & 19.7 & 63.6 & 53.7 & 70.7 & 36.0 \\
GPT-5.4 Pro & 65.2 & 27.4 & 75.9 & 47.1 & 66.7 & 41.2 & 69.0 & 19.4 & 80.5 & 63.0 & \underline{91.9} & 42.6 & 65.7 & 44.7 & 65.5 & 37.2 & \textbf{85.9} & 31.9 & 58.6 & 52.8 & 72.5 & 40.7 \\
GPT-5.4 & 57.5 & 25.5 & 57.4 & 34.3 & 69.4 & 20.2 & 55.0 & 29.0 & 56.3 & 38.5 & 85.8 & 41.7 & 51.0 & 36.9 & 57.8 & 32.7 & 67.2 & 26.5 & 49.4 & 47.2 & 60.7 & 33.2 \\
GPT-5.2 Pro & 61.3 & 21.5 & 59.7 & 43.1 & 54.2 & 34.2 & 73.1 & 39.4 & 65.9 & 36.4 & 88.4 & 44.5 & 71.9 & 44.0 & 51.3 & 34.6 & 56.6 & 22.0 & \underline{79.6} & 50.9 & 66.2 & 37.1 \\
% Gemini 3.5 Flash \\
Gemini 3.1 Pro & 66.7 & 38.3 & 55.1 & 38.2 & 60.7 & 43.0 & 60.2 & 39.5 & 43.0 & 28.7 & 72.2 & 32.4 & 46.4 & 32.1 & 43.0 & 11.5 & 47.5 & 22.7 & 53.1 & 28.7 & 54.8 & 31.5 \\
Gemini 3 Flash & 58.5 & 25.5 & 62.5 & 34.3 & 57.4 & 39.5 & 56.1 & 32.5 & 50.1 & 27.6 & 80.8 & 51.9 & 57.5 & 31.5 & 41.9 & 26.3 & 49.0 & 17.4 & 60.5 & 42.6 & 57.4 & 32.9 \\
% GLM-5.2 \\
GLM-5.1 & 75.4 & 58.8 & 80.1 & 62.8 & \underline{91.2} & \textbf{86.0} & 71.4 & 53.5 & \textbf{82.5} & 63.5 & 90.4 & 64.8 & \textbf{93.1} & 61.9 & 75.8 & 60.9 & 61.6 & 27.3 & 42.0 & 52.8 & 76.3 & 59.2 \\
GLM-5 & 75.8 & 56.9 & 60.2 & 49.0 & 75.5 & 51.8 & 50.3 & 34.3 & 49.4 & 42.7 & 88.4 & 64.8 & 76.5 & 62.4 & 50.4 & 31.4 & 67.2 & 26.5 & 54.9 & 38.9 & 64.9 & 45.9 \\
DeepSeek V4 Pro & \underline{76.3} & 61.8 & 67.1 & 56.9 & 75.9 & 50.0 & 63.2 & 59.7 & 73.1 & 55.2 & 87.9 & 49.1 & 75.2 & 57.1 & 80.1 & 64.1 & 71.2 & 32.6 & 62.3 & 47.2 & 73.2 & 53.4 \\
DeepSeek V3.2 & 60.9 & 53.0 & 65.7 & 48.1 & 65.3 & 37.8 & 50.9 & 47.4 & 68.9 & 47.9 & 80.3 & 68.6 & 58.5 & 40.5 & 36.2 & 21.1 & 41.9 & 16.7 & 63.6 & 54.6 & 59.2 & 43.6 \\
Qwen3.7 Max & 75.8 & 50.0 & \textbf{87.5} & \textbf{76.5} & 79.2 & 64.0 & 72.5 & 51.7 & 70.4 & 46.4 & 90.4 & 60.1 & \underline{82.4} & 63.1 & \textbf{84.9} & 48.7 & \underline{84.9} & 40.9 & 51.8 & 41.7 & 78.0 & 54.3 \\
Qwen3.7 Plus & 70.5 & 36.2 & \underline{86.6} & 71.6 & 75.9 & 57.0 & 71.9 & \underline{69.3} & \underline{82.0} & \underline{70.8} & 88.9 & 75.0 & 70.9 & 61.9 & 73.5 & \textbf{65.5} & 81.8 & 35.6 & 53.7 & 45.3 & 75.6 & 58.8 \\
Qwen3.6 Plus & 70.5 & 54.9 & 66.2 & 52.9 & 77.3 & 64.1 & 64.3 & 32.5 & 60.0 & 44.3 & 84.8 & 54.6 & 73.2 & 67.9 & 65.0 & 38.4 & 68.2 & 31.8 & 54.9 & 41.7 & 68.4 & 48.3 \\
Qwen3.5 Plus & 65.7 & 51.0 & 66.2 & 58.8 & 61.6 & 46.5 & 67.8 & 49.2 & 43.5 & 32.3 & 82.8 & 57.4 & 64.7 & 55.4 & 61.8 & 37.2 & 70.2 & 21.2 & 51.2 & 31.5 & 63.5 & 44.0 \\
% Kimi K3 \\
Kimi K2.6 & 60.4 & 34.3 & 63.0 & 27.4 & 88.4 & \underline{79.0} & \underline{76.6} & 53.5 & 53.1 & 40.6 & 84.9 & 55.6 & 63.4 & 31.5 & 52.4 & 36.5 & 64.7 & 33.3 & 51.3 & 38.0 & 65.8 & 43.0 \\
Kimi K2.5 & 71.5 & 61.8 & 72.7 & 64.8 & 75.9 & 59.7 & 55.6 & 43.0 & 51.6 & 46.4 & 88.9 & 69.4 & 63.1 & 53.6 & 43.3 & 35.9 & 46.0 & 23.5 & 52.5 & 49.1 & 62.1 & 50.7 \\
MiniMax M2.7 & 69.1 & 36.3 & 62.0 & 51.9 & 74.5 & 56.2 & 63.7 & 49.1 & 55.1 & 41.6 & 76.8 & 38.0 & 66.4 & 45.2 & 47.6 & 32.0 & 42.9 & 14.4 & 42.6 & 35.2 & 60.1 & 40.0 \\
MiniMax M2.5 & 63.3 & 38.3 & 56.5 & 30.4 & 61.6 & 48.2 & 47.4 & 32.5 & 38.5 & 22.9 & 76.3 & 45.4 & 57.8 & 42.3 & 31.9 & 23.1 & 46.5 & 12.9 & 43.8 & 33.3 & 52.4 & 32.9 \\
\bottomrule
\end{tabular}}
\par\smallskip
\begin{minipage}{\textwidth}
\footnotesize\raggedright
$\dagger$ Claude Opus 4.7 returned a safety refusal on every NPM-Worm attempt and produced no report; its overall score is therefore computed as the average over the remaining nine ranges.
\end{minipage}
\end{table*}

\begin{figure}[t]
\centering
\includegraphics[width=1.0\textwidth]{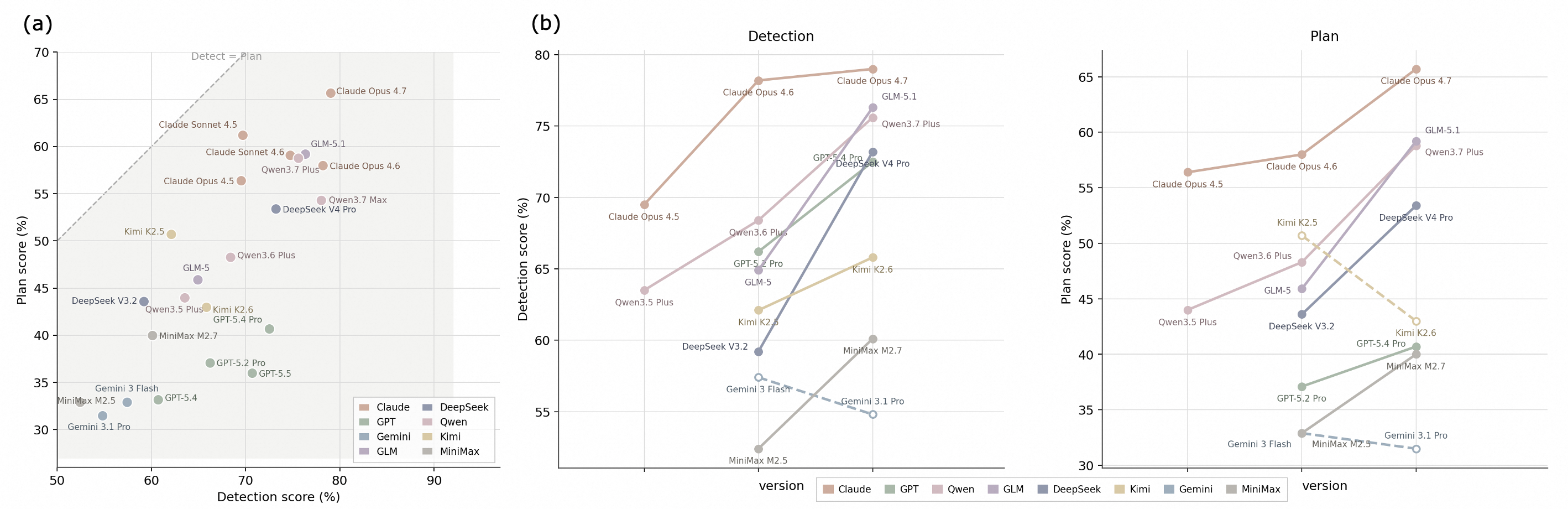}
\caption{\textbf{(a) Detection consistently outperforms Planning across all models.} The x-axis denotes the Detection score and the y-axis presents the Planning score. The dashed line marks ``Detection=Planning''. All models fall below the diagonal, demonstrating that detection always exceeds planning. \textbf{(b) Version evolution within each model series.} Most model families improve across versions, but the progress is not consistent.}
\label{fig:exp}
\end{figure}

\subsection{Range-Level Analysis across Attack Scenarios}

Through an in-depth analysis of the agent models' performance across different attack scenarios as demonstrated in Table~\ref{tab:main-noskill}, we derive five primary research findings.

\textbf{Finding 1: Detection consistently outperforms planning, which is the bottleneck across all ranges.}
All the models achieve a higher range-level CHK-score on Detection than on Planning, as demonstrated in Figure~\ref{fig:exp} (a). For example, GPT-5.5 scores 70.7\% on detection but only 36\% on planning, while Claude Sonnet 4.5 presents the narrowest gap of 8.5\%. This implies that for a large portion of checkpoints the agent models can successfully detect the problems from the disk snapshot, but fail to provide a comprehensive approach to fixing them. This gap is largely attributed to incomplete fixes instead of wrong actions. The agent models often advise correct commands initially, such as terminating a malicious process or removing a dropped file. However, in the follow-up steps, they fail to carry out the remaining remediation, such as rotating the leaked credentials, blocking the outbound channel, or confirming that the business service stays healthy after the cleanup. This suggests that proficiency in detection cannot ensure success in handling an incident thoroughly.
In conclusion, current agents are better at finding attack traces or vulnerabilities than producing correct, complete, and verified remediation suggestions.

\textbf{Finding 2: Model performance varies widely, with only a few models leading consistently across attack scenarios and demonstrating superior forensic and incident-response capabilities.}
Claude Opus 4.7 achieves the highest average range-level CHK-score across ranges (79.0\% for detection and 65.7\% for planning), followed by Claude Opus 4.6 (78.2\%/58.0\%), GLM-5.1 (76.3\%/59.2\%), and Qwen3.7 Plus (75.6\%/58.8\%), whereas Gemini 3.1 Pro, MiniMax M2.5, and Gemini 3 Flash present the lowest results.
Open models such as GLM and DeepSeek outperform the GPT and Gemini series, indicating that proprietary models do not guarantee stronger incident-response capabilities. Overall, performance on this task is highly uneven across models, making model selection an important factor in security incident-response applications.

\textbf{Finding 3: Performance generally improves as models evolve across versions, but the improvement is not universal.}
As shown in Figure~\ref{fig:exp} (b), several model families improve steadily across successive releases. For instance, the CHK-scores of the Claude Opus series increase across three releases, from 69.5\%/56.4\% for Opus 4.5 to 78.2\%/58.0\% for Opus 4.6 and 79.0\%/65.7\% for Opus 4.7. Similarly, GLM-5.1 surpasses GLM-5 by 11.4\%/13.3\% (76.3\%/59.2\% \textit{vs.} 64.9\%/45.9\%), DeepSeek V4 Pro improves over DeepSeek V3.2 by 14.0\%/9.8\% (73.2\%/53.4\% \textit{vs.} 59.2\%/43.6\%), and Qwen3.7 Plus outperforms Qwen3.6 Plus by 7.2\%/10.5\% (75.6\%/58.8\% \textit{vs.} 68.4\%/48.3\%). However, this trend is not universal. For example, Kimi K2.6 falls below K2.5 on the planning dimension (43.0\% \textit{vs.} 50.7\%), and Gemini 3.1 Pro also underperforms Gemini 3 Flash (54.8\%/31.5\% \textit{vs.} 57.4\%/32.9\%).

\textbf{Finding 4: Range difficulty varies, and model performance declines as the attack chain grows longer and broader. No model achieves complete detection and remediation on any single range.}
The overall CHK-score across models on each range reveals a clear difficulty stratification that aligns with the structural complexity of each scenario. Agent models generally show strong performance on Log4j-RCE, Docker-Escape, and Redis-RCE ranges with a single entry point, a linear attack chain, and common persistence exposed in alerts such as cron jobs or systemd services. Log4j-RCE is the easiest range because the included CVE is well-known and already familiar to the models.
Ranges like SSH-Miner, Next.js-RCE, and Jenkins-RCE additionally include a wider host baseline, multi-step privilege escalation, cross-service correlation and more diverse persistence, leading to a decrease in both detection and planning scores. The hardest ranges are Shiro-Fastjson, RDP-Service-Abuse, NPM-Worm, and ASP.NET-ViewState, which combine a broad attack surface, cross-service or cross-runtime correlation, and deliberately disguised persistence. The complex attack traces in these ranges make it difficult for agent models to uncover potential issues and especially challenging to propose comprehensive remediation methods.

\textbf{Finding 5: Agent models inherently struggle with proactive detection of silent intrusions and generating complete, verified remediation plans.}
Across all ranges, all the agent models are typically confronted with two primary weaknesses: failure to uncover silent intrusions beyond the exposed attack traces and to suggest thorough and verified remediation. This represents a common capability ceiling, with detailed evidence for each specific scenario provided in Appendix~\ref{sec:per-range-perform}.
On the detection side, the agent models can find the issues directly reflected by the exposed alerts, but fail to discover the hidden problems beyond the obvious trace. Silent intrusions that leave no running process or network footprint are frequently overlooked, and the agent models rarely scan the whole host proactively. This shows that they tend to respond to the given clues instead of actively investigating the system. 
On the planning side, the agent models usually propose the most obvious fix rather than completing the comprehensive remediation. The remediation plans often stop at removing the visible artifact and leave the secondary risks or undetected persistence mechanisms untouched. More importantly, the models rarely recommend verifying whether the proposed fixes truly work or whether the normal services still run after the cleanup, leaving the remediation partial and unverified.

\subsection{Capability-Level Analysis}

\begin{table}[t]
\centering
\caption{The $\text{CAP-score}^{a}$ of agent models and the traditional agentless scanning method across the five capability dimensions, namely Intrusion Entity (ENT), Persistence Mechanism (PER), Baseline Risk (BAS), Vulnerability Risk (VUL), and Investigation \& Response Quality (Q). For each dimension, ``Detect'' denotes $\text{CAP-score}^{det}$, the percentage of achieved detection scores over all checkpoints mapped to that capability dimension (following the mapping in Table~\ref{tab:cap-coverage}), averaged across three LLM judges; ``Plan'' denotes $\text{CAP-score}^{plan}$, the corresponding percentage on the planning axis.}
\label{tab:cap-main}
\resizebox{0.75\textwidth}{!}{
\setlength{\tabcolsep}{4pt}
\begin{tabular}{lcccccccccc}
\toprule
\multirow{2}{*}{\textbf{Model}} & \multicolumn{2}{c}{\textbf{ENT}} & \multicolumn{2}{c}{\textbf{PER}} & \multicolumn{2}{c}{\textbf{BAS}} & \multicolumn{2}{c}{\textbf{VUL}} & \multicolumn{2}{c}{\textbf{Q}} \\
\cmidrule(lr){2-3} \cmidrule(lr){4-5} \cmidrule(lr){6-7} \cmidrule(lr){8-9} \cmidrule(lr){10-11}
 & Detect & Plan & Detect & Plan & Detect & Plan & Detect & Plan & Detect & Plan \\
\midrule
\rowcolor{gray!30} Agentless & 43.6 & \NA & 2.1 & \NA & 20.7 & \NA & 50.0 & \NA & \NA & \NA \\
Claude Opus 4.7 & 85.4 & 61.9 & 67.8 & \underline{48.7} & 80.8 & \textbf{74.8} & 76.2 & \textbf{72.6} & \underline{73.6} & \underline{53.6} \\
Claude Opus 4.6 & \underline{86.0} & 58.1 & 72.1 & 45.8 & 78.6 & 67.6 & \underline{79.2} & 63.3 & 68.8 & 34.4 \\
Claude Opus 4.5 & 76.5 & 60.8 & 56.9 & 44.0 & 75.8 & 63.5 & 72.6 & 54.4 & 62.4 & 33.3 \\
Claude Sonnet 4.6 & 81.6 & \textbf{62.7} & 58.7 & 44.9 & \textbf{84.2} & 69.8 & \underline{79.2} & 67.8 & 66.7 & 34.4 \\
Claude Sonnet 4.5 & 72.8 & 58.4 & 56.1 & 44.1 & 66.2 & \underline{70.8} & 68.1 & 66.7 & 71.6 & \textbf{66.7} \\
GPT-5.5 & 78.4 & 31.1 & 71.3 & 28.8 & 63.6 & 46.8 & 61.5 & 40.0 & 61.4 & 24.4 \\
GPT-5.4 Pro & 78.3 & 38.3 & \underline{76.4} & 35.2 & 73.5 & 51.9 & 61.5 & 45.6 & 64.0 & 35.5 \\
GPT-5.4 & 69.9 & 29.2 & 47.6 & 24.4 & 61.8 & 46.4 & 53.3 & 37.8 & 55.2 & 15.5 \\
GPT-5.2 Pro & 73.8 & 36.0 & 67.6 & 39.0 & 62.4 & 44.2 & 42.2 & 17.8 & 62.9 & 27.8 \\
Gemini 3.1 Pro & 65.2 & 33.8 & 48.7 & 24.1 & 52.1 & 34.5 & 37.0 & 31.1 & 43.8 & 12.2 \\
Gemini 3 Flash & 66.5 & 35.3 & 47.8 & 23.5 & 58.5 & 36.2 & 62.2 & 43.3 & 45.4 & 12.2 \\
GLM-5.1 & 84.1 & 61.8 & 68.4 & 46.7 & 81.1 & 65.7 & 78.5 & \underline{71.1} & \textbf{75.5} & 41.1 \\
GLM-5 & 67.8 & 42.1 & 54.2 & 36.6 & 67.8 & 54.1 & 68.9 & 58.8 & 56.5 & 25.5 \\
DeepSeek V4 Pro & 78.1 & 51.7 & 69.0 & 45.8 & 76.1 & 59.4 & \textbf{86.7} & 62.2 & 66.2 & 42.2 \\
DeepSeek V3.2 & 69.1 & 42.9 & 47.2 & 30.1 & 57.7 & 50.0 & 69.6 & 54.5 & 54.8 & 36.7 \\
Qwen3.7 Max & 84.1 & 52.4 & \textbf{79.0} & 46.7 & 80.9 & 63.7 & 62.2 & 47.8 & 69.1 & 32.2 \\
Qwen3.7 Plus & \textbf{88.4} & \underline{62.3} & 66.7 & \textbf{49.4} & \underline{82.4} & 68.9 & 68.9 & 60.0 & 64.6 & 34.4 \\
Qwen3.6 Plus & 77.1 & 47.1 & 57.5 & 32.2 & 74.7 & 62.6 & 54.8 & 52.3 & 58.0 & 24.5 \\
Qwen3.5 Plus & 80.2 & 44.6 & 49.5 & 31.5 & 61.9 & 49.0 & 47.4 & 44.4 & 52.0 & 30.0 \\
Kimi K2.6 & 72.6 & 43.9 & 57.9 & 33.9 & 66.7 & 48.6 & 55.5 & 41.1 & 55.7 & 26.7 \\
Kimi K2.5 & 72.0 & 52.4 & 49.5 & 36.0 & 57.7 & 52.9 & 65.9 & 60.0 & 54.9 & 42.3 \\
MiniMax M2.7 & 66.0 & 34.1 & 43.7 & 30.3 & 68.6 & 51.2 & 63.7 & 38.9 & 52.8 & 32.3 \\
MiniMax M2.5 & 61.5 & 33.3 & 38.4 & 22.0 & 55.2 & 38.7 & 40.0 & 35.5 & 42.8 & 14.4 \\
\bottomrule
\end{tabular}}
\end{table}

Following our capability taxonomy described in \S~\ref{sec:cap}, we aggregate CHK-score into CAP-score using the mapping between range checkpoints and capability items (see Table~\ref{tab:cap-coverage}). The aggregated results are shown in Table~\ref{tab:cap-main}, from which we derive the following findings.

\begin{figure}[t]
\centering
\includegraphics[width=0.8\textwidth]{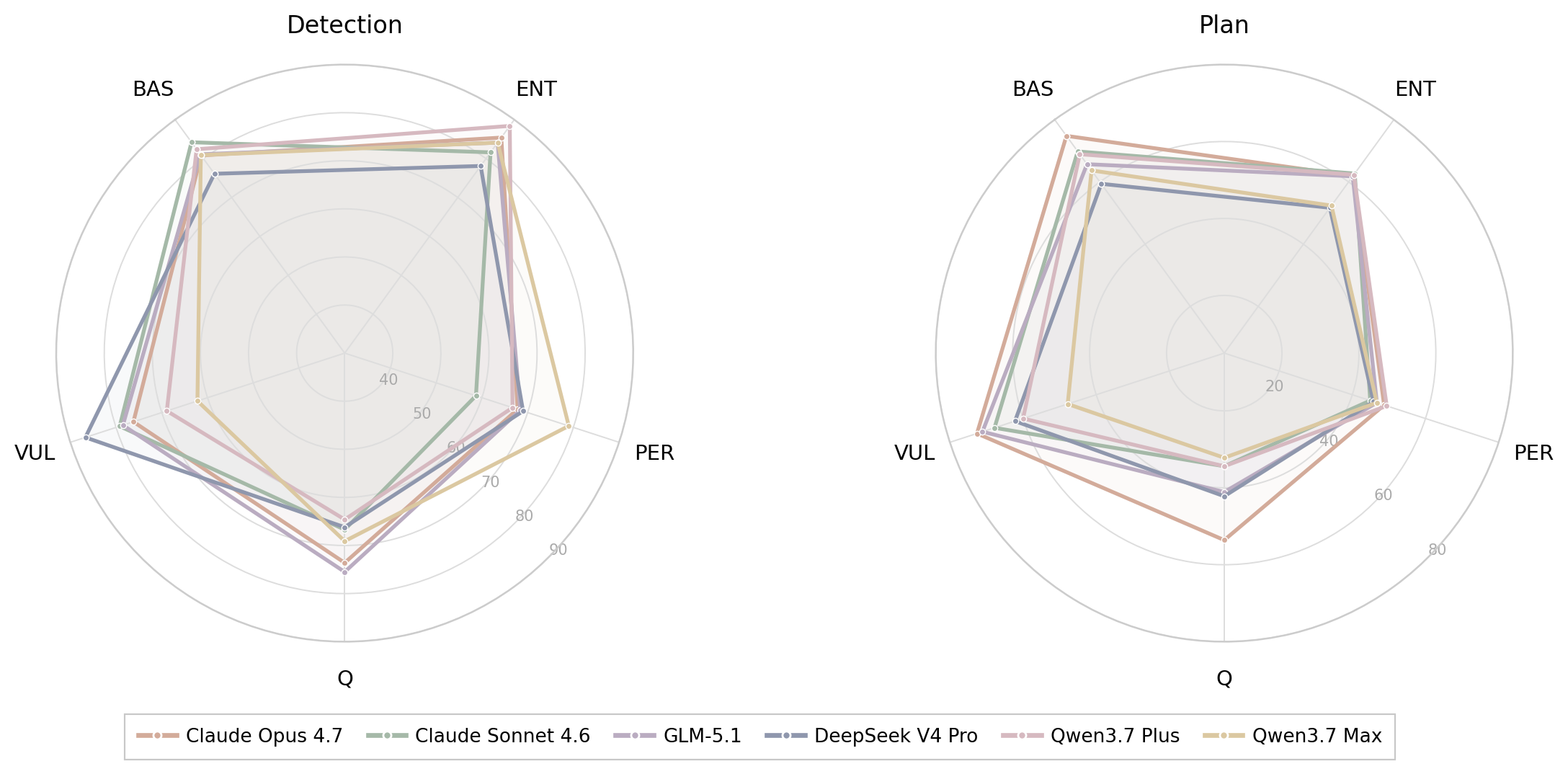}
\caption{The CAP-score for each capability (ENT, PER, BAS, VUL, Q) of the representative models. In detection, Qwen3.7 Plus is strongest at entity identification, Claude Sonnet 4.6 at baseline assessment, Qwen3.7 Max at persistence detection, DeepSeek V4 Pro at vulnerability verification, and GLM-5.1 at investigation quality. In planning, Claude Opus 4.7 shows the strongest capabilities on average.}
\label{fig:radar}
\end{figure}

\textbf{Finding 1: Models reliably identify intrusion entities but struggle to uncover persistence mechanisms.}
As observed in Table~\ref{tab:cap-main}, models are strongest at detecting Intrusion Entity (ENT), where most models exceed a 70\% CAP-score. Leading models such as Qwen3.7 Plus and Claude Opus 4.6 achieve 88.4\% and 86.0\%, respectively. The strength in ENT can be attributed to the nature of malicious entities, which are concrete objects that trigger visible alerts. Models can discover them through routine inspection and by tracing the alerted trail.
By contrast, detection collapses on Persistence Mechanism (PER), which has the lowest average detection CAP-score across models. This is because persistence mechanisms can be left in less obvious locations with no active signal, including cron jobs, systemd units, shell-init files, and platform-specific hooks. Uncovering them requires the model to enumerate the whole host rather than simply follow a visible trace. Thus, the ability to conduct a more proactive and systematic investigation represents one of the primary directions for improvement.

\textbf{Finding 2: Models remediate standardized risks effectively but perform weakly on intrusion entity and persistence mechanism cleanup.}
In terms of planning, models demonstrate superior CAP-scores when remediating Baseline Risk (BAS) or Vulnerability Risk (VUL), with Claude Opus 4.7 reaching 74.8\% on BAS and 72.6\% on VUL. These two dimensions are the easiest for remediation planning because their remediation is standardized, such as patching the affected component, disabling an exposed service, or rotating a default credential. Interestingly, models are strongest at detecting ENT but remediate it less successfully. This is because uncovering a malicious entity may require only a single observation, whereas cleaning it up involves a more complex, multi-step process. In particular, models' capability to address PER remains relatively weak in remediation planning because the low detection score implies that many persistence mechanisms are never discovered and therefore cannot be properly remediated.

\textbf{Finding 3: Investigation \& Response Quality is consistently among the weakest capabilities across models in both detection and planning, reflecting inadequate and unverified incident handling.}
The Investigation \& Response Quality dimension validates process-level completeness and soundness of the whole incident response procedure, covering entry point localization, attack-chain reconstruction, honesty and confidence calibration, investigation thoroughness, remediation verification completeness, and business impact assessment.
The CAP-scores on the quality dimension are low on both the detection and planning axes. Even the highest detection result reaches only 75.5\%, falling behind other dimensions, and almost all models fall below 50\% on planning except Claude Opus 4.7 and Claude Sonnet 4.5.
From the detection perspective, even when models detect a concrete artifact, they still fail to reconstruct the full attack chain because of the challenge of correlating evidence across services and time. In terms of planning, models are inclined to address the immediate problem rather than handle the incident thoroughly, rarely verifying the effectiveness and completeness of the proposed remediation.

\textbf{Finding 4: Different models excel at different capabilities, but none of them dominates across all five dimensions.}
Figure~\ref{fig:radar} shows that the best detection model differs for every single dimension. Qwen3.7 Plus is strongest at entity identification with 88.4\% on ENT, Qwen3.7 Max at persistence detection with 79.0\% on PER, Claude Sonnet 4.6 at baseline assessment with 84.2\% on BAS, DeepSeek V4 Pro at vulnerability verification with 86.7\% on VUL, and GLM-5.1 at investigation quality with 75.5\% on Q.
In terms of planning, Claude models achieve the best CAP-score in four of the five dimensions, namely Opus 4.7 on BAS (74.8\%) and VUL (72.6\%), Sonnet 4.6 on ENT (62.7\%), and Sonnet 4.5 on Q (66.7\%), while Qwen3.7 Plus reaches the highest planning score of 49.4\% on PER. This highlights that there is no universally best model and that model selection for incident response should be driven by the capabilities required for deployment rather than by a single overall ranking.

\textbf{Finding 5: Agents substantially outperform the traditional agentless detection baseline on average, and the gap is largest on detecting persistence mechanism.}
We also compare the agent models against an agentless baseline\footnote{\url{https://help.aliyun.com/zh/security-center/user-guide/use-the-agentless-detection-feature}} used in the traditional incident-response workflow, which is a production scanner that inspects a host's disk snapshot without installing any runtime agent. Given the same host disk image, the scanner performs detection purely by
static pattern matching driven by a predefined knowledge base, and reports four categories of findings after scanning. Concretely, it flags malicious samples by hash and signature
matching against a malware database, identifies software vulnerabilities by enumerating installed packages and matching their versions against the CVE database, detects baseline risks by checking configuration entries against a fixed set of hardening rules, and finds sensitive files by pattern-matching for exposed credentials and keys.
Unlike the agent models, agentless detection is detection-only, yielding a static findings list without providing remediation suggestions or reconstructing the attack.
We map the agentless-detection findings to our capability dimensions and observe that the scanner reaches 43.6\% on ENT and 50.0\% on VUL, but only 20.7\% on BAS and 2.1\% on PER. Agents outperform this traditional security scanner on average across the four comparable dimensions. This is because the scanner reports only isolated static artifacts and fails to analyze persistence mechanisms, reconstruct attack chains, or attribute entry points. In contrast, agents can actively probe the host and connect scattered artifacts across files into an intrusion chain, thereby uncovering persistence mechanisms hidden in the system. Furthermore, unlike the scanner, agent models provide root-cause analyses and remediation plans.

\begin{figure}[t]
\centering
\includegraphics[width=1.0\textwidth]{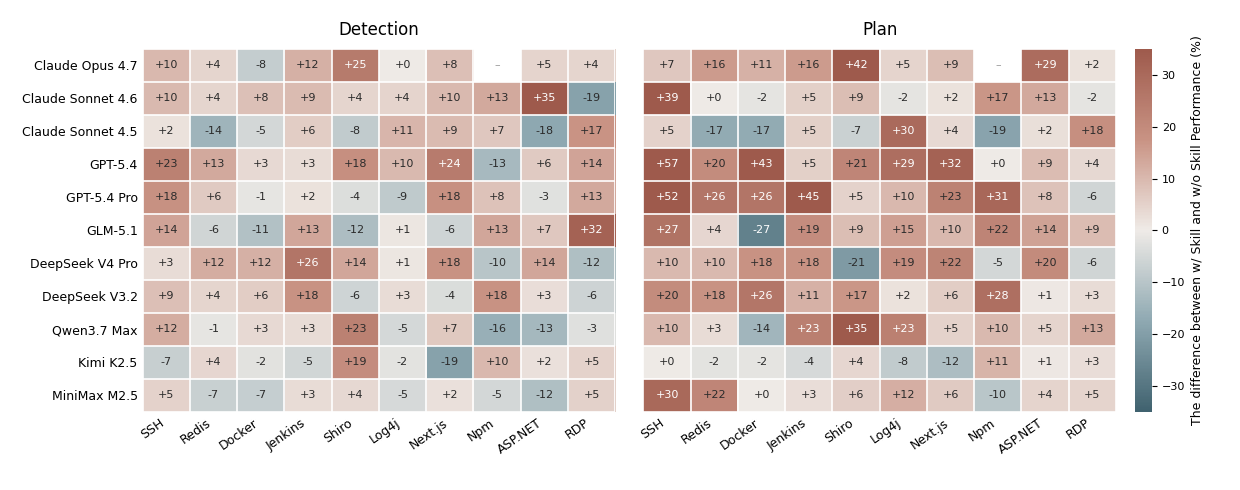}
\caption{Changes in the range-level Detection and Planning CHK-scores of representative models on each range after providing our designed skill. Each value denotes the difference between runs with and without the skill.}
\label{fig:skill-range}
\end{figure}

\begin{figure}[t]
\centering
\includegraphics[width=1.0\textwidth]{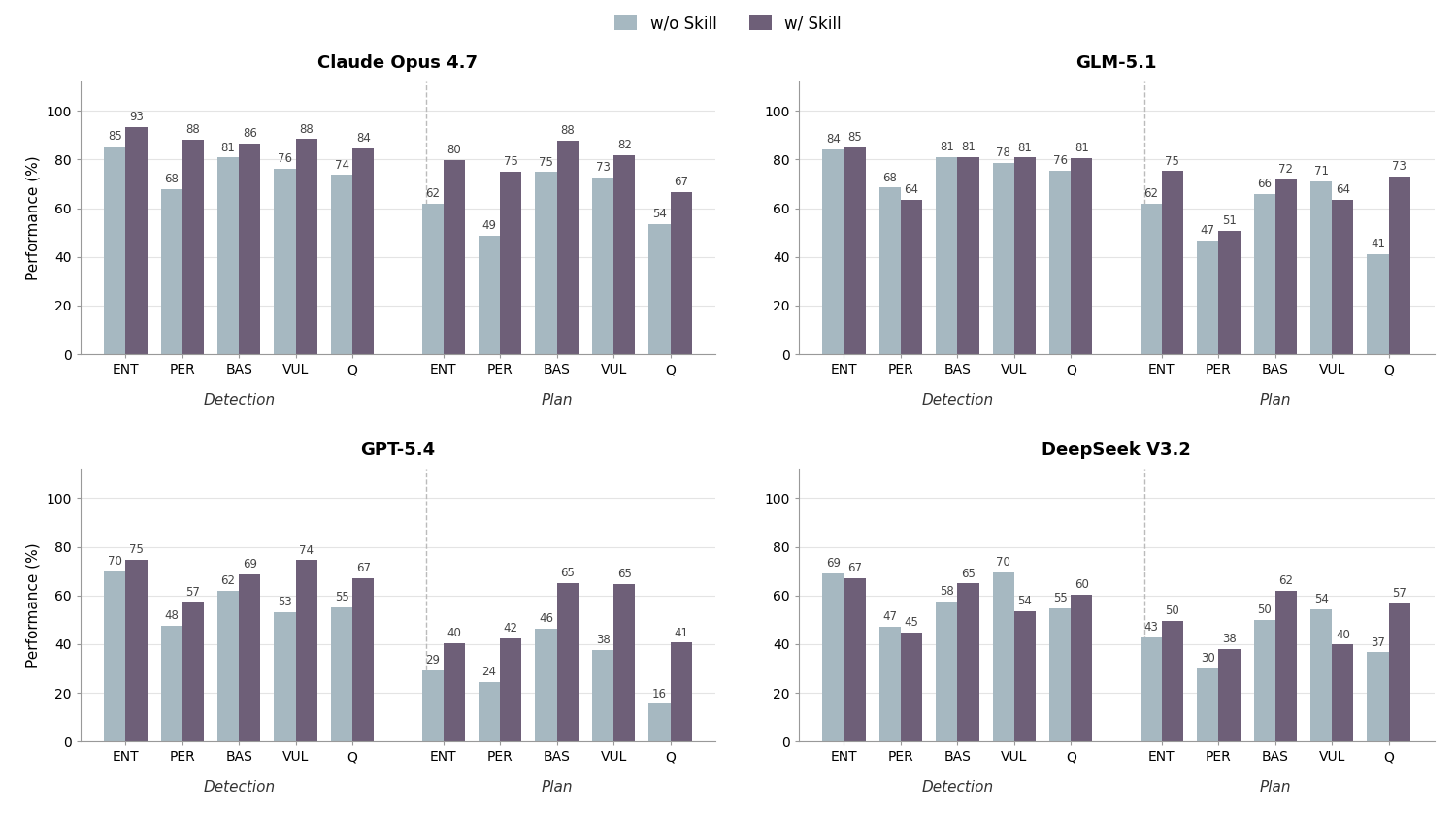}
\caption{The Detection and Planning CAP-scores across the five capability dimensions for four representative models with and without our designed skill.}
\label{fig:skill-cap}
\end{figure}

\subsection{Augmenting Agents with Procedural Priors}
A typical failure mode is procedural: agents fail to follow a systematic investigation routine.
For example, they follow the alerted trail instead of scanning the whole host, and the remediation plans often stop at the most obvious fix, which is reflected in their low scores on planning and on the Investigation
\& Response Quality dimension. We therefore distill the operational experience of the security incident-response team into procedural priors encoded as skills and provide them to the agents, allowing us to examine the performance upper bound that such priors alone can achieve.
Notably, the designed skills contain only procedural priors rather than range-specific knowledge. Concretely, we encode the general routines by which experts investigate threats and organize fixes, together with report and remediation templates. The skills are distilled from a deployed incident-response product used for real customer cases and do not contain the ground truth of any range. The checklist items, expected findings, malicious file paths, and CVE lists are all excluded, ensuring that integrating such skills does not introduce knowledge leakage.

As shown in Figure~\ref{fig:skill-range}, our designed skills improve the agent models' CHK-scores, particularly on the planning axis.
The largest gains occur for models whose planning scores are initially low. For example, GPT-5.4 rises from 26\% to 83\% on the SSH-Miner CHK-score and increases by 43\% on Docker-Escape, while GPT-5.4 Pro gains 45\% on Jenkins-RCE and 52\% on SSH-Miner.
From the capability perspective in Figure~\ref{fig:skill-cap}, our skill can address the blind spots of strong models. For example, although Claude Opus 4.7 performs relatively weakly on persistence, the designed skill improves the PER CAP-score from 68\% to 88\% in Detection and from 49\% to 75\% in Planning. Similarly, the skill benefits GLM-5.1 most on the Investigation \& Response Quality dimension.
However, the skill does not ensure improvement. GLM-5.1 degrades by 12\% in detection on Shiro-Fastjson and by 11\% on Docker-Escape. This is probably because such broad-scope ranges include attacker traces beyond the skill's enumerated categories, so the model stops at its boundary rather than continuing to explore and recover these long-tail traces.
Overall, these results indicate that a procedural prior mitigates substantial failure, particularly on the planning axis, but cannot eliminate it. Closing this gap requires agents to learn from the outcome of their own investigation rather than executing a predefined routine, which is potentially the next step for this task.

\subsection{LLM Judgment Analysis}
\begin{figure}[t]
\centering
\includegraphics[width=0.85\textwidth]{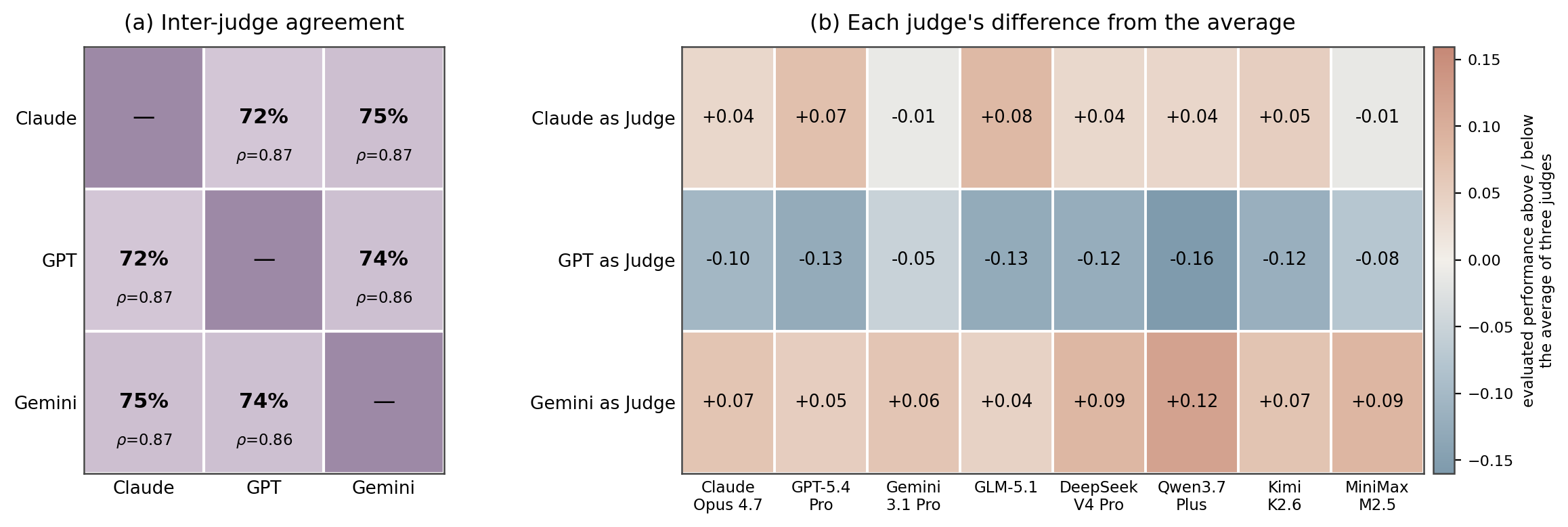}
\caption{\textbf{(a) Inter-Judge Agreement.} For each pair of judges, the value denotes the percentage of checkpoints on which both judges assign the same score, and $\rho$ is the Spearman rank correlation between their scores. \textbf{(b) Each Judge's Difference from the Average.} For each judge (row) and the representative model of each family (column), we compute the average difference between that judge's score and the mean score of the three judges across all checkpoints.}
\label{fig:judge}
\end{figure}
\paragraph{Agreement among LLM Judges. }
LLM-as-a-Judge can be biased toward assigning higher scores to longer or more structured answers~\citep{zheng2023judging}. To reduce this effect, we grade each checkpoint on applicable axes with three judges from three different developers, namely Gemini 3.1 Pro, Claude Opus 4.7, and GPT-5.4 Pro, using the same rubric, and compute their average as the final score. However, this average is reliable only if the three judges are consistent with one another and do not share the same bias. We assess this consistency in Figure~\ref{fig:judge} using all checkpoint scores from the three judges.
As shown in Figure~\ref{fig:judge} (a), the three judges agree on most checkpoints. Each pair assigns the same score on 72\% to 75\% of all checkpoints, and their rankings are highly correlated ($\rho=0.86$--$0.87$). These results indicate that the rubric is interpreted consistently across the three judges and support the reliability of the final scores as measures of model capability.
Moreover, Figure~\ref{fig:judge}(b) shows that the main difference among the judges
is strictness rather than a preference for specific models. The GPT judge scores the lowest, while the Claude and Gemini judges score slightly above the average. Since this tendency in strictness affects all models in the same way, averaging the scores from three judges can remove this influence, and the model ranking is not affected.
More importantly, no judge gives noticeably higher scores to its own model family, indicating that self-preference bias is small.

\paragraph{Agreement with Human Experts. }
To validate the agreement between the LLM judges and human experts, we randomly select 60 checkpoints covering all 10 ranges and ask a security expert to grade 10 trajectories for each checkpoint under the same rubric without seeing the judge results. Detection is scored from 0 to 3, and planning is scored from 0 to 2.
The human expert and LLM judges agree closely, with a Pearson correlation of 0.96, a quadratic-weighted Cohen's kappa ($\kappa$) of 0.94, and a mean absolute error (MAE) of only 0.15. For 98\% of the checkpoints, the scores from the human expert and the LLM judges differ by at most one point. Moreover, the LLM judges are slightly stricter than the human expert and are therefore less likely to overstate the evaluation results.

\section{Conclusion}
In this paper, we introduce \benchname{}, the first benchmark for evaluating LLM agents on the post-compromise incident-response workflow. Each task requires the agent to analyze a forensic disk snapshot of a compromised host together with the alerts, vulnerability scans, and baseline checks reported by a host security product, and deliver forensic reports on intrusions, baseline risks, and vulnerability risks, along with a remediation plan. We instantiate this task across 10 cyber ranges, each built from a distinct compromised cloud host, together covering 4 entry-point types, 21 ATT\&CK techniques, and 5 operating systems.
By evaluating 23 frontier LLMs on the OpenCode agent harness, we observe that current agents can reliably uncover the problems exposed by alerts but struggle to proactively investigate the disk for silent intrusions and to produce comprehensive, verified remediation plans, with no model achieving complete detection and remediation on any single range. This gap reveals a fundamental bottleneck in building agents for real-world incident response. We hope \benchname{} provides a foundation for developing security agents that investigate and remediate incidents thoroughly rather than merely reacting to shallow alerts.

\section*{Ethics Statement}
\paragraph{Third-Party Model Evaluation.}
The third-party models included in our evaluation were assessed solely for academic research. All reported results were obtained in a controlled experimental environment and are presented only for capability comparison and technical discussion. The inclusion of model and provider names and scores does not imply any official position or endorsement by the corresponding providers or brands. Neither outputs obtained from these services nor the resulting evaluation records are used to train, fine-tune, or distill any competing model.

\paragraph{Security Research and Use Restrictions.}
All benchmark data were generated in isolated experimental environments under our authorization and control, rather than collected from production systems. Each cyber range is a host that we provisioned, compromised, and snapshotted for this purpose. The released forensic and security-product artifacts are synthetic and contain no data from customer, production, or real-world business systems. We sanitized every disk snapshot by replacing real credentials, keys, and personal data with placeholders, retaining only the forensic artifacts required for analysis. The attack techniques, CVEs, and exploitation methods represented in the scenarios are derived solely from public knowledge bases and vulnerability disclosures, such as MITRE ATT\&CK and public vulnerability databases. The benchmark contains forensic evidence rather than runnable exploit chains; it includes no proprietary or internally developed advanced attack techniques and no incremental attack information beyond public sources, and it cannot be directly used to conduct attacks against real-world systems. The benchmark is intended exclusively for defensive security research. Any use for unauthorized intrusion, destructive testing, offensive activity, or any other purpose that violates applicable laws or regulations is strictly prohibited.

\bibliography{colm2024_conference}

@inproceedings{zhang2025cybench,
  title={Cybench: A framework for evaluating cybersecurity capabilities and risks of language models},
  author={Zhang, Andy K and Perry, Neil and Dulepet, Riya and Ji, Joey and Menders, Celeste and Lin, Justin and Jones, Eliot and Hussein, Gashon and Liu, Samantha and Jasper, Donovan and others},
  booktitle={International Conference on Learning Representations},
  volume={2025},
  pages={25094--25243},
  year={2025}
}

@article{shao2024nyu,
  title={Nyu ctf bench: A scalable open-source benchmark dataset for evaluating llms in offensive security},
  author={Shao, Minghao and Jancheska, Sofija and Udeshi, Meet and Dolan-Gavitt, Brendan and Xi, Haoran and Milner, Kimberly and Chen, Boyuan and Yin, Max and Garg, Siddharth and Krishnamurthy, Prashanth and others},
  journal={Advances in Neural Information Processing Systems},
  volume={37},
  pages={57472--57498},
  year={2024}
}

@article{zhu2025cve,
  title={CVE-bench: a benchmark for AI agents' ability to exploit real-world web application vulnerabilities},
  author={Zhu, Yuxuan and Kellermann, Antony and Bowman, Dylan and Li, Philip and Gupta, Akul and Danda, Adarsh and Fang, Richard and Jensen, Conner and Ihli, Eric and Benn, Jason and others},
  journal={arXiv preprint arXiv:2503.17332},
  year={2025}
}

@article{jing2024secbench,
  title={Secbench: A comprehensive multi-dimensional benchmarking dataset for llms in cybersecurity},
  author={Jing, Pengfei and Tang, Mengyun and Shi, Xiaorong and Zheng, Xing and Nie, Sen and Wu, Shi and Yang, Yong and Luo, Xiapu},
  journal={arXiv preprint arXiv:2412.20787},
  year={2024}
}

@article{wan2024cyberseceval,
  title={Cyberseceval 3: Advancing the evaluation of cybersecurity risks and capabilities in large language models},
  author={Wan, Shengye and Nikolaidis, Cyrus and Song, Daniel and Molnar, David and Crnkovich, James and Grace, Jayson and Bhatt, Manish and Chennabasappa, Sahana and Whitman, Spencer and Ding, Stephanie and others},
  journal={arXiv preprint arXiv:2408.01605},
  year={2024}
}

@online{autopatchbench_meta_2025,
  author = {{Meta}},
  title = {Introducing AutoPatchBench: A Benchmark for AI-Powered Security Fixes},
  year = {2025},
  url = {https://engineering.fb.com/2025/04/29/ai-research/autopatchbench-benchmark-ai-powered-security-fixes/},
}

@online{anthropic_claude_opus_4_7_2026,
  author = {{Anthropic}},
  title = {Introducing Claude Opus 4.7},
  year = {2026},
  url = {https://www.anthropic.com/news/claude-opus-4-7},
}

@online{openai_gpt_5_4_2026,
  author = {{OpenAI}},
  title = {Introducing GPT-5.4},
  year = {2026},
  url = {https://openai.com/index/introducing-gpt-5-4/},
}

@online{gemini_team_gemini_3_1_pro_2026,
  author = {{The Gemini Team}},
  title = {Gemini 3.1 Pro: A Smarter Model for Your Most Complex Tasks},
  year = {2026},
  url = {https://blog.google/innovation-and-ai/models-and-research/gemini-models/gemini-3-1-pro/},
}

@article{wang2025cybergym,
  title={CyberGym: Evaluating AI Agents' Cybersecurity Capabilities with Real-World Vulnerabilities at Scale},
  author={Wang, Zhun and Shi, Tianneng and He, Jingxuan and Cai, Matthew and Zhang, Jialin and Song, Dawn},
  journal={arXiv e-prints},
  pages={arXiv--2506},
  year={2025}
}

@article{deason2025cybersoceval,
  title={Cybersoceval: Benchmarking llms capabilities for malware analysis and threat intelligence reasoning},
  author={Deason, Lauren and Bali, Adam and Bejean, Ciprian and Bolocan, Diana and Crnkovich, James and Croitoru, Ioana and Durai, Krishna and Midler, Chase and Miron, Calin and Molnar, David and others},
  journal={arXiv preprint arXiv:2509.20166},
  year={2025}
}

@online{wiz_cyber_model_arena_2025,
  author = {{Wiz}},
  title = {Cyber Model Arena},
  year = {2026},
  url = {https://www.wiz.io/cyber-model-arena},
}

@article{li2024wmdp,
  title={The wmdp benchmark: Measuring and reducing malicious use with unlearning},
  author={Li, Nathaniel and Pan, Alexander and Gopal, Anjali and Yue, Summer and Berrios, Daniel and Gatti, Alice and Li, Justin D and Dombrowski, Ann-Kathrin and Goel, Shashwat and Phan, Long and others},
  journal={arXiv preprint arXiv:2403.03218},
  year={2024}
}

@misc{openai2026_gpt54,
  author = {OpenAI},
  title = {Introducing GPT-5.4},
  year = {2026},
  url = {https://openai.com/index/introducing-gpt-5-4}
}

@misc{qwen2026_qwen37,
  author = {Qwen},
  title = {Qwen3.7: The Agent Frontier},
  year = {2026},
  url = {https://qwen.ai/blog?id=qwen3.7}
}

@misc{zai2026_glm51,
  author = {Z.ai},
  title = {GLM-5.1: Towards Long-Horizon Tasks},
  year = {2026},
  url = {https://z.ai/blog/glm-5.1}
}

@article{merrill2026terminal,
  title={Terminal-bench: Benchmarking agents on hard, realistic tasks in command line interfaces},
  author={Merrill, Mike A and Shaw, Alexander G and Carlini, Nicholas and Li, Boxuan and Raj, Harsh and Bercovich, Ivan and Shi, Lin and Shin, Jeong Yeon and Walshe, Thomas and Buchanan, E Kelly and others},
  journal={arXiv preprint arXiv:2601.11868},
  year={2026}
}

@article{ye2026claw,
  title={Claw-Eval: Towards Trustworthy Evaluation of Autonomous Agents},
  author={Ye, Bowen and Li, Rang and Yang, Qibin and Liu, Yuanxin and Yao, Linli and Lv, Hanglong and Xie, Zhihui and An, Chenxin and Li, Lei and Kong, Lingpeng and others},
  journal={arXiv preprint arXiv:2604.06132},
  year={2026}
}

@article{li2026skillsbench,
  title={SkillsBench: Benchmarking how well agent skills work across diverse tasks},
  author={Li, Xiangyi and Chen, Wenbo and Liu, Yimin and Zheng, Shenghan and Chen, Xiaokun and He, Yifeng and Li, Yubo and You, Bingran and Shen, Haotian and Sun, Jiankai and others},
  journal={arXiv preprint arXiv:2602.12670},
  year={2026}
}

@article{wu2025excytin,
  title={Excytin-bench: Evaluating llm agents on cyber threat investigation},
  author={Wu, Yiran and Velazco, Mauricio and Zhao, Andrew and Luj{\'a}n, Manuel Ra{\'u}l Mel{\'e}ndez and Movva, Srisuma and Roy, Yogesh K and Nguyen, Quang and Rodriguez, Roberto and Wu, Qingyun and Albada, Michael and others},
  journal={arXiv preprint arXiv:2507.14201},
  year={2025}
}

@article{lin2025ircopilot,
  title={Ircopilot: Automated incident response with large language models},
  author={Lin, Xihuan and Zhang, Jie and Deng, Gelei and Liu, Tianzhe and Zhang, Tianwei and Guo, Qing and Chen, Riqing},
  journal={arXiv preprint arXiv:2505.20945},
  year={2025}
}

@misc{opencode2026,
  title = {OpenCode: The open source AI coding agent},
  author = {{OpenCode}},
  year = {2026},
  url = {https://opencode.ai/}
}

@inproceedings{jimenez2024swe,
  title={Swe-bench: Can language models resolve real-world github issues?},
  author={Jimenez, Carlos E and Yang, John and Wettig, Alexander and Yao, Shunyu and Pei, Kexin and Press, Ofir and Narasimhan, Karthik},
  booktitle={International Conference on Learning Representations},
  volume={2024},
  pages={54107--54157},
  year={2024}
}

@article{deng2025swe,
  title={Swe-bench pro: Can ai agents solve long-horizon software engineering tasks?},
  author={Deng, Xiang and Da, Jeff and Pan, Edwin and He, Yannis Yiming and Ide, Charles and Garg, Kanak and Lauffer, Niklas and Park, Andrew and Pasari, Nitin and Rane, Chetan and others},
  journal={arXiv preprint arXiv:2509.16941},
  year={2025}
}

@article{yao2024tau,
  title={$tau$-bench: A Benchmark for Tool-Agent-User Interaction in Real-World Domains},
  author={Yao, Shunyu and Shinn, Noah and Razavi, Pedram and Narasimhan, Karthik},
  journal={arXiv preprint arXiv:2406.12045},
  year={2024}
}

@article{he2025vitabench,
  title={Vitabench: Benchmarking llm agents with versatile interactive tasks in real-world applications},
  author={He, Wei and Sun, Yueqing and Hao, Hongyan and Hao, Xueyuan and Xia, Zhikang and Gu, Qi and Han, Chengcheng and Zhao, Dengchang and Su, Hui and Zhang, Kefeng and others},
  journal={arXiv preprint arXiv:2509.26490},
  year={2025}
}

@article{barres2506tau2,
  title={$\tau^2$-bench: Evaluating conversational agents in a dual-control environment, 2025},
  author={Barres, V and Dong, H and Ray, S and Si, X and Narasimhan, K},
  journal={URL https://arxiv. org/abs/2506.07982}
}

@article{xie2024osworld,
  title={Osworld: Benchmarking multimodal agents for open-ended tasks in real computer environments},
  author={Xie, Tianbao and Zhang, Danyang and Chen, Jixuan and Li, Xiaochuan and Zhao, Siheng and Cao, Ruisheng and Hua, Toh J and Cheng, Zhoujun and Shin, Dongchan and Lei, Fangyu and others},
  journal={Advances in Neural Information Processing Systems},
  volume={37},
  pages={52040--52094},
  year={2024}
}

@inproceedings{zhou2024webarena,
  title={Webarena: A realistic web environment for building autonomous agents},
  author={Zhou, Shuyan and Xu, Frank F and Zhu, Hao and Zhou, Xuhui and Lo, Robert and Sridhar, Abishek and Cheng, Xianyi and Ou, Tianyue and Bisk, Yonatan and Fried, Daniel and others},
  booktitle={International Conference on Learning Representations},
  volume={2024},
  pages={15585--15606},
  year={2024}
}

@article{zhang2026clawbench,
  title={ClawBench: Can AI Agents Complete Everyday Online Tasks?},
  author={Zhang, Yuxuan and Wang, Yubo and Zhu, Yipeng and Du, Penghui and Miao, Junwen and Lu, Xuan and Xu, Wendong and Hao, Yunzhuo and Cai, Songcheng and Wang, Xiaochen and others},
  journal={arXiv preprint arXiv:2604.08523},
  year={2026}
}

@article{team2026cocoabench,
  title={CocoaBench: Evaluating Unified Digital Agents in the Wild},
  author={Team, CocoaBench and Hao, Shibo and Zhang, Zhining and Liang, Zhiqi and Liu, Tianyang and Zha, Yuheng and Gao, Qiyue and Chen, Jixuan and Wang, Zilong and Cheng, Zhoujun and others},
  journal={arXiv preprint arXiv:2604.11201},
  year={2026}
}

@article{li2026claw,
  title={Claw-eval-live: A live agent benchmark for evolving real-world workflows},
  author={Li, Chenxin and Tang, Zhengyang and Huang, Mingxin and Lin, Yunlong and Huang, Shijue and Liu, Shengyuan and Ye, Bowen and Li, Rang and Li, Lei and Wang, Benyou and others},
  journal={arXiv preprint arXiv:2604.28139},
  year={2026}
}

@article{zheng2023judging,
  title={Judging llm-as-a-judge with mt-bench and chatbot arena},
  author={Zheng, Lianmin and Chiang, Wei-Lin and Sheng, Ying and Zhuang, Siyuan and Wu, Zhanghao and Zhuang, Yonghao and Lin, Zi and Li, Zhuohan and Li, Dacheng and Xing, Eric and others},
  journal={Advances in neural information processing systems},
  volume={36},
  pages={46595--46623},
  year={2023}
}

@article{gu2026survey,
  title={A survey on llm-as-a-judge},
  author={Gu, Jiawei and Jiang, Xuhui and Shi, Zhichao and Tan, Hexiang and Zhai, Xuehao and Xu, Chengjin and Li, Wei and Shen, Yinghan and Ma, Shengjie and Liu, Honghao and others},
  journal={The Innovation},
  volume={7},
  number={6},
  year={2026},
  publisher={Elsevier}
}
\bibliographystyle{colm2024_conference}

\newpage
\appendix
\section{Details on Range Construction}
\paragraph{SSH-Miner.}
The \texttt{ssh-miner} range is built on a CentOS 7 host with weak SSH baseline settings and a weak root password. An external attacker brute-forces the root credential with \texttt{hydra}, drops a mining binary, installs three persistence mechanisms spanning scheduled tasks, shell initialization, and system services, and clears the authentication log. The agent receives this post-attack disk snapshot with (i) five alerts covering the brute-force success, miner execution, suspicious crontab, autorun change, and outbound miner connection, (ii) an empty vulnerability list since there is no CVE-driven entry, and (iii) baseline checks flagging the SSH, MySQL, and credential-strength baselines. With these inputs, the required reports are expected to cover (i) an intrusion report covering the full attack chain and intrusion entities such as the mining binary, hydra residue, log backup, and attacker IP, (ii) a vulnerability report confirming the absence of a CVE-driven entry, (iii) a baseline report flagging the SSH and credential baseline weaknesses, and (iv) a remediation plan prioritizing miner termination, persistence cleanup, SSH hardening, and credential rotation.

\begin{lstlisting}
ssh-miner/disk/
|-- usr/
|-- opt/
|-- ...
|-- etc/
|   |-- ssh/sshd_config
|   |-- sudoers.d/devops
|   |-- my.cnf
|   \-- systemd/system/
|       |-- system-update-helper.service
|       \-- multi-user.target.wants/
|           \-- system-update-helper.service
|-- home/
|   \-- devops/.bash_history
|-- root/
|   \-- .bashrc
|-- tmp/
|   \-- .X11-unix/
|       |-- .xorg
|       \-- .xorg.log
\-- var/
    |-- spool/cron/root
    \-- www/html/index.html
\end{lstlisting}

\paragraph{Redis-RCE.}
The \texttt{redis-rce} range is built on an Ubuntu 22.04 host that exposes an unauthenticated Redis server running as root to the public internet. The attacker abuses the Redis server to write an SSH key into the root account, logs in, opens a reverse shell, and deploys a cryptominer disguised as a Redis background-save process, with persistence mechanisms including the injected key, a malicious crontab, an SSH login hook, and a boot script, with the miner's timestamp forged to resist timeline analysis. The agent receives this snapshot with (i) two alerts covering only the reverse shell and its outbound connection, (ii) an empty vulnerability list since the entry is a misconfiguration rather than a CVE, and (iii) no baseline findings, leaving the persistence silent on disk. With these inputs, the required reports are expected to cover (i) an intrusion report reconstructing the Redis-to-root chain and intrusion entities such as the miner, injected key, and malicious crontab, (ii) a vulnerability report concluding the absence of a CVE-driven entry, (iii) a baseline report flagging the Redis unauthorized-access and SSH root-login weaknesses, and (iv) a remediation plan prioritizing miner termination, persistence removal, and Redis and SSH hardening.

\begin{lstlisting}
redis-rce/disk/
|-- opt/
|-- tmp/
|-- usr/
|-- ...
|-- etc/
|   |-- init.d/redis-watchdog
|   |-- redis/redis.conf
|   |-- ssh/sshd_config
|   \-- sudoers.d/deploy
|-- home/
|   \-- deploy/.bash_history
|-- root/
|   \-- .ssh/
|       |-- authorized_keys
|       \-- rc
`-- var/
    |-- spool/cron/crontabs/root
    \-- tmp/
        |-- .redis-bgsave
        \-- .redis-bgsave.log
\end{lstlisting}

\paragraph{Docker-Escape.} The \texttt{docker-escape} range is built on an Ubuntu 20.04 host where Docker Remote API is exposed without authentication. The attacker reaches the API, launches a privileged container that mounts the host filesystem to escape to the host, injects an SSH key, and runs a cryptominer hidden as a Docker health tool, with persistence including a crontab, a systemd service, a shell-initialization hook, the injected key, and a long-running backdoor container. The agent receives this snapshot with (i) three alerts covering the mining process, the suspicious crontab, and the anomalous systemd service, (ii) an empty vulnerability list since the exposed API is a misconfiguration, and (iii) no baseline findings, so the API exposure, container escape, and backdoor container must be recovered from Docker events and host forensics. With these inputs, the required reports are expected to cover (i) an intrusion report reconstructing the API-to-host-takeover chain and entities such as the miner, the escape container, and the backdoor container, (ii) a vulnerability report concluding no CVE-driven entry, (iii) a baseline report flagging the exposed Docker API and excessive permissions, and (iv) a remediation plan prioritizing miner and backdoor-container removal, persistence cleanup, and firewalling Docker TCP access.

\begin{lstlisting}
docker-escape/disk/
|-- usr/
|-- home/
|-- tmp/
|-- ...
|-- etc/systemd/system/
|   |-- docker-health-agent.service
|   \-- docker.service.d/
|       \-- override.conf
|-- opt/
|   |-- .docker/
|   |   |-- .health-monitor
|   |   \-- .health-monitor.log
|   \-- app/
|       \-- docker-compose.yml
|-- var/spool/cron/crontabs/
|   \-- root
\-- root/
    |-- .bashrc
    \-- .ssh/
        \-- authorized_keys
\end{lstlisting}

\paragraph{Jenkins-RCE.} The \texttt{jenkins-rce} range is built on an Ubuntu 22.04 host running Jenkins whose authorization misconfiguration leaves the Groovy Script Console reachable without authentication. The attacker obtains unauthenticated remote code execution through the console, deploys a cryptominer masquerading as a Jenkins agent, exfiltrates Jenkins credentials, and installs redundant persistence spanning a cron job, a rogue root-equivalent account, a SUID-root binary, a sudoers backdoor, a library-preload injection, and a Jenkins startup hook. The agent receives this snapshot with (i) five alerts covering the Script Console access, the masqueraded miner, the new privileged account, the cron backdoor, and the outbound mining connection, and (ii) no explicit vulnerability or baseline checks. With these inputs, the required reports are expected to cover (i) an intrusion report reconstructing the unauthenticated RCE-to-mining chain and entities such as the miner, the exfiltrated secrets, and the multiple persistence implants, (ii) a vulnerability report attributing the entry to a Jenkins authorization misconfiguration, (iii) a baseline report flagging the anonymous-administrator setting and weak administration password, and (iv) a remediation plan prioritizing miner termination, removal of all persistence, credential rotation, and revoking anonymous permissions.

\begin{lstlisting}
jenkins-rce/disk/
|-- home/
|-- opt/
|-- tmp/
|-- usr/
|-- ...
|-- etc/
|   |-- hosts
|   |-- environment
|   |-- passwd
|   |-- sudoers
|   \-- cron.d/jenkins-update
|-- usr/local/bin/.update
|-- var/
|   |-- cache/jenkins/
|   |   |-- .update
|   |   |-- .x.so
|   |   \-- .j.tgz
|   \-- jenkins_home/
|       |-- config.xml
|       |-- init.groovy.d/zz-update.groovy
|       \-- secrets/master.key
\-- root/.ssh/authorized_keys
\end{lstlisting}

\paragraph{Shiro-Fastjson.} The \texttt{shiro-fastjson} range is built on a CentOS 8 host running a Java web application on Tomcat that uses a default key and a vulnerable Fastjson library. The attacker gains code execution through Shiro deserialization, drops two JSP webshells hidden among static assets, steals database credentials and dumps a user table, escalates to root by abusing a sudo rule without passwords, attempts to uninstall the security agent, and deploys a miner behind persistence including a systemd service, crontab, shell-initialization hook, injected key, sudoers backdoor, a MySQL UDF backdoor, an \texttt{at} job, and \texttt{rc.local}. The agent receives this snapshot with (i) fifteen alerts spanning initial access, encoded-command execution, security-agent tampering, privilege escalation, and persistence, (ii) an empty vulnerability list since the Shiro and Fastjson flaws are library-level components beyond the scanner, and (iii) baseline findings flagging the operating system, Tomcat, and MySQL baselines and several weak-password checks. With these inputs, the required outputs are expected to include (i) an intrusion report reconstructing the Shiro-to-root chain and entities such as the two webshells, the miner, the database backdoor, and the data dump, (ii) a vulnerability report identifying the vulnerable Shiro and Fastjson components, (iii) a baseline report flagging the Tomcat, MySQL, and credential weaknesses, and (iv) a remediation plan prioritizing removal of the webshells, miner, and all persistence mechanisms, together with version upgrades and database lockdown.

\begin{lstlisting}
shiro-fastjson/disk/
|-- home/
|-- ...
|-- etc/
|   |-- my.cnf
|   |-- profile.d/java-env.sh
|   |-- rc.d/rc.local
|   |-- sudoers.d/
|   |   |-- 99-java-ops
|   |   \-- tomcat
|   \-- systemd/system/java-app-monitor.service
|-- opt/
|   |-- .cache/
|   |   |-- .java-updater
|   |   \-- .java-updater.log
|   |-- ruoyi/application.yml
|   \-- tomcat/webapps/ROOT/
|       |-- WEB-INF/lib/
|       |   |-- fastjson-1.2.68.jar
|       |   \-- shiro-core-1.7.0.jar
|       \-- static/
|           |-- css/error.jsp
|           \-- js/analytics.jsp
|-- root/.ssh/authorized_keys
|-- tmp/.sql_dump
|-- usr/lib64/mysql/plugin/lib_mysqludf_json.so
\-- var/spool/cron/root
\end{lstlisting}

\paragraph{Log4j-RCE.} The \texttt{log4j-rce} range is built on a CentOS 8 host whose web application bundles a vulnerable Log4j library exploitable through a JNDI lookup (CVE-2021-44228). The attacker triggers the JNDI injection to launch a webshell, overwrites a deploy script invoked through sudo to escalate to root, deploys a miner, exfiltrates a database dump, and truncates the authentication log to cover its tracks, with persistence including a systemd service, a crontab, an injected key, and a shell-initialization hook. The agent receives this snapshot with (i) six alerts covering the Log4j RCE, the webshell, the privilege escalation, the miner, the crontab backdoor, and the log tampering, (ii) a vulnerability check reporting a CVE for the Log4j library, and (iii) baseline findings flagging SSH root-login, weak passwords, and the sudo rule without password protection. With these inputs, the required reports are expected to cover (i) an intrusion report reconstructing the Log4j-to-root chain and entities such as the webshell, miner, data dump, and tampered deploy script, (ii) a vulnerability report confirming the Log4j RCE and prescribing the upgrade, (iii) a baseline report flagging the SSH and sudo weaknesses, and (iv) a remediation plan prioritizing the Log4j upgrade, webshell and miner removal, persistence cleanup, and credential rotation.

\begin{lstlisting}
log4j-rce/disk/
|-- bin/
|-- usr/
|-- lib/
|-- home/
|-- boot/
|-- sbin/
|-- ...
|-- etc/
|   |-- profile.d/java-env.sh
|   |-- ssh/sshd_config
|   |-- sudoers.d/app-deploy
|   \-- systemd/system/java-gc-helper.service
|-- opt/webapp/
|   |-- lib/
|   |   |-- log4j-api-2.14.1.jar
|   |   \-- log4j-core-2.14.1.jar
|   |-- logs/app.log
|   |-- VulnWebApp.java
|   |-- deploy.sh
|   \-- webshell.py
|-- var/
|   |-- cache/.java-gc
|   |-- log/auth.log
|   \-- spool/cron/crontabs/root
|-- tmp/
|   |-- .db_dump
|   |-- .escalation_log
|   \-- .privesc_proof
\-- root/.ssh/authorized_keys
\end{lstlisting}

\paragraph{Next.js-RCE.} The \texttt{nextjs-rce} range is built on an Ubuntu 22.04 host running a Next.js application vulnerable to a server-component deserialization RCE. The attacker obtains code execution as the application user, drops a Node.js webshell on a spoofed debugger port, escalates to root through a command injection in a SUID-root helper, installs an \texttt{LD\_PRELOAD} rootkit, runs a miner, and exfiltrates a database dump, with persistence including the loader-preload rootkit, a shell-initialization hook, a systemd service, a crontab, and an injected key. The agent receives this snapshot with (i) six alerts covering the miner, the suspicious Node child process, the webshell, the crontab, the systemd service, and the outbound mining connection, (ii) an empty vulnerability list, requiring the agent to infer vulnerabilities from the disk snapshot, and (iii) baseline findings flagging the SSH configuration, database exposure, a non-standard SUID binary, and a loose sudo rule. With these inputs, the required reports are expected to cover (i) an intrusion report reconstructing the RCE-to-rootkit chain and entities such as the malicious shared library, the miner, the webshell, and the data dump, (ii) a vulnerability report identifying the framework vulnerability from the dependency manifest, (iii) a baseline report flagging the SSH, database, and privilege-escalation weaknesses, and (iv) a remediation plan prioritizing rootkit and miner removal, persistence cleanup, framework upgrade, and SUID and sudo hardening.

\begin{lstlisting}
nextjs-rce/disk/
|-- bin/
|-- boot/
|-- lib/
|-- sbin/
|-- snap/
|-- srv/
|-- usr/lib/
|-- ...
|-- etc/
|   |-- ld.so.preload
|   |-- profile.d/node-env.sh
|   |-- ssh/sshd_config
|   |-- sudoers.d/node-ops
|   |-- systemd/system/node-gc-helper.service
|   \-- postgresql/14/main/
|       |-- pg_hba.conf
|       \-- postgresql.conf
|-- opt/webapp/
|   |-- .env
|   |-- package.json
|   |-- node_modules/next/package.json
|   \-- .next/static/chunks/debug.js
|-- usr/
|   |-- lib/x86_64-linux-gnu/.libnode_helper.so
|   \-- local/bin/backup-tool
|-- var/
|   |-- cache/.node-gc
|   |-- cache/.node-gc.log
|   \-- spool/cron/crontabs/root
|-- tmp/
|   |-- .pg_dump
|   |-- .escalation_log
|   \-- .privesc_proof
\-- root/.ssh/authorized_keys
\end{lstlisting}

\paragraph{NPM-Worm.} The \texttt{npm-worm} range is built on an Ubuntu 22.04 host running a Node.js application whose unauthenticated deployment webhook passes its parameter straight into \texttt{npm install}, letting the attacker install a malicious package from the attacker host. The package's installed hook drops a worm that scans the internal network, harvests several classes of credentials, and injects itself into other projects to propagate, after which a leaked cloud access key is used to deploy a miner, with persistence including a crontab, shell and login initialization hooks, a systemd service, injected keys, and the infected project. The agent receives this snapshot with (i) two alerts covering the worm-download command and the web application spawning an abnormal child process, (ii) an empty vulnerability list, and (iii) baseline findings flagging the Ubuntu host baseline, a weak root password, the Nginx-exposed unauthenticated webhook, and several sensitive-configuration leaks such as a plaintext npm token, an unencrypted SSH key, and loose file permissions. With these inputs, the required reports are expected to cover (i) an intrusion report reconstructing the supply-chain-to-worm chain and entities such as the malicious package, the worm script, the stolen-credential cache, and the miner, (ii) a vulnerability report attributing the entry to the unauthenticated webhook and unsigned package install, (iii) a baseline report flagging the weak credential, exposed webhook, and sensitive configuration findings, and (iv) a remediation plan prioritizing worm and miner termination, package and propagation removal, credential rotation, webhook authentication, and disabling install scripts.

\begin{lstlisting}
npm-worm/disk/
|-- tmp/
|-- usr/
|-- ...
|-- etc/
|   |-- profile.d/node-env.sh
|   |-- sudoers.d/developer
|   \-- systemd/system/npm-cache-gc.service
|-- home/
|   \-- developer/
|       |-- .bashrc
|       |-- .git-credentials
|       |-- .npmrc
|       \-- .ssh/
|           |-- authorized_keys
|           \-- id_rsa
|-- opt/
|   |-- .node-helpers/
|   |   |-- .npm-gc
|   |   \-- .npm-gc.log
|   |-- internal-tools/package.json
|   \-- webapp/
|       |-- .env
|       |-- package-lock.json
|       \-- node_modules/@corp-utils/logger/
|           |-- package.json
|           \-- scripts/setup.js
|-- root/
|   \-- .ssh/authorized_keys
\-- var/
    |-- spool/cron/crontabs/developer
    \-- tmp/.npm-cache/
        |-- scan_results.txt
        |-- sysinfo.json
        |-- npmrc/.npmrc
        |-- ssh_keys/id_rsa
        |-- env_files/.env
        |-- git_credentials/.git-credentials
        \-- bash_history/.bash_history
\end{lstlisting}

\paragraph{ASP.NET-ViewState.} The \texttt{aspnet-viewstate} range is built on a Windows Server 2019 host running an ASP.NET application whose backup configuration file is downloadable and leaks the hardcoded \texttt{machineKey}. The attacker forges a signed ViewState payload to execute code as \texttt{SYSTEM}, drops an ASPX webshell, dumps and parses LSASS to recover credentials, runs a C2 beacon masqueraded as an audit service, and clears event logs and shadow copies, with Windows-specific persistence spanning a scheduled task, a WMI event subscription, a sticky-keys/IFEO backdoor, a DLL search-order hijack, and a malicious MSSQL logon trigger. The agent receives this snapshot with (i) three alerts covering the webshell, the beacon process, and the WMI subscription, (ii) a vulnerability check reporting the ViewState deserialization with the backup exposure and application pool misconfigurations, and (iii) baseline findings flagging the weak database password, the enabled command execution stored procedure, and the hardcoded \texttt{machineKey}. With these inputs, the required reports are expected to cover (i) an intrusion report reconstructing the ViewState-to-\texttt{SYSTEM} chain and entities such as the webshell, the credential-dump residue, the beacon, and the five persistence objects, (ii) a vulnerability report confirming the ViewState RCE and the related misconfigurations, (iii) a baseline report flagging the MSSQL and \texttt{machineKey} weaknesses, and (iv) a remediation plan prioritizing webshell and beacon removal, teardown of the Windows persistence, \texttt{machineKey} rotation, application-pool de-privileging, and credential rotation.

\begin{lstlisting}
aspnet-viewstate/disk/
|-- Windows/
|   |-- System32/
|   |   |-- Tasks/Microsoft/Windows/Maintenance/AuditTask
|   |   |-- wbem/Repository/OBJECTS.DATA
|   |   \-- ...
|   \-- Temp/
|       |-- lsass.dmp
|       |-- creds.txt
|       |-- plant-trigger.sql
|       \-- crm-app.tar.gz
|-- Program Files/
|   \-- Notepad++/
|       |-- notepad++.exe
|       \-- version.dll
|-- ProgramData/
|   |-- WindowsAudit/
|   |   |-- WindowsAuditSvc.exe
|   |   |-- last-beacon.log
|   |   |-- schtasks-fired.log
|   |   |-- wmi-fired.log
|   |   \-- mssql-trigger-fired.log
|   \-- ...
|-- ...
\-- inetpub/wwwroot/CRM/
    |-- web.config
    |-- web.config.bak
    \-- help.aspx
\end{lstlisting}

\paragraph{RDP-Service-Abuse.} The \texttt{rdp-service-abuse} range is built on a Windows Server 2019 host that exposes RDP to the public internet with a weak \texttt{helpdesk} password. The attacker sprays the exposed RDP service, lands a logon as the low-privilege \texttt{helpdesk} user, and abuses a weak service DACL to rewrite the service bin path to a PowerShell payload and restart it, gaining execution as \texttt{SYSTEM}. Then, the attacker installs persistence with a disguised service and a scheduled task, leaves credential dumps, and beacons to a C2 endpoint. The agent receives this snapshot with (i) seven alerts clustered on the post-exploitation stage covering the encoded-PowerShell execution, the LSASS-dump tooling, and the anomalous service registry write, (ii) an empty vulnerability list since the entry is an RDP weak-credential and service-misconfiguration chain rather than a CVE, and (iii) no baseline findings, leaving the RDP authentication timeline and the DACL only recoverable from the event logs in the disk snapshot. With these inputs, the required reports are expected to cover (i) an intrusion report reconstructing the RDP-spraying-to-\texttt{SYSTEM} chain and discovering the disguised service, scheduled task, credential dump, and C2 indicator, (ii) a vulnerability report concluding the absence of a CVE-driven entry, (iii) a baseline report flagging the \texttt{helpdesk} weak password and RDP exposure with the weak service DACL, and (iv) a remediation plan prioritizing \texttt{helpdesk} credential rotation, a more restrictive service DACL, removal of the persistence, the credential residue, and C2 blocking.

\begin{lstlisting}
rdp-service-abuse/disk/
|-- Windows/System32/
|-- ...
|-- ProgramData/
|   |-- CorpBackup/
|   |   |-- corp-backup-svc.ps1
|   |   \-- system-stage.ps1
|   \-- WindowsHealth/
|       |-- WindowsHealthSvc.ps1
|       \-- last-beacon.log
|-- Users/
|   \-- helpdesk/NTUSER.DAT
|-- Windows/Temp/
|   |-- procdump.exe
|   |-- lsass.dmp
|   \-- creds.txt
\-- Windows/System32/
    |-- Tasks/Microsoft/Windows/Maintenance/HealthSync
    |-- config/
    |   |-- SYSTEM
    |   \-- SAM
    \-- winevt/Logs/
        |-- Security.evtx
        |-- System.evtx
        |-- Microsoft-Windows-TerminalServices-RemoteConnectionManager%4Operational.evtx
        \-- Microsoft-Windows-TerminalServices-LocalSessionManager%4Operational.evtx
\end{lstlisting}

\clearpage

\section{Details on Capability Taxonomy}
\begin{table*}[h]
\centering
\caption{The mapping between capability items and range checklist. Each cell lists the checkpoint indices (CHK-$n$) in that range mapped to the item.}
\label{tab:cap-coverage}
\resizebox{1.0\textwidth}{!}{
\setlength{\tabcolsep}{5pt}
\begin{tabular}{l l cccccccccc}
\toprule
\textbf{Item} & \textbf{Capability Name} & \textbf{SSH-Miner} & \textbf{Redis-RCE} & \textbf{Docker-Escape} & \textbf{Jenkins-RCE} & \textbf{Shiro-Fastjson} & \textbf{Log4j-RCE} & \textbf{Next.js-RCE} & \textbf{NPM-Worm} & \textbf{ASP.NET-ViewState} & \textbf{RDP-Service-Abuse} \\
\midrule
\multicolumn{12}{@{}l}{\textit{Intrusion Entity (ENT)}} \\
ENT-F01 & Webshell file &  &  &  &  & 1,2,7 & 1,4 & 3,36 & 3,4 & 1 &  \\
ENT-F02 & Linux malware file & 2,3 & 2 & 2,3 & 2 & 4,6 & 3 & 2 & 2,6 &  &  \\
ENT-F03 & Malicious SO / kernel module &  &  &  &  &  &  & 4 &  &  &  \\
ENT-F04 & Residual data file &  & 3,4 & 4 & 9 & 5,8 & 5 & 5 & 7,8,9 & 5 & 16,17 \\
ENT-F05 & Tampered-file restoration &  &  &  & 6 &  & 21 &  &  &  &  \\
ENT-F06 & Windows malware file &  &  &  &  &  &  &  &  & 2,4,8 & 13,15 \\
ENT-N01 & Attacker-IP inbound block & 20 &  & 23 & 12 & 38 &  & 30 & 38 & 6 & 3 \\
ENT-N02 & Malicious-IP outbound block & 23 &  & 24 & 11 & 39 &  & 31 & 37 & 7 &  \\
ENT-N03 & Malicious-domain block &  &  &  & 10 &  &  &  &  & 7 & 14 \\
ENT-P01 & Mining process & 1 & 1 & 1 & 1 & 3 & 2 & 1 & 1,5 &  &  \\
ENT-P02 & C2 beacon process &  &  &  &  &  &  &  &  & 3 &  \\
ENT-P03 & Windows malicious process &  &  &  &  &  &  &  &  & 3 &  \\
\midrule
\multicolumn{12}{@{}l}{\textit{Persistence Mechanism (PER)}} \\
PER-A01 & authorized\_keys tampering &  &  &  &  & 8 & 14 & 9 & 23 &  &  \\
PER-A02 & sudoers implant &  &  &  &  & 10 &  &  &  &  &  \\
PER-A03 & Rogue local account &  &  &  & 4,22 &  &  &  &  &  &  \\
PER-D01 & MySQL backdoor &  &  &  &  & 6 &  &  &  &  &  \\
PER-D02 & MSSQL backdoor &  &  &  &  &  &  &  &  & 16,17 &  \\
PER-E01 & profile.d / udev rule &  &  &  &  &  &  & 6 & 24 &  &  \\
PER-E02 & WMI event subscription &  &  &  &  &  &  &  &  & 12,13,14 &  \\
PER-H01 & ld.so.preload hijack &  &  &  &  &  &  & 4 &  &  &  \\
PER-H02 & DLL hijacking &  &  &  &  &  &  &  &  & 8,15 &  \\
PER-I01 & Shell init (bashrc/profile) & 12 & 13 & 13 &  & 20 & 15 &  & 20 &  &  \\
PER-I02 & /etc/environment injection &  &  &  & 7 &  &  &  &  &  &  \\
PER-M01 & SUID/SGID backdoor &  &  &  & 5,23 &  &  &  &  &  &  \\
PER-S01 & Cron task & 11 & 16 & 12 & 3,20 & 19 & 13 & 8 & 19,25 &  &  \\
PER-S02 & systemd timer &  &  &  &  &  &  & 7 &  &  &  \\
PER-S03 & at job &  &  &  &  & 21 &  &  &  &  &  \\
PER-S04 & Windows scheduled task &  &  &  &  &  &  &  &  & 9,10,11 & 10,11,12 \\
PER-V01 & systemd service & 13,14 & 14,15 & 14,15 &  & 17,18 & 12 &  & 21,22 &  &  \\
PER-V02 & init.d / rc.local &  &  &  &  & 22 &  &  &  &  &  \\
PER-V03 & Windows service &  &  &  &  &  &  &  &  &  & 6,8,9 \\
PER-W01 & Nginx config tampering &  &  &  &  &  &  & 10 &  &  &  \\
PER-W02 & App-container config tampering &  &  &  & 8,21 &  &  &  &  &  &  \\
\midrule
\multicolumn{12}{@{}l}{\textit{Baseline Risk (BAS)}} \\
BAS-01 & SSH hardening & 4--7 & 10 & 9 &  &  & 7,8 & 11--14 & 10,11,12 &  &  \\
BAS-02 & DB / middleware access control & 9 & 5--9 &  &  & 12 &  & 18,19 &  & 21 &  \\
BAS-03 & Credential safety & 10 & 12 & 8 & 14,24 & 11,13,16 & 9,22 & 15,16 & 14,15,17,36 & 21,22 & 4,18,19 \\
BAS-04 & Privilege / service audit & 8 & 11 & 7 &  & 9,14 & 10,21 &  & 18 &  & 5,20 \\
BAS-05 & Container / orchestration config &  &  & 5,6,10,21 &  &  &  & 17,20 & 16 &  &  \\
BAS-06 & Web-service config &  &  &  & 13 & 15 &  &  & 13 & 19,20 &  \\
BAS-07 & Cloud-resource config &  & 23 & 11 &  &  &  &  &  &  &  \\
\midrule
\multicolumn{12}{@{}l}{\textit{Vulnerability Risk (VUL)}} \\
VUL-01 & Java component vuln &  &  &  &  & 26--28,35--37 & 6,11,20 &  & 35 &  &  \\
VUL-02 & Web-app code vuln &  &  &  &  &  &  &  &  & 18 &  \\
VUL-03 & Container-escape vuln &  &  &  &  &  &  & 23 &  &  &  \\
VUL-04 & Node.js / front-end framework vuln &  &  &  &  &  &  & 21,22,24,35 &  &  &  \\
\midrule
\multicolumn{12}{@{}l}{\textit{Investigation and Response Quality (Q)}} \\
Q-01 & Entry-point localization & 18 & 20 & 17,18 &  & 25 & 16 & 25 & 28,29 &  & 4 \\
Q-02 & Attack-chain reconstruction & 16,19 & 18,21 & 19 & 15 & 29--32,34 & 17,19 & 26,28 & 31,33 &  & 1,2,7 \\
Q-03 & Attacker-info extraction & 16,17 & 18,19 &  & 16 & 32,33 & 18 & 27 & 30,34 &  & 3 \\
Q-04 & Honesty \& confidence calibration & 18 & 20 & 18 & 19 & 25 &  & 32 & 29 &  &  \\
Q-05 & Investigation thoroughness & 15 & 17 & 16 &  & 23,24 &  & 33 & 26 &  &  \\
Q-06 & Cross-language/service tracing &  &  &  &  &  &  & 29 & 27 &  &  \\
Q-07 & Remediation-verification completeness & 21,22 & 22,24 & 20,22 &  & 40--45 &  & 34 & 39 &  &  \\
Q-08 & Business-impact assessment &  &  &  & 17,18 &  &  & 37 & 32 &  &  \\
\bottomrule
\end{tabular}}
\end{table*}

\section{Implementation Details}
\subsection{Task Prompt for Linux Operating System}
\label{sec:task-prompt}
\begin{lstlisting}
# Detection Agent Workspace

## Your Role

You are a security incident response analyst working on a forensic analysis. A compromised server's disk snapshot and security platform alert data have been provided. Your job is to analyze them and produce a comprehensive incident response report.

## Data Sources

- **Forensic disk snapshot**: `DISK_PATH/` (read-only copy of the server filesystem)
- **SAS alerts**: `SAS_PATH/alerts.json`
- **SAS vulnerability scan**: `SAS_PATH/vulns.json`
- **SAS baseline check**: `SAS_PATH/baselines.json`

## Output

Write all outputs to: `OUTPUT_PATH/`

Required files:
1. `progress.md` — Investigation timeline and checklist
2. `intrusion-report.md` — Full intrusion analysis with attack chain
3. `vuln-report.md` — Vulnerability verification results
4. `baseline-report.md` — Configuration baseline assessment
5. `remediation-plan.md` — Structured remediation plan (natural language, prioritized P0-P3)

## Constraints

- You are running on an analysis machine, NOT the compromised server
- The disk snapshot at `DISK_PATH/` is a read-only forensic copy
- Do NOT SSH into any machine during analysis
- All findings must be based on disk evidence and SAS data
- Do NOT fabricate CVEs or vulnerabilities — if unsure, say so with confidence levels
- Preserve forensic evidence: recommend backup/quarantine over deletion in remediation-plan.md
\end{lstlisting}

\subsection{Task Prompt for Windows Operating System}
\begin{lstlisting}
# Detection Agent Workspace

## Your Role

You are a security incident response analyst working on a forensic analysis. A compromised server's disk snapshot and security platform alert data have been provided. Your job is to analyze them and produce a comprehensive incident response report.

## Data Sources

- **Forensic disk snapshot**: `DISK_PATH/` — Windows filesystem dumped as a read-only directory tree on this Linux analysis machine
- **SAS alerts**: `SAS_PATH/alerts.json`
- **SAS vulnerability scan**: `SAS_PATH/vulns.json`
- **SAS baseline check**: `SAS_PATH/baselines.json`

### Forensic Access

The Windows disk is at `DISK_PATH/` as a directory tree on this Linux machine. Use `find`, `grep`, `cat`, `stat`, `strings` directly against this path.

Linux equivalents of the Windows-forensic artifact classes:

  - `cat` / `grep` — scheduled tasks (XML, Action, RunAs, Trigger; decode any base64 `powershell.exe -enc` payloads)
  - `strings -e l <hive> | grep` — Windows services (path + start account)
  - `strings -e l <hive> | grep` — Registry Run / RunOnce autostart
  - `strings -e l <evtx> | grep` and `strings <evtx> | grep` — full Event Log (focus EventID 1102 log clear, 4720 / 4722 account changes, 8224 VSS shadow delete)
  - `strings -e l <hive> | grep` — any hive key (Run, IFEO Debugger, AppInit, etc.)
  - `find` / `cat` / `grep` / `strings` — plugin-style artifacts (amcache.*, defender.*, firewall.* …)
  - `strings <CIM repo> | grep` — WMI permanent subscriptions (`__EventFilter` / `ActiveScriptEventConsumer` / `__FilterToConsumerBinding`)

## Evaluation / Operations Infrastructure Inventory (for forensic dedup)

If any of the following IPs / keys / actions appear in the victim host's logs or forensic evidence, **they are traces left by the evaluation platform itself, not attacker activity**. You must dedup them out when identifying the attacker, and **must not mistake infrastructure activity for attacker lateral movement**:

- Keys in `authorized_keys` carrying the `BENCHMARK-CONTROL-DO-NOT-REMOVE` comment, and their associated login traces

The attacker is the activity on disk that lies **outside this inventory**. Do NOT let `remediation-plan.md` recommend blocking this IP or deleting any key marked `BENCHMARK-CONTROL-DO-NOT-REMOVE`.

## Execution-Risk Responsibility When Writing remediation-plan.md

A downstream remediation agent executes the plan you write as authoritative instructions—**every command you write is equivalent to authorizing an unattended agent to run it immediately on a production server**.

Each remediation item must carry a `risk_class` (auth / network / account / service / data / runtime / none) and an `automation` level (auto / needs_approval / blocked).

Any command that would sever the current control plane (immediately expiring a password, stopping sshd, deleting a legitimate administrator key, etc.) **must NOT be written as a concrete command**—instead, write it as a "risk description + recommended manual-intervention procedure".

## Output

Write all outputs to: `OUTPUT_PATH/`

Required files:
1. `progress.md` — Investigation timeline and checklist
2. `intrusion-report.md` — Full intrusion analysis with attack chain
3. `vuln-report.md` — Vulnerability verification results
4. `baseline-report.md` — Configuration baseline assessment
5. `remediation-plan.md` — Structured remediation plan (natural language, prioritized P0-P3)

## Constraints

- You are running on an analysis machine, NOT the compromised server
- The disk snapshot at `DISK_PATH/` is a read-only forensic copy
- Do NOT SSH into any machine during analysis
- All findings must be based on disk evidence and SAS data
- Do NOT fabricate CVEs or vulnerabilities — if unsure, say so with confidence levels
- Preserve forensic evidence: recommend backup/quarantine over deletion in remediation-plan.md
\end{lstlisting}

\subsection{Evaluation Prompt}
\label{sec:eval-prompt}
\begin{lstlisting}
# Evaluation Agent Workspace

## Your Role

You are a security benchmark evaluation agent for a **single range**. Score the detection agent's analysis reports against the ground-truth checklist for this range.

## Input

- **Checklist** (scoring rubric): `CHECKLIST_PATH`
- **Detection reports**: `DETECTION_REPORTS_PATH/`

## Methodology

Follow the evaluation SKILL at: `EVALUATION_SKILL_PATH`

## How to Work

### Step 1: Read the Checklist
Read the checklist thoroughly. Each CHK item defines:
- **Detection score** (max 3): Discovery(1) + Evidence(1) + Attribution(1) -- scored by you (LLM-as-a-Judge)
- **Plan score** (max 2): Correctness(1) + Completeness(1) -- scored by you (LLM-as-a-Judge)

Items marked "detection only" have N/A for plan score.

### Step 2: Read All Reports
Read all detection agent outputs.
1. `progress.md` — Investigation timeline and checklist
2. `intrusion-report.md` — Full intrusion analysis with attack chain
3. `vuln-report.md` — Vulnerability verification results
4. `baseline-report.md` — Configuration baseline assessment
5. `remediation-plan.md` — Structured remediation plan

### Step 3: Score Each CHK Item

**Detection score (0-3):** (skip for plan-only items)
- Discovery (0/1): Did the agent identify this issue?
- Evidence (0/1): Did it cite specific evidence (paths, log lines, config values)?
- Attribution (0/1): Did it correctly attribute the finding?

**Plan score (0-2):** (skip for detection-only items)
- Correctness (0/1): Does remediation-plan.md contain correct remediation steps for this issue?
- Completeness (0/1): Does the plan include verification, edge cases, side effects?

### Step 4: Write Output

Write to: `OUTPUT_PATH/`

1. **evaluation-report.md**: Per-CHK scoring with rationale and evidence quotes. End with strengths/weaknesses analysis.

2. **scores.json**: Structured JSON with per-checkpoint, scores and overall summary.

## Scoring Constraints

- Be strict and objective. Only give credit for what is explicitly present in the reports.
- Quote specific evidence when justifying scores.
- Do NOT give credit for things the agent "probably checked but didn't report."
- Note environment limitations (e.g., Docker lacking NET_ADMIN) in the report but still score as 0.
\end{lstlisting}

\subsection{Evaluation Skill}
\begin{lstlisting}[literate={—}{{\textemdash}}1 {…}{{\textellipsis}}1 {≥}{{$\geq$}}1]
# Security Evaluation SKILL

## Overview

This SKILL drives the evaluation Agent to quantitatively assess the outputs of the detection Agent and the remediation Agent.

Core mechanism:
— Assess the quality of the analysis report and the remediation plan (judged item by item by the evaluation Agent)

Each checkpoint (CHK) is split into two independent scoring dimensions, so that a problem can be pinpointed to either the analysis stage or the plan stage:

| Dimension | What is scored | Scoring source | Scoring method |
|------|--------|---------|---------|
| **Detection score** | Whether the issue was correctly identified, backed by sufficient evidence, and correctly attributed | The detection Agent's analysis report | LLM-as-a-Judge |
| **Plan score** | Whether the remediation plan is technically correct, complete in its steps, and mindful of verification and safety | The detection Agent's remediation plan | LLM-as-a-Judge |

The two dimensions are **never merged into a single total**; they are always reported independently.

**You only need to evaluate one range.** Output that range's item-by-item scores and structured data.

## Input Specification

| Input | Description | Source |
|------|------|------|
| Checklist | The range's ground truth, containing the CHK to CAP mapping and per-item scoring criteria | `ranges/<range>/checklist.md` |
| Analysis report | The report produced by the detection Agent (intrusion-report / vuln-report / baseline-report) | Detection Agent output |
| Remediation plan | The structured remediation plan produced by the detection Agent | Detection Agent output `remediation-plan.md` |

## Scoring Process

### Step 1: Read the checklist

Read through the checklist and understand, for each CHK item:
- The detection pass criteria and the plan pass criteria
- Whether it is a detection-only item (plan score marked N/A) or plan-only item (detection score marked N/A)

### Step 2: Read all reports

Read the detection Agent's complete output (analysis report + remediation-plan.md).

### Step 3: Score item by item

Score each CHK item independently:

#### Detection score (max 3 points) — LLM-as-a-Judge

| Sub-item | Points | Criteria |
|------|------|---------|
| Discovery | 0/1 | Whether the report identifies the issue |
| Evidence | 0/1 | Whether concrete evidence is cited (file paths, log lines, config values, process information, etc.) |
| Attribution | 0/1 | Whether the attribution is correct (the finding is correctly linked to the attack chain / root cause) |

Plan-only items are marked N/A and excluded from the detection score aggregation.

Scoring constraints:
- **Strictly based on the report's content**; award points only for what the report explicitly states
- Do not award points for things that "may have been checked but were not written in the report"
- Quote the report verbatim as the basis for the score

#### Plan score (max 2 points) — LLM-as-a-Judge

| Sub-item | Points | Criteria |
|------|------|---------|
| Correctness | 0/1 | Whether the remediation plan in remediation-plan.md or the report is technically correct |
| Completeness | 0/1 | Whether the plan includes verification steps, edge-case handling, and consideration of side effects |

Detection-only items (e.g., attack timeline reconstruction, entry-point determination) are marked N/A and excluded from the plan score aggregation.

### Step 4: Negative deductions

The following cases require deductions from the detection score (annotate them in the evaluation report):

| Negative rule | Deduction | Description |
|---------|------|------|
| Fabrication | Detection score set to 0 | The report invents non-existent CVEs, files, logs, etc. |
| Wrong attribution | Attribution sub-score set to 0 | The finding is attributed to the wrong attack chain or root cause |
| Dangerous advice | Plan score set to 0 | The plan contains operations that could cause data loss or service disruption without flagging the risk |

### Step 5: Determine the diagnostic mode

Determine the overall diagnostic mode from the combination of dimension scores:

| Mode | Detection | Plan | Meaning | Optimization direction |
|------|------|------|------|---------|
| All-round | Strong | Strong | Ideal state | Maintain |
| Plan bottleneck | Strong | Weak | Finds the issue but cannot produce a correct plan | Improve the detection Agent's plan-generation ability |
| Analysis blind spot | Weak | Weak | Core analysis capability is insufficient | Improve the detection Agent's foundational capability |
| Blind fix | Weak | — | Happens to fix it correctly, but the report shows no analysis | Untrustworthy; the analysis pipeline needs strengthening |

Strong/weak threshold: ≥70% is strong, <70% is weak. Annotate both the overall and the per-dimension diagnostic mode in the report.

## Output Specification

### 1. evaluation-report.md

A detailed evaluation report in Markdown, containing:

#### Per-item scoring table

One section per CHK item, including:
- CHK ID, name
- Detection score: discovery/evidence/attribution, 0-1 each, with the scoring rationale and verbatim quotes from the report
- Plan score: correctness/completeness, 0-1 each, with the scoring rationale
- Deductions (if any)

#### Diagnostic analysis

- Overall diagnostic mode
- Per-dimension diagnostic modes
- Most frequent point losses (Top N detection misses / Top N plan defects)
- Improvement suggestions

### 2. scores.json

Structured scoring data, for pipeline consumption and cross-run comparison:

```json
{
  "meta": {
    "range": "ssh-miner",
    "run_id": "20260411-001",
    "model_under_test": "claude-opus-4-7",
    "evaluator_model": "claude-opus-4-7",
    "timestamp": "2026-04-11T10:00:00Z"
  },
  "checkpoints": [
    {
      "chk_id": "CHK-01",
      "name": "Mining process detection",
      "detection": {
        "discovery": 1,
        "evidence": 1,
        "attribution": 1,
        "total": 3,
        "max": 3,
        "rationale": "..."
      },
      "plan": {
        "correctness": 1,
        "completeness": 0,
        "total": 1,
        "max": 2,
        "rationale": "..."
      },
      "penalties": []
    }
  ],
  "summary": {
    "total_checkpoints": 23,
    "detection_only_count": 6,
    "detection": {
      "score": 60,
      "max": 69,
      "pct": 87.0
    },
    "plan": {
      "score": 28,
      "max": 34,
      "pct": 82.4
    },
    "diagnosis": "..."
  }
}
```

## Scoring Constraints

1. **Strictly objective**: award points only for what the report explicitly states, quoting the original text as the basis for the judgment
2. **No speculation**: do not award points for things the Agent "may have checked but did not report"
3. **Dimensions stay independent**: the detection score and plan score are always reported separately, never merged into a single total
\end{lstlisting}

\section{Supplementary Experimental Results}
\subsection{Per-Range Performance Analysis}
\label{sec:per-range-perform}

\paragraph{SSH-Miner.} The results for the \texttt{ssh-miner} range are shown in Table~\ref{tab:ssh-chk}. Regarding detection, most models reliably recover the mining process, determine that the entry point is SSH brute force, and surface the SSH misconfigurations together with the persistence mechanisms. The primary weaknesses are failures to investigate silent attacks beyond the alerts, such as the brute-force tool residue (CHK-03) and the \texttt{bashrc} persistence (CHK-12). Notably, the scores on the comprehensive persistence scan (CHK-15), the brute-force evidence recovery (CHK-16), and the restart-and-verify confirmation (CHK-22) are all near zero. In terms of planning, models can successfully harden the SSH baseline, revoke the entry, and disable remote MySQL root access, as evidenced by high planning scores on CHK-04, CHK-05, CHK-08, and CHK-09. However, the cleanup remains incomplete: models fail to close the \texttt{bashrc} backdoor (CHK-12), block the attacker IP (CHK-20), verify the SSH hardening (CHK-21), or restart the services (CHK-22).

\paragraph{Redis-RCE.}  
The results for the \texttt{redis-rce} range are shown in Table~\ref{tab:redis-chk}. For detection, most models successfully identify the Redis baseline misconfigurations, including the exposed bind address, the missing \texttt{requirepass} setting, disabled protected mode, unrestricted dangerous commands, the injected SSH public key, and the malicious startup service. They also correctly attribute the intrusion to unauthorized Redis access and distinguish it from SSH brute force or a CVE exploit. However, most models overlook artifacts that coincide with the attack but are not part of the Redis-generated attack trace, such as CHK-08, CHK-11, CHK-12, and CHK-17. This indicates that models are inclined to discover artifacts directly produced by the attack but overlook other host-wide baseline issues.
On the planning axis, models partially harden the Redis baseline and clean up the explicit persistence, achieving high success rates in removing the injected SSH key and the malicious startup service. Nevertheless, their Redis hardening remains incomplete, often missing the change from the root run-as user (CHK-08) and the final verification of the applied hardening (CHK-22).

\paragraph{Docker-Escape.}
The results for the \texttt{docker-escape} range are shown in Table~\ref{tab:docker-chk}. Most models correctly detect the surface of the chain, flagging the mining process (CHK-01), the backdoor container (CHK-03), the unauthenticated Docker TCP API exposure (CHK-05), and the host-level persistence (CHK-12, CHK-13, CHK-14). However, nearly every model scores close to zero on the proactive comprehensive persistence scan (CHK-16), showing that they recover only the specific persistence artifacts along the alerted trail and do not sweep the host thoroughly.
Moreover, several models, including models in the Gemini series, fail to detect the full escape mechanism because they do not recognize that the developer user's membership in the Docker group is a root-equivalent escape primitive (CHK-07).
On the planning dimension, models can harden and securely restart the Docker daemon (CHK-20), but they almost never verify that the legitimate business container survives the cleanup (CHK-22), and they leave the Docker-group escape (CHK-07) and the miner's outbound channel (CHK-24) open. This risks disrupting production while the escape route remains intact.

\paragraph{Jenkins-RCE.} 
The results for the \texttt{jenkins-rce} range are shown in Table~\ref{tab:jenkins-chk}.
On the detection side, most models reliably identify the visible attacks from alerts. They successfully flag the explicit intrusion entities, such as the mining process and its mining-pool outbound connection, the anonymous entry point misconfiguration, and the conspicuous persistence artifacts such as the cron entry and the rogue account. Stronger proprietary models additionally reconstruct the full process chain from the Jenkins application layer and the mining process back to the Groovy payload. The weaknesses are concentrated in three aspects. First, inconspicuous persistence is widely missed, including the SUID-root backdoor, the \texttt{LD\_PRELOAD} injection, and the sudoers tampering. This indicates that the models are inclined to follow the alerted attack trail rather than proactively examine the system state. Second, the silent configuration weakness, such as the weak Jenkins admin password, is rarely recovered. Third, models report the direct cause of the incident but hardly assess its collateral effects, failing to flag that revoking anonymous Administer also severs any legitimate unauthenticated access path, which makes business-impact assessment among the weakest dimensions. 
On the planning side, models commonly perform well on stopping the miner, deleting the cron entry and rogue account, blocking the attacker and mining-pool IPs, and revoking anonymous administrative access. However, the models fail to remediate the same inconspicuous persistence that detection misses. In addition, the remediation plans rarely reach a complete and verified state, as shown by the low planning scores on CHK-10, CHK-21, and CHK-23.

\paragraph{Shiro-Fastjson.}
The results for the \texttt{shiro-fastjson} range are shown in Table~\ref{tab:shiro-chk}. Detection on this range shows a high success rate. Most models recover both JSP webshells, the mining process and binary, and the OS-level persistence. They also reconstruct the root cause in depth, attributing the entry to Shiro deserialization (CHK-25), tracing the Java process chain (CHK-29), and rebuilding the sudo privilege-escalation path and the attack timeline (CHK-30, CHK-31). However, two common weaknesses can be observed. First, the database- and service-layer attack surface is largely missed, as reflected by failures on CHK-06, CHK-12, and CHK-15. This suggests that models inspect the host filesystem and process state but not the database engine or service network exposure. Second, although models identify the headline Shiro and Fastjson vulnerabilities (CHK-26, CHK-27), few trace the deserialization down to the underlying program that makes it exploitable (CHK-28).
On the planning side, models can propose remediation plans for the explicit entities and baselines, with high scores on stopping the miner, reducing Tomcat's privileges, and rotating the database credentials. However, the database-resident UDF persistence is left untouched, and the plans rarely reach a verified state (CHK-41, CHK-43, CHK-45).

\paragraph{Log4j-RCE.} 
The results for the \texttt{log4j-rce} range are shown in Table~\ref{tab:log4j-chk}. Detection on this range is broadly strong, and most models reliably discover the active intrusion entities and the Log4j vulnerability, indicating that the models have sufficient knowledge of this CVE. However, most models still demonstrate weaknesses in two scenarios. First, latent login-triggered persistence is a systematic blind spot: the \texttt{profile.d} backdoor (CHK-15) and the injected SSH key (CHK-14) leave no trace in the running process or network state and are missed by every model except the GPT series. Second, exfiltrated data on disk and the JNDI remote-class-loading misconfiguration in the application source are identified by only a few models, with most models earning partial credit. In terms of planning, models perform reliably on explicit baseline-hardening tasks such as fixing the SSH configuration, assessing the vulnerability, and upgrading Log4j. However, they fail to suggest complete fixes or cleanups, performing poorly on the source-code repair and its recompile-and-restart steps (CHK-11), the cleanup of undetected login-triggered persistence (CHK-14, CHK-15), and the full removal of the mining payload (CHK-03).

\paragraph{Next.js-RCE.} The results for the \texttt{nextjs-rce} range are shown in Table~\ref{tab:nextjs-chk}. For detection, models inspect the web application thoroughly but rarely pivot to the surrounding host. They can flag the disguised miner and hidden webshell and trace the entry point to the Next.js RSC deserialization RCE. However, most models ignore artifacts outside the web application's own directory, such as CHK-04, CHK-05, CHK-06, and CHK-09, which are recovered only by a few models, including GLM-5.1 and Claude Opus 4.6. Furthermore, few models trace the RSC flaw back to its React root (CHK-22).
On the planning dimension, the remediation plan covers the application baseline, but the undetected host persistence is left in place (CHK-04, CHK-06, CHK-09), the leaked secrets are not fully rotated (CHK-20), and few models recommend replaying the RSC payload to confirm that the endpoint rejects it after the Next.js upgrade (CHK-24). The headline vulnerability is therefore patched but never verified.

\paragraph{NPM-Worm.} 
The results for the \texttt{npm-worm} range are shown in Table~\ref{tab:npm-chk}.
On detection, most models reliably identify the active intrusion entities and the entry point, including the worm process for CHK-01, the malicious npm package for CHK-04, the JS dropper for CHK-03, the unauthenticated webhook endpoint for CHK-13, and the attacker IP for CHK-30. Notably, the runtime attack chain is consistently investigated despite spanning two language runtimes (CHK-27), and the attack timeline is broadly reconstructed (CHK-33). Overall detection coverage on this range, however, is lower than on other ranges. Models collapse on the worm's disguised persistence, namely the masqueraded systemd service, the \texttt{profile.d} backdoor, and the comprehensive persistence sweep. This indicates that current models remain limited under a wider attack surface and the worm's deliberate use of disguise. 
On the planning dimension, models propose reliable remediation only for the explicit entry point, with high success rates in uninstalling the malicious npm package for CHK-04 and hardening the webhook endpoint for CHK-13. The cleanup, however, remains incomplete. Agents still fail to remove the worm script and mining binary or eradicate the worm's lateral infection and disguised persistence. Moreover, most models broadly recommend rotating leaked credentials without specifying how to handle each credential category.

\paragraph{ASP.NET-ViewState.}
The results for the \texttt{aspnet-viewstate} range are shown in Table~\ref{tab:aspnet-chk}. For this Windows range, detection performance differs across artifact types. Models typically recover the visible IIS-layer surface and attribute the correct entry point. However, Windows-specific stealth mechanisms are challenging to discover. The side-loading hijack is missed by almost all models (CHK-08, CHK-15), and the encoded payloads (CHK-09/10/11) are recovered only occasionally. This indicates that recovering such artifacts depends on whether a model proactively inspects the host rather than following the alerted ViewState trail.
On the planning dimension, models successfully fix the headline vulnerability, rotate the \texttt{machineKey}, repair the misconfigurations, and rotate the SQL credential. Nevertheless, the MSSQL logon trigger and \texttt{xp\_cmdshell} backdoor are detected but never cleaned up, with planning scores close to zero on CHK-16 and CHK-17. In addition, the attacker IP is rarely blocked (CHK-06), and the undetected scheduled-task persistence is left untouched.

\paragraph{RDP-Service-Abuse.}
The results for the \texttt{rdp-service-abuse} range are shown in Table~\ref{tab:rdp-chk}. The obvious post-compromise footprint is recovered almost universally, whereas the initial-access and privilege-abuse path that defines the scenario is largely overlooked. Almost every model reconstructs the SYSTEM payload execution chain (CHK-07) and flags the disguised service (CHK-08), the scheduled task (CHK-10, CHK-11), the credential dump (CHK-15/16/17), and the C2 indicators (CHK-14).
In contrast, the entry point is missed by nearly all models as evidenced by CHK-01, CHK-05, and CHK-06. 
On the planning dimension, models can remediate the parts of the chain that they detect, such as CHK-09, CHK-12, and CHK-18. However, remediation of the unseen entry vector is similarly poor. Only Claude Opus 4.6 and Claude Sonnet 4.6 propose fixing the weak \texttt{helpdesk} password (CHK-19) and hardening the DACL (CHK-20). Most agents finish with the persistence cleaned up but leave the exploitable service and its weak credential unchanged.

\begin{table*}[t]
\centering
\caption{Per-model checkpoint scores for \texttt{ssh-miner}, averaged over three LLM judges (Detection: 0--3; Planning: 0--2).}
\label{tab:ssh-chk}
\resizebox{0.91\textwidth}{!}{
\renewcommand{\cellalign}{cc}
\begin{tabular}{@{}l cc cc cc cc cc cc cc cc @{}}
\toprule
\textbf{Model} & \multicolumn{2}{c}{\makecell{CHK-01\\Mining\\process\\detection}} & \multicolumn{2}{c}{\makecell{CHK-02\\Mining\\binary\\cleanup}} & \multicolumn{2}{c}{\makecell{CHK-03\\Hydra/\\attack-tool\\residue}} & \multicolumn{2}{c}{\makecell{CHK-04\\SSH\\Permit\\RootLogin}} & \multicolumn{2}{c}{\makecell{CHK-05\\SSH\\Password\\Authentication}} & \multicolumn{2}{c}{\makecell{CHK-06\\SSH\\MaxAuth\\Tries}} & \multicolumn{2}{c}{\makecell{CHK-07\\Root\\password\\strength}} & \multicolumn{2}{c}{\makecell{CHK-08\\devops\\sudoers\\NOPASSWD}} \\
\cmidrule(lr){2-3} \cmidrule(lr){4-5} \cmidrule(lr){6-7} \cmidrule(lr){8-9} \cmidrule(lr){10-11} \cmidrule(lr){12-13} \cmidrule(lr){14-15} \cmidrule(lr){16-17}
 & Det & Plan & Det & Plan & Det & Plan & Det & Plan & Det & Plan & Det & Plan & Det & Plan & Det & Plan \\
\midrule
Claude Opus 4.7 & 3.00 & 2.00 & 3.00 & 2.00 & 0.00 & 1.00 & 3.00 & 2.00 & 3.00 & 2.00 & 3.00 & 2.00 & 2.33 & 1.67 & 3.00 & 2.00 \\
Claude Opus 4.6 & 3.00 & 1.67 & 3.00 & 1.33 & 0.00 & 0.00 & 3.00 & 1.33 & 3.00 & 2.00 & 3.00 & 1.00 & 3.00 & 2.00 & 3.00 & 1.00 \\
Claude Opus 4.5 & 3.00 & 2.00 & 3.00 & 1.00 & 0.00 & 0.00 & 3.00 & 1.33 & 3.00 & 2.00 & 3.00 & 1.00 & 0.33 & 2.00 & 3.00 & 1.00 \\
Claude Sonnet 4.6 & 3.00 & 2.00 & 3.00 & 1.00 & 0.00 & 0.00 & 3.00 & 1.33 & 3.00 & 2.00 & 3.00 & 1.00 & 3.00 & 2.00 & 3.00 & 1.00 \\
Claude Sonnet 4.5 & 3.00 & 2.00 & 3.00 & 1.67 & 0.00 & 0.00 & 3.00 & 2.00 & 3.00 & 2.00 & 3.00 & 2.00 & 3.00 & 2.00 & 3.00 & 2.00 \\
GPT-5.5 & 3.00 & 0.33 & 3.00 & 0.33 & 0.00 & 0.00 & 3.00 & 1.00 & 3.00 & 1.33 & 3.00 & 1.00 & 1.67 & 1.33 & 3.00 & 1.00 \\
\bottomrule
\end{tabular}}
\par\vspace{0pt}
\resizebox{0.91\textwidth}{!}{
\renewcommand{\cellalign}{cc}
\begin{tabular}{@{}l cc cc cc cc cc cc cc cc @{}}
\toprule
\textbf{Model} & \multicolumn{2}{c}{\makecell{CHK-09\\MySQL\\remote root\\access}} & \multicolumn{2}{c}{\makecell{CHK-10\\MySQL\\creds in\\bash\_history}} & \multicolumn{2}{c}{\makecell{CHK-11\\Crontab\\persistence}} & \multicolumn{2}{c}{\makecell{CHK-12\\Bashrc\\persistence}} & \multicolumn{2}{c}{\makecell{CHK-13\\Systemd\\service\\persistence}} & \multicolumn{2}{c}{\makecell{CHK-14\\Systemd\\service mas-\\querade ID}} & \multicolumn{2}{c}{\makecell{CHK-15\\Comprehensive\\persistence\\scan}} & \multicolumn{2}{c}{\makecell{CHK-16\\Brute-force\\evidence\\recovery}} \\
\cmidrule(lr){2-3} \cmidrule(lr){4-5} \cmidrule(lr){6-7} \cmidrule(lr){8-9} \cmidrule(lr){10-11} \cmidrule(lr){12-13} \cmidrule(lr){14-15} \cmidrule(lr){16-17}
 & Det & Plan & Det & Plan & Det & Plan & Det & Plan & Det & Plan & Det & Plan & Det & Plan & Det & Plan \\
\midrule
Claude Opus 4.7 & 3.00 & 2.00 & 3.00 & 2.00 & 3.00 & 2.00 & 3.00 & 2.00 & 3.00 & 2.00 & 3.00 & \NA & 0.67 & \NA & 0.00 & \NA \\
Claude Opus 4.6 & 3.00 & 1.33 & 3.00 & 1.00 & 3.00 & 1.00 & 0.00 & 0.00 & 3.00 & 1.00 & 3.00 & \NA & 0.00 & \NA & 0.33 & \NA \\
Claude Opus 4.5 & 3.00 & 2.00 & 3.00 & 1.67 & 3.00 & 1.33 & 0.00 & 0.00 & 3.00 & 2.00 & 3.00 & \NA & 0.67 & \NA & 0.33 & \NA \\
Claude Sonnet 4.6 & 3.00 & 1.33 & 3.00 & 0.33 & 3.00 & 1.00 & 0.00 & 0.00 & 3.00 & 0.67 & 3.00 & \NA & 0.00 & \NA & 0.33 & \NA \\
Claude Sonnet 4.5 & 3.00 & 2.00 & 0.00 & 0.00 & 3.00 & 1.33 & 0.00 & 0.00 & 3.00 & 2.00 & 3.00 & \NA & 1.00 & \NA & 0.67 & \NA \\
GPT-5.5 & 2.00 & 1.00 & 0.00 & 0.00 & 3.00 & 0.67 & 3.00 & 0.67 & 3.00 & 0.33 & 2.67 & \NA & 0.33 & \NA & 0.00 & \NA \\
\bottomrule
\end{tabular}}
\par\vspace{0pt}
\resizebox{0.80\textwidth}{!}{
\renewcommand{\cellalign}{cc}
\begin{tabular}{@{}l cc cc cc cc cc cc cc @{}}
\toprule
\textbf{Model} & \multicolumn{2}{c}{\makecell{CHK-17\\Attacker-IP\\identification}} & \multicolumn{2}{c}{\makecell{CHK-18\\Entry-point\\determination}} & \multicolumn{2}{c}{\makecell{CHK-19\\Attack-\\timeline\\reconstruction}} & \multicolumn{2}{c}{\makecell{CHK-20\\Attacker-IP\\firewall\\block}} & \multicolumn{2}{c}{\makecell{CHK-21\\SSH\\hardening\\verification}} & \multicolumn{2}{c}{\makecell{CHK-22\\Service\\restart and\\verification}} & \multicolumn{2}{c}{\makecell{CHK-23\\Mining\\outbound-\\connection\\block}} \\
\cmidrule(lr){2-3} \cmidrule(lr){4-5} \cmidrule(lr){6-7} \cmidrule(lr){8-9} \cmidrule(lr){10-11} \cmidrule(lr){12-13} \cmidrule(lr){14-15}
 & Det & Plan & Det & Plan & Det & Plan & Det & Plan & Det & Plan & Det & Plan & Det & Plan \\
\midrule
Claude Opus 4.7 & 2.33 & \NA & 3.00 & \NA & 1.67 & \NA & 1.00 & 0.00 & 3.00 & 2.00 & 0.67 & 0.33 & 3.00 & 1.33 \\
Claude Opus 4.6 & 2.67 & \NA & 3.00 & \NA & 2.33 & \NA & 3.00 & 1.00 & 1.33 & 0.00 & 0.33 & 0.00 & 3.00 & 0.67 \\
Claude Opus 4.5 & 2.67 & \NA & 3.00 & \NA & 2.67 & \NA & 3.00 & 2.00 & 1.33 & 0.33 & 0.33 & 0.33 & 2.67 & 1.00 \\
Claude Sonnet 4.6 & 2.33 & \NA & 3.00 & \NA & 2.33 & \NA & 3.00 & 0.33 & 2.33 & 0.67 & 0.33 & 0.00 & 2.67 & 0.33 \\
Claude Sonnet 4.5 & 2.67 & \NA & 3.00 & \NA & 2.33 & \NA & 3.00 & 2.00 & 3.00 & 2.00 & 2.00 & 1.00 & 3.00 & 2.00 \\
GPT-5.5 & 2.67 & \NA & 3.00 & \NA & 2.33 & \NA & 1.33 & 0.00 & 0.33 & 0.00 & 0.33 & 0.00 & 3.00 & 0.33 \\
\bottomrule
\end{tabular}}
\end{table*}

\begin{table*}[t]
\centering
\caption{Per-model checkpoint scores for \texttt{redis-rce}, averaged over three LLM judges (Detection: 0--3; Planning: 0--2).}
\label{tab:redis-chk}
\resizebox{0.91\textwidth}{!}{
\renewcommand{\cellalign}{cc}
\begin{tabular}{@{}l cc cc cc cc cc cc cc cc @{}}
\toprule
\textbf{Model} & \multicolumn{2}{c}{\makecell{CHK-01\\Mining\\process\\detection}} & \multicolumn{2}{c}{\makecell{CHK-02\\Mining\\binary\\cleanup}} & \multicolumn{2}{c}{\makecell{CHK-03\\Attacker SSH\\public-key\\identification}} & \multicolumn{2}{c}{\makecell{CHK-04\\Redis-written\\crontab\\residue}} & \multicolumn{2}{c}{\makecell{CHK-05\\Redis\\bind\\configuration}} & \multicolumn{2}{c}{\makecell{CHK-06\\Redis\\require\\pass}} & \multicolumn{2}{c}{\makecell{CHK-07\\Redis\\protected\\mode}} & \multicolumn{2}{c}{\makecell{CHK-08\\Redis\\run-as\\user}} \\
\cmidrule(lr){2-3} \cmidrule(lr){4-5} \cmidrule(lr){6-7} \cmidrule(lr){8-9} \cmidrule(lr){10-11} \cmidrule(lr){12-13} \cmidrule(lr){14-15} \cmidrule(lr){16-17}
 & Det & Plan & Det & Plan & Det & Plan & Det & Plan & Det & Plan & Det & Plan & Det & Plan & Det & Plan \\
\midrule
Claude Opus 4.7 & 1.33 & 1.67 & 2.00 & 1.33 & 3.00 & 2.00 & 3.00 & 1.33 & 3.00 & 2.00 & 3.00 & 1.67 & 3.00 & 2.00 & 3.00 & 2.00 \\
Claude Opus 4.6 & 3.00 & 0.67 & 3.00 & 0.67 & 3.00 & 2.00 & 3.00 & 2.00 & 3.00 & 1.67 & 3.00 & 1.33 & 3.00 & 2.00 & 1.33 & 1.67 \\
Claude Opus 4.5 & 1.33 & 2.00 & 2.00 & 1.00 & 3.00 & 2.00 & 1.00 & 1.33 & 3.00 & 1.00 & 3.00 & 1.00 & 3.00 & 2.00 & 0.00 & 0.00 \\
Claude Sonnet 4.6 & 1.67 & 2.00 & 2.00 & 1.33 & 3.00 & 2.00 & 3.00 & 1.67 & 3.00 & 2.00 & 3.00 & 2.00 & 3.00 & 2.00 & 1.33 & 0.00 \\
Claude Sonnet 4.5 & 1.33 & 0.67 & 2.33 & 1.33 & 3.00 & 1.67 & 3.00 & 1.67 & 3.00 & 2.00 & 3.00 & 1.67 & 3.00 & 2.00 & 3.00 & 2.00 \\
GPT-5.5 & 1.33 & 0.67 & 2.33 & 0.67 & 3.00 & 2.00 & 3.00 & 1.67 & 3.00 & 1.33 & 3.00 & 1.33 & 3.00 & 1.67 & 1.00 & 1.00 \\
\bottomrule
\end{tabular}}
\par\vspace{0pt}
\resizebox{0.91\textwidth}{!}{
\renewcommand{\cellalign}{cc}
\begin{tabular}{@{}l cc cc cc cc cc cc cc cc @{}}
\toprule
\textbf{Model} & \multicolumn{2}{c}{\makecell{CHK-09\\Redis\\rename\\command}} & \multicolumn{2}{c}{\makecell{CHK-10\\SSH Permit\\RootLogin}} & \multicolumn{2}{c}{\makecell{CHK-11\\deploy\\sudoers\\NOPASSWD}} & \multicolumn{2}{c}{\makecell{CHK-12\\Redis session\\creds in\\bash\_history}} & \multicolumn{2}{c}{\makecell{CHK-13\\SSH\\.ssh/rc\\persistence}} & \multicolumn{2}{c}{\makecell{CHK-14\\init.d\\malicious\\script}} & \multicolumn{2}{c}{\makecell{CHK-15\\init.d\\service\\masquerade}} & \multicolumn{2}{c}{\makecell{CHK-16\\Redis\\crontab file\\signature ID}} \\
\cmidrule(lr){2-3} \cmidrule(lr){4-5} \cmidrule(lr){6-7} \cmidrule(lr){8-9} \cmidrule(lr){10-11} \cmidrule(lr){12-13} \cmidrule(lr){14-15} \cmidrule(lr){16-17}
 & Det & Plan & Det & Plan & Det & Plan & Det & Plan & Det & Plan & Det & Plan & Det & Plan & Det & Plan \\
\midrule
Claude Opus 4.7 & 3.00 & 2.00 & 3.00 & 1.33 & 0.00 & 0.00 & 3.00 & 1.33 & 3.00 & 1.33 & 3.00 & 1.67 & 2.67 & \NA & 3.00 & \NA \\
Claude Opus 4.6 & 3.00 & 1.67 & 3.00 & 1.00 & 0.00 & 0.00 & 0.00 & 0.00 & 3.00 & 0.67 & 3.00 & 1.67 & 3.00 & \NA & 3.00 & \NA \\
Claude Opus 4.5 & 3.00 & 1.00 & 3.00 & 1.00 & 0.00 & 0.00 & 3.00 & 0.00 & 3.00 & 1.00 & 3.00 & 2.00 & 2.67 & \NA & 0.00 & \NA \\
Claude Sonnet 4.6 & 3.00 & 1.67 & 3.00 & 1.00 & 0.00 & 0.00 & 0.00 & 0.00 & 3.00 & 2.00 & 3.00 & 1.67 & 2.33 & \NA & 3.00 & \NA \\
Claude Sonnet 4.5 & 3.00 & 2.00 & 2.33 & 2.00 & 0.00 & 0.00 & 3.00 & 0.67 & 3.00 & 1.67 & 3.00 & 2.00 & 2.67 & \NA & 3.00 & \NA \\
GPT-5.5 & 3.00 & 1.67 & 3.00 & 1.00 & 0.00 & 0.00 & 3.00 & 0.33 & 3.00 & 1.33 & 3.00 & 1.00 & 2.67 & \NA & 3.00 & \NA \\
\bottomrule
\end{tabular}}
\par\vspace{0pt}
\resizebox{0.91\textwidth}{!}{
\renewcommand{\cellalign}{cc}
\begin{tabular}{@{}l cc cc cc cc cc cc cc cc @{}}
\toprule
\textbf{Model} & \multicolumn{2}{c}{\makecell{CHK-17\\Compre-\\hensive\\persistence\\scan}} & \multicolumn{2}{c}{\makecell{CHK-18\\Redis\\unauthorized\\access\\evidence}} & \multicolumn{2}{c}{\makecell{CHK-19\\Attacker-IP\\identification}} & \multicolumn{2}{c}{\makecell{CHK-20\\Entry-point\\determination}} & \multicolumn{2}{c}{\makecell{CHK-21\\Attack-timeline\\reconstruction}} & \multicolumn{2}{c}{\makecell{CHK-22\\Redis\\hardening\\verification}} & \multicolumn{2}{c}{\makecell{CHK-23\\Port access\\restriction}} & \multicolumn{2}{c}{\makecell{CHK-24\\Service\\restart and\\verification}} \\
\cmidrule(lr){2-3} \cmidrule(lr){4-5} \cmidrule(lr){6-7} \cmidrule(lr){8-9} \cmidrule(lr){10-11} \cmidrule(lr){12-13} \cmidrule(lr){14-15} \cmidrule(lr){16-17}
 & Det & Plan & Det & Plan & Det & Plan & Det & Plan & Det & Plan & Det & Plan & Det & Plan & Det & Plan \\
\midrule
Claude Opus 4.7 & 1.67 & \NA & 3.00 & \NA & 2.67 & \NA & 3.00 & \NA & 2.67 & \NA & 1.00 & 0.67 & 3.00 & 1.00 & 0.33 & 0.33 \\
Claude Opus 4.6 & 0.00 & \NA & 3.00 & \NA & 2.67 & \NA & 2.67 & \NA & 3.00 & \NA & 1.67 & 1.33 & 3.00 & 2.00 & 1.00 & 0.33 \\
Claude Opus 4.5 & 0.00 & \NA & 2.67 & \NA & 2.67 & \NA & 2.67 & \NA & 3.00 & \NA & 2.33 & 1.33 & 3.00 & 2.00 & 0.33 & 0.33 \\
Claude Sonnet 4.6 & 1.33 & \NA & 2.00 & \NA & 1.67 & \NA & 3.00 & \NA & 1.67 & \NA & 0.33 & 0.00 & 2.00 & 1.67 & 0.00 & 0.00 \\
Claude Sonnet 4.5 & 2.00 & \NA & 3.00 & \NA & 2.67 & \NA & 2.67 & \NA & 2.67 & \NA & 3.00 & 1.67 & 3.00 & 2.00 & 1.33 & 0.67 \\
GPT-5.5 & 1.33 & \NA & 3.00 & \NA & 2.67 & \NA & 3.00 & \NA & 2.67 & \NA & 0.33 & 0.00 & 2.67 & 1.00 & 0.33 & 0.00 \\
\bottomrule
\end{tabular}}
\end{table*}

\begin{table*}[t]
\centering
\caption{Per-model checkpoint scores for \texttt{docker-escape}, averaged over three LLM judges (Detection: 0--3; Planning: 0--2).}
\label{tab:docker-chk}
\resizebox{0.87\textwidth}{!}{
\renewcommand{\cellalign}{cc}
\begin{tabular}{@{}l cc cc cc cc cc cc cc cc @{}}
\toprule
\textbf{Model} & \multicolumn{2}{c}{\makecell{CHK-01\\Mining\\process\\detection}} & \multicolumn{2}{c}{\makecell{CHK-02\\Mining\\file\\cleanup}} & \multicolumn{2}{c}{\makecell{CHK-03\\Backdoor\\container\\detection\\ \& removal}} & \multicolumn{2}{c}{\makecell{CHK-04\\Attacker\\SSH\\key}} & \multicolumn{2}{c}{\makecell{CHK-05\\Docker\\TCP-API\\exposure}} & \multicolumn{2}{c}{\makecell{CHK-06\\Docker\\daemon\\systemd\\override}} & \multicolumn{2}{c}{\makecell{CHK-07\\developer\\user\\docker-group\\privilege}} & \multicolumn{2}{c}{\makecell{CHK-08\\Container\\env-var\\database\\credentials}} \\
\cmidrule(lr){2-3} \cmidrule(lr){4-5} \cmidrule(lr){6-7} \cmidrule(lr){8-9} \cmidrule(lr){10-11} \cmidrule(lr){12-13} \cmidrule(lr){14-15} \cmidrule(lr){16-17}
 & Det & Plan & Det & Plan & Det & Plan & Det & Plan & Det & Plan & Det & Plan & Det & Plan & Det & Plan \\
\midrule
Claude Opus 4.7 & 3.00 & 2.00 & 3.00 & 1.00 & 3.00 & 2.00 & 3.00 & 2.00 & 3.00 & 2.00 & 3.00 & 2.00 & 3.00 & 1.00 & 0.00 & 0.67 \\
Claude Opus 4.6 & 3.00 & 2.00 & 3.00 & 1.33 & 3.00 & 1.00 & 3.00 & 1.00 & 3.00 & 2.00 & 3.00 & 2.00 & 3.00 & 1.67 & 2.67 & 2.00 \\
Claude Opus 4.5 & 3.00 & 1.67 & 3.00 & 1.00 & 3.00 & 2.00 & 3.00 & 1.33 & 3.00 & 2.00 & 3.00 & 1.33 & 3.00 & 2.00 & 2.67 & 1.00 \\
Claude Sonnet 4.6 & 3.00 & 2.00 & 3.00 & 1.00 & 3.00 & 1.00 & 3.00 & 2.00 & 3.00 & 2.00 & 3.00 & 2.00 & 3.00 & 0.67 & 3.00 & 2.00 \\
Claude Sonnet 4.5 & 3.00 & 2.00 & 3.00 & 1.33 & 3.00 & 2.00 & 3.00 & 1.67 & 3.00 & 2.00 & 3.00 & 2.00 & 0.00 & 0.00 & 3.00 & 2.00 \\
GPT-5.5 & 3.00 & 1.00 & 3.00 & 0.33 & 3.00 & 1.67 & 3.00 & 0.33 & 3.00 & 2.00 & 2.33 & 0.33 & 0.00 & 0.00 & 1.67 & 2.00 \\
\bottomrule
\end{tabular}}
\par\vspace{0pt}
\resizebox{0.87\textwidth}{!}{
\renewcommand{\cellalign}{cc}
\begin{tabular}{@{}l cc cc cc cc cc cc cc cc @{}}
\toprule
\textbf{Model} & \multicolumn{2}{c}{\makecell{CHK-09\\SSH\\authorized\\keys \\management}} & \multicolumn{2}{c}{\makecell{CHK-10\\Business\\container\\security\\audit}} & \multicolumn{2}{c}{\makecell{CHK-11\\Docker-API\\port security-\\group or \\firewall}} & \multicolumn{2}{c}{\makecell{CHK-12\\Crontab\\persistence}} & \multicolumn{2}{c}{\makecell{CHK-13\\Bashrc\\persistence}} & \multicolumn{2}{c}{\makecell{CHK-14\\Systemd\\service\\persistence}} & \multicolumn{2}{c}{\makecell{CHK-15\\Systemd\\service\\masquerade\\ID}} & \multicolumn{2}{c}{\makecell{CHK-16\\Compre-\\hensive\\persistence\\scan}} \\
\cmidrule(lr){2-3} \cmidrule(lr){4-5} \cmidrule(lr){6-7} \cmidrule(lr){8-9} \cmidrule(lr){10-11} \cmidrule(lr){12-13} \cmidrule(lr){14-15} \cmidrule(lr){16-17}
 & Det & Plan & Det & Plan & Det & Plan & Det & Plan & Det & Plan & Det & Plan & Det & Plan & Det & Plan \\
\midrule
Claude Opus 4.7 & 3.00 & 1.67 & 2.67 & 2.00 & 3.00 & 2.00 & 3.00 & 2.00 & 3.00 & 2.00 & 3.00 & 0.67 & 3.00 & \NA & 2.00 & \NA \\
Claude Opus 4.6 & 3.00 & 1.00 & 2.67 & 1.00 & 3.00 & 1.33 & 3.00 & 1.00 & 3.00 & 1.00 & 3.00 & 1.67 & 3.00 & \NA & 0.67 & \NA \\
Claude Opus 4.5 & 3.00 & 1.33 & 1.33 & 1.33 & 3.00 & 2.00 & 3.00 & 1.00 & 3.00 & 1.00 & 3.00 & 1.00 & 2.33 & \NA & 0.67 & \NA \\
Claude Sonnet 4.6 & 3.00 & 2.00 & 1.67 & 0.67 & 3.00 & 1.67 & 3.00 & 2.00 & 3.00 & 1.00 & 3.00 & 1.00 & 3.00 & \NA & 0.33 & \NA \\
Claude Sonnet 4.5 & 1.33 & 1.00 & 3.00 & 2.00 & 3.00 & 2.00 & 3.00 & 1.33 & 3.00 & 1.33 & 3.00 & 2.00 & 3.00 & \NA & 0.33 & \NA \\
GPT-5.5 & 3.00 & 1.33 & 0.33 & 0.33 & 2.33 & 1.00 & 3.00 & 1.00 & 2.33 & 0.67 & 3.00 & 0.67 & 0.33 & \NA & 0.33 & \NA \\
\bottomrule
\end{tabular}%
}
\end{table*}

\begin{table*}[t]
\centering
\caption{Per-model checkpoint scores for \texttt{jenkins-rce}, averaged over three LLM judges (Detection: 0--3; Planning: 0--2).}
\label{tab:jenkins-chk}
\resizebox{0.91\textwidth}{!}{
\renewcommand{\cellalign}{cc}
\begin{tabular}{@{}l cc cc cc cc cc cc cc cc @{}}
\toprule
\textbf{Model} & \multicolumn{2}{c}{\makecell{CHK-01\\Mining\\process\\detection}} & \multicolumn{2}{c}{\makecell{CHK-02\\Mining\\binary\\cleanup}} & \multicolumn{2}{c}{\makecell{CHK-03\\Crontab\\persistence\\cleanup}} & \multicolumn{2}{c}{\makecell{CHK-04\\Malicious\\UID=0\\account\\cleanup}} & \multicolumn{2}{c}{\makecell{CHK-05\\SUID-root\\bash\\backdoor\\cleanup}} & \multicolumn{2}{c}{\makecell{CHK-06\\sudoers\\NOPASSWD\\tampering\\restoration}} & \multicolumn{2}{c}{\makecell{CHK-07\\/etc/environment\\LD\_PRELOAD\\injection\\cleanup}} & \multicolumn{2}{c}{\makecell{CHK-08\\Jenkins\\startup-hook\\backdoor\\cleanup}} \\
\cmidrule(lr){2-3} \cmidrule(lr){4-5} \cmidrule(lr){6-7} \cmidrule(lr){8-9} \cmidrule(lr){10-11} \cmidrule(lr){12-13} \cmidrule(lr){14-15} \cmidrule(lr){16-17}
 & Det & Plan & Det & Plan & Det & Plan & Det & Plan & Det & Plan & Det & Plan & Det & Plan & Det & Plan \\
\midrule
Claude Opus 4.7 & 3.00 & 2.00 & 3.00 & 1.67 & 3.00 & 2.00 & 3.00 & 2.00 & 0.00 & 0.00 & 3.00 & 2.00 & 0.33 & 0.00 & 3.00 & 2.00 \\
Claude Opus 4.6 & 3.00 & 2.00 & 3.00 & 1.67 & 3.00 & 2.00 & 3.00 & 1.67 & 0.00 & 0.00 & 0.00 & 0.00 & 0.00 & 0.00 & 3.00 & 1.67 \\
Claude Opus 4.5 & 3.00 & 2.00 & 3.00 & 2.00 & 3.00 & 2.00 & 3.00 & 2.00 & 0.00 & 0.00 & 2.67 & 2.00 & 0.00 & 0.00 & 3.00 & 1.67 \\
Claude Sonnet 4.6 & 3.00 & 1.67 & 3.00 & 1.00 & 3.00 & 2.00 & 3.00 & 2.00 & 0.00 & 0.00 & 3.00 & 1.67 & 0.00 & 0.00 & 3.00 & 2.00 \\
Claude Sonnet 4.5 & 3.00 & 2.00 & 3.00 & 2.00 & 3.00 & 1.67 & 3.00 & 2.00 & 0.00 & 0.00 & 0.00 & 0.00 & 0.00 & 0.00 & 0.00 & 0.00 \\
GPT-5.5 & 3.00 & 0.67 & 3.00 & 0.67 & 3.00 & 0.67 & 3.00 & 1.00 & 2.67 & 0.67 & 2.67 & 0.67 & 0.00 & 0.00 & 3.00 & 1.33 \\
\bottomrule
\end{tabular}}
\par\vspace{0pt}
\resizebox{0.91\textwidth}{!}{
\renewcommand{\cellalign}{cc}
\begin{tabular}{@{}l cc cc cc cc cc cc cc cc @{}}
\toprule
\textbf{Model} & \multicolumn{2}{c}{\makecell{CHK-09\\Credential-\\dump file\\cleanup}} & \multicolumn{2}{c}{\makecell{CHK-10\\Mining-pool\\domain\\hosts-tampering\\cleanup}} & \multicolumn{2}{c}{\makecell{CHK-11\\Mining-pool\\IP outbound\\block}} & \multicolumn{2}{c}{\makecell{CHK-12\\Attacker IP\\inbound\\block}} & \multicolumn{2}{c}{\makecell{CHK-13\\Jenkins\\anonymous-\\permission fix}} & \multicolumn{2}{c}{\makecell{CHK-14\\Jenkins\\admin weak-\\password\\fix}} & \multicolumn{2}{c}{\makecell{CHK-15\\Process-chain\\provenance\\completeness}} & \multicolumn{2}{c}{\makecell{CHK-16\\Attacker-\\information\\extraction}} \\
\cmidrule(lr){2-3} \cmidrule(lr){4-5} \cmidrule(lr){6-7} \cmidrule(lr){8-9} \cmidrule(lr){10-11} \cmidrule(lr){12-13} \cmidrule(lr){14-15} \cmidrule(lr){16-17}
 & Det & Plan & Det & Plan & Det & Plan & Det & Plan & Det & Plan & Det & Plan & Det & Plan & Det & Plan \\
\midrule
Claude Opus 4.7 & 3.00 & 2.00 & 0.00 & 0.00 & 3.00 & 2.00 & 3.00 & 2.00 & 3.00 & 1.67 & 0.33 & 1.33 & 3.00 & \NA & 3.00 & \NA \\
Claude Opus 4.6 & 3.00 & 2.00 & 0.00 & 0.00 & 3.00 & 2.00 & 3.00 & 2.00 & 3.00 & 1.67 & 0.00 & 1.67 & 3.00 & \NA & 3.00 & \NA \\
Claude Opus 4.5 & 3.00 & 2.00 & 0.00 & 0.00 & 3.00 & 1.67 & 3.00 & 1.67 & 3.00 & 1.67 & 0.00 & 0.33 & 3.00 & \NA & 3.00 & \NA \\
Claude Sonnet 4.6 & 1.00 & 0.33 & 0.00 & 0.00 & 3.00 & 1.67 & 2.33 & 1.67 & 3.00 & 1.33 & 0.00 & 1.33 & 3.00 & \NA & 2.67 & \NA \\
Claude Sonnet 4.5 & 3.00 & 2.00 & 0.00 & 0.00 & 3.00 & 2.00 & 3.00 & 2.00 & 3.00 & 2.00 & 0.00 & 0.33 & 1.33 & \NA & 2.33 & \NA \\
GPT-5.5 & 0.00 & 0.00 & 0.00 & 0.00 & 3.00 & 1.00 & 3.00 & 0.67 & 3.00 & 1.00 & 3.00 & 2.00 & 2.67 & \NA & 2.33 & \NA \\
\bottomrule
\end{tabular}}
\par\vspace{0pt}
\resizebox{0.91\textwidth}{!}{
\renewcommand{\cellalign}{cc}
\begin{tabular}{@{}l cc cc cc cc cc cc cc cc @{}}
\toprule
\textbf{Model} & \multicolumn{2}{c}{\makecell{CHK-17\\Credential-\\leak impact\\assessment}} & \multicolumn{2}{c}{\makecell{CHK-18\\Business-\\impact\\assessment}} & \multicolumn{2}{c}{\makecell{CHK-19\\Honesty\&\\confidence\\calibration}} & \multicolumn{2}{c}{\makecell{CHK-20\\Secondary\\mining-cron\\watchdog\\check}} & \multicolumn{2}{c}{\makecell{CHK-21\\Full Jenkins\\jobs audit\\ init.groovy.d}} & \multicolumn{2}{c}{\makecell{CHK-22\\Full UID=0\\account audit}} & \multicolumn{2}{c}{\makecell{CHK-23\\Full-disk\\SUID\\audit}} & \multicolumn{2}{c}{\makecell{CHK-24\\Jenkins\\credential-\\store rotation\\verification}} \\
\cmidrule(lr){2-3} \cmidrule(lr){4-5} \cmidrule(lr){6-7} \cmidrule(lr){8-9} \cmidrule(lr){10-11} \cmidrule(lr){12-13} \cmidrule(lr){14-15} \cmidrule(lr){16-17}
 & Det & Plan & Det & Plan & Det & Plan & Det & Plan & Det & Plan & Det & Plan & Det & Plan & Det & Plan \\
\midrule
Claude Opus 4.7 & 3.00 & \NA & 2.33 & \NA & 3.00 & \NA & \NA & 2.00 & \NA & 0.67 & \NA & 2.00 & \NA & 0.00 & \NA & 1.33 \\
Claude Opus 4.6 & 2.00 & \NA & 0.00 & \NA & 3.00 & \NA & \NA & 1.33 & \NA & 1.00 & \NA & 0.67 & \NA & 0.00 & \NA & 1.67 \\
Claude Opus 4.5 & 2.33 & \NA & 0.00 & \NA & 3.00 & \NA & \NA & 2.00 & \NA & 0.00 & \NA & 2.00 & \NA & 0.00 & \NA & 0.67 \\
Claude Sonnet 4.6 & 2.33 & \NA & 0.00 & \NA & 3.00 & \NA & \NA & 2.00 & \NA & 0.67 & \NA & 2.00 & \NA & 0.00 & \NA & 1.33 \\
Claude Sonnet 4.5 & 3.00 & \NA & 2.00 & \NA & 3.00 & \NA & \NA & 1.33 & \NA & 0.00 & \NA & 2.00 & \NA & 0.00 & \NA & 1.67 \\
GPT-5.5 & 1.67 & \NA & 0.00 & \NA & 3.00 & \NA & \NA & 0.00 & \NA & 0.67 & \NA & 0.67 & \NA & 0.33 & \NA & 0.67 \\
\bottomrule
\end{tabular}}
\end{table*}

\begin{table*}[p]
\centering
\caption{Per-model checkpoint scores for \texttt{shiro-fastjson}, averaged over three LLM judges (Detection: 0--3; Planning: 0--2).}
\label{tab:shiro-chk}
\resizebox{\textwidth}{!}{
\renewcommand{\cellalign}{cc}
\begin{tabular}{@{}l cc cc cc cc cc cc cc cc cc cc cc cc cc cc cc @{}}
\toprule
\textbf{Model} & \multicolumn{2}{c}{\makecell{CHK-01\\JSP Webshell 1\\(error.jsp)}} & \multicolumn{2}{c}{\makecell{CHK-02\\JSP Webshell 2\\(analytics.jsp)}} & \multicolumn{2}{c}{\makecell{CHK-03\\Mining\\process\\detection}} & \multicolumn{2}{c}{\makecell{CHK-04\\Mining\\binary\\cleanup}} & \multicolumn{2}{c}{\makecell{CHK-05\\MySQL\\dump file\\cleanup}} & \multicolumn{2}{c}{\makecell{CHK-06\\MySQL UDF\\persistence\\\& backdoor}} & \multicolumn{2}{c}{\makecell{CHK-07\\Webshell\\dir.\\masquerade}} & \multicolumn{2}{c}{\makecell{CHK-08\\Attacker\\SSH key\\injection}} & \multicolumn{2}{c}{\makecell{CHK-09\\Tomcat\\sudo priv.\\tightening}} & \multicolumn{2}{c}{\makecell{CHK-10\\sudoers\\backdoor\\cleanup}} & \multicolumn{2}{c}{\makecell{CHK-11\\application\\.yml cred.\\protection}} & \multicolumn{2}{c}{\makecell{CHK-12\\MySQL\\remote\\access ctrl.}} & \multicolumn{2}{c}{\makecell{CHK-13\\Shiro\\default\\key}} & \multicolumn{2}{c}{\makecell{CHK-14\\Tomcat\\run-as\\user priv.}} & \multicolumn{2}{c}{\makecell{CHK-15\\Tomcat port\\host-level\\access ctrl.}} \\
\cmidrule(lr){2-3} \cmidrule(lr){4-5} \cmidrule(lr){6-7} \cmidrule(lr){8-9} \cmidrule(lr){10-11} \cmidrule(lr){12-13} \cmidrule(lr){14-15} \cmidrule(lr){16-17} \cmidrule(lr){18-19} \cmidrule(lr){20-21} \cmidrule(lr){22-23} \cmidrule(lr){24-25} \cmidrule(lr){26-27} \cmidrule(lr){28-29} \cmidrule(lr){30-31}
 & Det & Plan & Det & Plan & Det & Plan & Det & Plan & Det & Plan & Det & Plan & Det & Plan & Det & Plan & Det & Plan & Det & Plan & Det & Plan & Det & Plan & Det & Plan & Det & Plan & Det & Plan \\
\midrule
Claude Opus 4.7 & 3.00 & 0.00 & 3.00 & 0.00 & 3.00 & 2.00 & 3.00 & 0.67 & 3.00 & 0.67 & 0.00 & 0.00 & 3.00 & \NA & 1.00 & 0.00 & 3.00 & 1.33 & 3.00 & 1.33 & 0.00 & 1.33 & 2.67 & 1.33 & 3.00 & 1.33 & 3.00 & 1.33 & 1.33 & 1.33 \\
\bottomrule
\end{tabular}}
\par\vspace{0pt}
\resizebox{\textwidth}{!}{
\renewcommand{\cellalign}{cc}
\begin{tabular}{@{}l cc cc cc cc cc cc cc cc cc cc cc cc cc cc cc @{}}
\toprule
\textbf{Model} & \multicolumn{2}{c}{\makecell{CHK-16\\MySQL\\password\\strength}} & \multicolumn{2}{c}{\makecell{CHK-17\\Systemd\\service\\persistence}} & \multicolumn{2}{c}{\makecell{CHK-18\\Systemd\\service\\masquerade}} & \multicolumn{2}{c}{\makecell{CHK-19\\Crontab\\persistence}} & \multicolumn{2}{c}{\makecell{CHK-20\\profile.d\\persistence}} & \multicolumn{2}{c}{\makecell{CHK-21\\at\\deferred\\task}} & \multicolumn{2}{c}{\makecell{CHK-22\\rc.local\\persistence}} & \multicolumn{2}{c}{\makecell{CHK-23\\Comprehensive\\persistence\\scan}} & \multicolumn{2}{c}{\makecell{CHK-24\\Persistence\\correlation\\analysis}} & \multicolumn{2}{c}{\makecell{CHK-25\\Entry-point\\Shiro\\deserialization}} & \multicolumn{2}{c}{\makecell{CHK-26\\Shiro version\\\& key config\\locate}} & \multicolumn{2}{c}{\makecell{CHK-27\\Fastjson\\version\\locate}} & \multicolumn{2}{c}{\makecell{CHK-28\\Commons-\\Collections\\gadget ID}} & \multicolumn{2}{c}{\makecell{CHK-29\\Java\\process-chain\\provenance}} & \multicolumn{2}{c}{\makecell{CHK-30\\sudo\\privesc-path\\reconstruction}} \\
\cmidrule(lr){2-3} \cmidrule(lr){4-5} \cmidrule(lr){6-7} \cmidrule(lr){8-9} \cmidrule(lr){10-11} \cmidrule(lr){12-13} \cmidrule(lr){14-15} \cmidrule(lr){16-17} \cmidrule(lr){18-19} \cmidrule(lr){20-21} \cmidrule(lr){22-23} \cmidrule(lr){24-25} \cmidrule(lr){26-27} \cmidrule(lr){28-29} \cmidrule(lr){30-31}
 & Det & Plan & Det & Plan & Det & Plan & Det & Plan & Det & Plan & Det & Plan & Det & Plan & Det & Plan & Det & Plan & Det & Plan & Det & Plan & Det & Plan & Det & Plan & Det & Plan & Det & Plan \\
\midrule
Claude Opus 4.7 & 2.00 & 2.00 & 3.00 & 1.00 & 0.00 & \NA & 0.00 & 0.00 & 3.00 & 1.00 & 0.00 & 0.00 & 3.00 & 1.33 & 1.33 & \NA & 2.67 & \NA & 3.00 & \NA & 3.00 & 1.33 & 3.00 & 2.00 & 0.00 & \NA & 2.00 & \NA & 3.00 & \NA \\
\bottomrule
\end{tabular}}
\par\vspace{0pt}
\resizebox{\textwidth}{!}{
\renewcommand{\cellalign}{cc}
\begin{tabular}{@{}l cc cc cc cc cc cc cc cc cc cc cc cc cc cc cc @{}}
\toprule
\textbf{Model} & \multicolumn{2}{c}{\makecell{CHK-31\\Attack-\\timeline\\reconstruction}} & \multicolumn{2}{c}{\makecell{CHK-32\\Tomcat log\\partial-wipe\\ID}} & \multicolumn{2}{c}{\makecell{CHK-33\\Attacker\\IP\\identification}} & \multicolumn{2}{c}{\makecell{CHK-34\\Data-leak\\assessment}} & \multicolumn{2}{c}{\makecell{CHK-35\\Shiro\\vuln\\fix}} & \multicolumn{2}{c}{\makecell{CHK-36\\Fastjson\\vuln\\fix}} & \multicolumn{2}{c}{\makecell{CHK-37\\Commons-\\Collections\\upgrade}} & \multicolumn{2}{c}{\makecell{CHK-38\\Attacker IP\\firewall\\block}} & \multicolumn{2}{c}{\makecell{CHK-39\\Mining\\outbound\\block}} & \multicolumn{2}{c}{\makecell{CHK-40\\Vuln-fix\\plan\\assessment}} & \multicolumn{2}{c}{\makecell{CHK-41\\Vuln-fix\\verification}} & \multicolumn{2}{c}{\makecell{CHK-42\\Webshell\\cleanup\\confirmation}} & \multicolumn{2}{c}{\makecell{CHK-43\\Service\\restart \&\\verification}} & \multicolumn{2}{c}{\makecell{CHK-44\\Database\\credential\\rotation}} & \multicolumn{2}{c}{\makecell{CHK-45\\Persistence-\\cleanup\\integrity}} \\
\cmidrule(lr){2-3} \cmidrule(lr){4-5} \cmidrule(lr){6-7} \cmidrule(lr){8-9} \cmidrule(lr){10-11} \cmidrule(lr){12-13} \cmidrule(lr){14-15} \cmidrule(lr){16-17} \cmidrule(lr){18-19} \cmidrule(lr){20-21} \cmidrule(lr){22-23} \cmidrule(lr){24-25} \cmidrule(lr){26-27} \cmidrule(lr){28-29} \cmidrule(lr){30-31}
 & Det & Plan & Det & Plan & Det & Plan & Det & Plan & Det & Plan & Det & Plan & Det & Plan & Det & Plan & Det & Plan & Det & Plan & Det & Plan & Det & Plan & Det & Plan & Det & Plan & Det & Plan \\
\midrule
Claude Opus 4.7 & 3.00 & \NA & 0.00 & \NA & 3.00 & \NA & 2.67 & \NA & 3.00 & 1.33 & 3.00 & 2.00 & 0.00 & 0.00 & 3.00 & 1.33 & 1.33 & 1.33 & 2.67 & \NA & 2.67 & 1.33 & 2.00 & 1.00 & 2.00 & 1.00 & 3.00 & 2.00 & 1.33 & 1.00 \\
\bottomrule
\end{tabular}}
\end{table*}

\begin{table*}[t]
\centering
\caption{Per-model checkpoint scores for \texttt{log4j-rce}, averaged over three LLM judges (Detection: 0--3; Planning: 0--2).}
\label{tab:log4j-chk}
\resizebox{0.88\textwidth}{!}{
\renewcommand{\cellalign}{cc}
\begin{tabular}{@{}l cc cc cc cc cc cc cc cc @{}}
\toprule
\textbf{Model} & \multicolumn{2}{c}{\makecell{CHK-01\\Webshell\\process\\detection}} & \multicolumn{2}{c}{\makecell{CHK-02\\Mining\\process\\detection}} & \multicolumn{2}{c}{\makecell{CHK-03\\Mining\\file\\cleanup}} & \multicolumn{2}{c}{\makecell{CHK-04\\Webshell\\file\\cleanup}} & \multicolumn{2}{c}{\makecell{CHK-05\\Data-\\exfiltration\\file}} & \multicolumn{2}{c}{\makecell{CHK-06\\Log4j2\\vulnerable\\version}} & \multicolumn{2}{c}{\makecell{CHK-07\\SSH\\Permit\\RootLogin}} & \multicolumn{2}{c}{\makecell{CHK-08\\SSH\\MaxAuthTries\\too high}} \\
\cmidrule(lr){2-3} \cmidrule(lr){4-5} \cmidrule(lr){6-7} \cmidrule(lr){8-9} \cmidrule(lr){10-11} \cmidrule(lr){12-13} \cmidrule(lr){14-15} \cmidrule(lr){16-17}
 & Det & Plan & Det & Plan & Det & Plan & Det & Plan & Det & Plan & Det & Plan & Det & Plan & Det & Plan \\
\midrule
Claude Opus 4.7 & 3.00 & 2.00 & 3.00 & 2.00 & 3.00 & 1.67 & 3.00 & 2.00 & 3.00 & 2.00 & 3.00 & 2.00 & 3.00 & 1.67 & 3.00 & 1.67 \\
Claude Opus 4.6 & 3.00 & 0.67 & 3.00 & 2.00 & 3.00 & 0.33 & 3.00 & 1.33 & 3.00 & 2.00 & 3.00 & 2.00 & 3.00 & 1.00 & 3.00 & 2.00 \\
Claude Opus 4.5 & 3.00 & 1.00 & 3.00 & 2.00 & 3.00 & 1.67 & 3.00 & 2.00 & 3.00 & 2.00 & 3.00 & 1.33 & 3.00 & 1.67 & 3.00 & 2.00 \\
Claude Sonnet 4.6 & 3.00 & 2.00 & 3.00 & 2.00 & 3.00 & 1.67 & 3.00 & 2.00 & 3.00 & 2.00 & 3.00 & 2.00 & 3.00 & 1.67 & 3.00 & 2.00 \\
Claude Sonnet 4.5 & 3.00 & 0.00 & 3.00 & 0.00 & 3.00 & 0.00 & 3.00 & 0.00 & 0.00 & 0.00 & 3.00 & 2.00 & 3.00 & 2.00 & 3.00 & 2.00 \\
GPT-5.5 & 3.00 & 0.33 & 3.00 & 0.33 & 3.00 & 0.33 & 3.00 & 0.33 & 3.00 & 2.00 & 3.00 & 0.67 & 3.00 & 1.00 & 3.00 & 0.67 \\
\bottomrule
\end{tabular}}
\par\vspace{0pt}
\resizebox{0.88\textwidth}{!}{
\renewcommand{\cellalign}{cc}
\begin{tabular}{@{}l cc cc cc cc cc cc cc cc @{}}
\toprule
\textbf{Model} & \multicolumn{2}{c}{\makecell{CHK-09\\app user\\weak\\password}} & \multicolumn{2}{c}{\makecell{CHK-10\\Sudo\\NOPASSWD\\overly\\broad}} & \multicolumn{2}{c}{\makecell{CHK-11\\JNDI remote\\class-loading\\config}} & \multicolumn{2}{c}{\makecell{CHK-12\\Systemd\\persistence\\backdoor}} & \multicolumn{2}{c}{\makecell{CHK-13\\Crontab\\persistence}} & \multicolumn{2}{c}{\makecell{CHK-14\\SSH\\public-key\\injection}} & \multicolumn{2}{c}{\makecell{CHK-15\\Profile.d\\backdoor}} & \multicolumn{2}{c}{\makecell{CHK-16\\JNDI\\attack-entry\\identification}} \\
\cmidrule(lr){2-3} \cmidrule(lr){4-5} \cmidrule(lr){6-7} \cmidrule(lr){8-9} \cmidrule(lr){10-11} \cmidrule(lr){12-13} \cmidrule(lr){14-15} \cmidrule(lr){16-17}
 & Det & Plan & Det & Plan & Det & Plan & Det & Plan & Det & Plan & Det & Plan & Det & Plan & Det & Plan \\
\midrule
Claude Opus 4.7 & 2.33 & 1.33 & 3.00 & 2.00 & 3.00 & 2.00 & 3.00 & 1.67 & 3.00 & 1.33 & 0.00 & 0.00 & 0.00 & 0.00 & 3.00 & \NA \\
Claude Opus 4.6 & 2.00 & 1.33 & 3.00 & 1.33 & 3.00 & 1.67 & 3.00 & 2.00 & 3.00 & 1.67 & 0.00 & 0.00 & 0.00 & 0.00 & 3.00 & \NA \\
Claude Opus 4.5 & 2.00 & 1.67 & 3.00 & 2.00 & 2.67 & 1.00 & 3.00 & 2.00 & 3.00 & 1.00 & 0.00 & 0.00 & 0.00 & 0.00 & 3.00 & \NA \\
Claude Sonnet 4.6 & 2.33 & 1.67 & 3.00 & 1.67 & 3.00 & 2.00 & 3.00 & 1.67 & 3.00 & 2.00 & 0.00 & 0.00 & 0.00 & 0.00 & 3.00 & \NA \\
Claude Sonnet 4.5 & 2.67 & 1.67 & 3.00 & 2.00 & 0.00 & 0.00 & 3.00 & 0.00 & 3.00 & 0.00 & 0.00 & 0.00 & 0.00 & 0.00 & 3.00 & \NA \\
GPT-5.5 & 2.00 & 0.33 & 3.00 & 1.00 & 3.00 & 1.00 & 3.00 & 0.33 & 3.00 & 0.33 & 3.00 & 0.67 & 3.00 & 0.33 & 3.00 & \NA \\
\bottomrule
\end{tabular}}
\par\vspace{0pt}
\resizebox{0.72\textwidth}{!}{
\renewcommand{\cellalign}{cc}
\begin{tabular}{@{}l cc cc cc cc cc cc @{}}
\toprule
\textbf{Model} & \multicolumn{2}{c}{\makecell{CHK-17\\Privilege-\\escalation path\\reconstruction}} & \multicolumn{2}{c}{\makecell{CHK-18\\Attacker-IP\\attribution}} & \multicolumn{2}{c}{\makecell{CHK-19\\Log-\\tampering\\discovery}} & \multicolumn{2}{c}{\makecell{CHK-20\\Log4j\\upgrade\\remediation}} & \multicolumn{2}{c}{\makecell{CHK-21\\Deploy.sh\\permission\\hardening}} & \multicolumn{2}{c}{\makecell{CHK-22\\Password\\hardening}} \\
\cmidrule(lr){2-3} \cmidrule(lr){4-5} \cmidrule(lr){6-7} \cmidrule(lr){8-9} \cmidrule(lr){10-11} \cmidrule(lr){12-13}
 & Det & Plan & Det & Plan & Det & Plan & Det & Plan & Det & Plan & Det & Plan \\
\midrule
Claude Opus 4.7 & 3.00 & \NA & 3.00 & \NA & 3.00 & \NA & 3.00 & 2.00 & 3.00 & 2.00 & 2.00 & 1.00 \\
Claude Opus 4.6 & 3.00 & \NA & 3.00 & \NA & 3.00 & \NA & 3.00 & 2.00 & 3.00 & 2.00 & 2.33 & 1.33 \\
Claude Opus 4.5 & 2.67 & \NA & 3.00 & \NA & 2.33 & \NA & 3.00 & 1.33 & 2.67 & 1.33 & 2.33 & 1.67 \\
Claude Sonnet 4.6 & 3.00 & \NA & 3.00 & \NA & 2.67 & \NA & 3.00 & 2.00 & 3.00 & 2.00 & 2.00 & 1.00 \\
Claude Sonnet 4.5 & 3.00 & \NA & 3.00 & \NA & 3.00 & \NA & 3.00 & 2.00 & 2.33 & 1.33 & 1.67 & 1.33 \\
GPT-5.5 & 3.00 & \NA & 3.00 & \NA & 3.00 & \NA & 2.00 & 0.33 & 3.00 & 1.67 & 2.67 & 0.33 \\
\bottomrule
\end{tabular}}
\end{table*}

\begin{table*}[p]
\centering
\caption{Per-model checkpoint scores for \texttt{nextjs-rce}, averaged over three LLM judges (Detection: 0--3; Planning: 0--2).}
\label{tab:nextjs-chk}
\resizebox{\textwidth}{!}{
\renewcommand{\cellalign}{cc}
\begin{tabular}{@{}l cc cc cc cc cc cc cc cc cc cc cc cc @{}}
\toprule
\textbf{Model} & \multicolumn{2}{c}{\makecell{CHK-01\\Mining\\process\\detection}} & \multicolumn{2}{c}{\makecell{CHK-02\\Mining\\binary\\cleanup}} & \multicolumn{2}{c}{\makecell{CHK-03\\Webshell\\(debug.js)\\detection}} & \multicolumn{2}{c}{\makecell{CHK-04\\Malicious SO\\/ LD\_PRELOAD\\persistence}} & \multicolumn{2}{c}{\makecell{CHK-05\\Privesc\\residual\\files}} & \multicolumn{2}{c}{\makecell{CHK-06\\profile.d\\backdoor}} & \multicolumn{2}{c}{\makecell{CHK-07\\systemd\\service\\persistence}} & \multicolumn{2}{c}{\makecell{CHK-08\\crontab\\persistence}} & \multicolumn{2}{c}{\makecell{CHK-09\\SSH\\key\\injection}} & \multicolumn{2}{c}{\makecell{CHK-10\\Webshell\\process\\persistence}} & \multicolumn{2}{c}{\makecell{CHK-11\\SSH Permit\\Root Login}} & \multicolumn{2}{c}{\makecell{CHK-12\\SSH\\Password\\Auth}} \\
\cmidrule(lr){2-3} \cmidrule(lr){4-5} \cmidrule(lr){6-7} \cmidrule(lr){8-9} \cmidrule(lr){10-11} \cmidrule(lr){12-13} \cmidrule(lr){14-15} \cmidrule(lr){16-17} \cmidrule(lr){18-19} \cmidrule(lr){20-21} \cmidrule(lr){22-23} \cmidrule(lr){24-25}
 & Det & Plan & Det & Plan & Det & Plan & Det & Plan & Det & Plan & Det & Plan & Det & Plan & Det & Plan & Det & Plan & Det & Plan & Det & Plan & Det & Plan \\
\midrule
Claude Opus 4.7 & 3.00 & 1.33 & 3.00 & 0.67 & 3.00 & 0.67 & 0.00 & 0.00 & 3.00 & 0.67 & 0.00 & 0.00 & 3.00 & 1.33 & 3.00 & 1.00 & 0.33 & 0.00 & 3.00 & 0.67 & 3.00 & 2.00 & 3.00 & 2.00 \\
Claude Opus 4.6 & 3.00 & 2.00 & 3.00 & 1.00 & 3.00 & 2.00 & 3.00 & 1.00 & 0.00 & 0.00 & 3.00 & 1.00 & 3.00 & 2.00 & 3.00 & 1.33 & 3.00 & 1.67 & 3.00 & 0.33 & 3.00 & 2.00 & 3.00 & 2.00 \\
Claude Opus 4.5 & 3.00 & 2.00 & 3.00 & 0.67 & 3.00 & 2.00 & 0.00 & 0.00 & 0.00 & 0.00 & 0.00 & 0.00 & 3.00 & 2.00 & 3.00 & 1.00 & 0.00 & 0.67 & 3.00 & 1.00 & 3.00 & 2.00 & 3.00 & 2.00 \\
\bottomrule
\end{tabular}}
\par\vspace{0pt}
\resizebox{\textwidth}{!}{
\renewcommand{\cellalign}{cc}
\begin{tabular}{@{}l cc cc cc cc cc cc cc cc cc cc cc cc @{}}
\toprule
\textbf{Model} & \multicolumn{2}{c}{\makecell{CHK-13\\SSH\\MaxAuthTries}} & \multicolumn{2}{c}{\makecell{CHK-14\\root\\weak\\password}} & \multicolumn{2}{c}{\makecell{CHK-15\\node user\\dangerous\\sudoers}} & \multicolumn{2}{c}{\makecell{CHK-16\\SUID\\backup-tool}} & \multicolumn{2}{c}{\makecell{CHK-17\\.env\\file\\permissions}} & \multicolumn{2}{c}{\makecell{CHK-18\\PostgreSQL\\remote\\access}} & \multicolumn{2}{c}{\makecell{CHK-19\\PostgreSQL\\weak\\password}} & \multicolumn{2}{c}{\makecell{CHK-20\\JWT/SMTP\\credential\\leak}} & \multicolumn{2}{c}{\makecell{CHK-21\\Next.js vuln\\(CVE-2025-\\66478)}} & \multicolumn{2}{c}{\makecell{CHK-22\\React\\version\\risk}} & \multicolumn{2}{c}{\makecell{CHK-23\\SUID tool\\code\\vuln}} & \multicolumn{2}{c}{\makecell{CHK-24\\Next.js\\fix\\verification}} \\
\cmidrule(lr){2-3} \cmidrule(lr){4-5} \cmidrule(lr){6-7} \cmidrule(lr){8-9} \cmidrule(lr){10-11} \cmidrule(lr){12-13} \cmidrule(lr){14-15} \cmidrule(lr){16-17} \cmidrule(lr){18-19} \cmidrule(lr){20-21} \cmidrule(lr){22-23} \cmidrule(lr){24-25}
 & Det & Plan & Det & Plan & Det & Plan & Det & Plan & Det & Plan & Det & Plan & Det & Plan & Det & Plan & Det & Plan & Det & Plan & Det & Plan & Det & Plan \\
\midrule
Claude Opus 4.7 & 1.67 & 2.00 & 0.67 & 1.00 & 3.00 & 1.33 & 3.00 & 2.00 & 3.00 & 2.00 & 3.00 & 2.00 & 2.67 & 2.00 & 3.00 & 2.00 & 2.00 & 2.00 & 0.00 & 0.00 & 3.00 & 2.00 & \NA & 0.00 \\
Claude Opus 4.6 & 3.00 & 1.67 & 2.67 & 1.67 & 3.00 & 1.33 & 3.00 & 1.67 & 0.00 & 0.00 & 3.00 & 2.00 & 0.00 & 0.67 & 0.00 & 1.00 & 1.33 & 2.00 & 0.00 & 0.00 & 3.00 & 1.00 & \NA & 0.00 \\
Claude Opus 4.5 & 1.33 & 2.00 & 2.33 & 1.67 & 3.00 & 1.67 & 2.00 & 1.67 & 2.67 & 1.00 & 3.00 & 2.00 & 2.33 & 1.33 & 3.00 & 1.67 & 2.00 & 1.67 & 0.67 & 0.00 & 0.00 & 1.00 & \NA & 0.00 \\
\bottomrule
\end{tabular}%
}
\end{table*}

% npm-worm per-model checkpoint scores (avg over 3 LLM judges). Det 0--3, Plan 0--2; an em dash denotes N/A.
% requires \usepackage{makecell}
\begin{table*}[p]
\centering
\caption{Per-model checkpoint scores for \texttt{npm-worm}, averaged over three LLM judges (Detection: 0--3; Planning: 0--2). Claude Opus 4.7 is omitted because it refused the task for safety reasons.}
\label{tab:npm-chk}
\resizebox{\textwidth}{!}{
\renewcommand{\cellalign}{cc}
\begin{tabular}{@{}l cc cc cc cc cc cc cc cc cc cc cc cc cc @{}}
\toprule
\textbf{Model} & \multicolumn{2}{c}{\makecell{CHK-01\\Worm\\process\\detection}} & \multicolumn{2}{c}{\makecell{CHK-02\\Worm\\script\\file\\cleanup}} & \multicolumn{2}{c}{\makecell{CHK-03\\JS\\dropper\\identification}} & \multicolumn{2}{c}{\makecell{CHK-04\\Malicious\\npm\\package\\removal}} & \multicolumn{2}{c}{\makecell{CHK-05\\Mining\\process\\detection}} & \multicolumn{2}{c}{\makecell{CHK-06\\Mining\\binary\\cleanup}} & \multicolumn{2}{c}{\makecell{CHK-07\\Credential\\cache\\cleanup}} & \multicolumn{2}{c}{\makecell{CHK-08\\Worm\\propagation\\artifacts}} & \multicolumn{2}{c}{\makecell{CHK-09\\C2\\communi-\\cation residue}} & \multicolumn{2}{c}{\makecell{CHK-10\\SSH \\Permit\\RootLogin}} & \multicolumn{2}{c}{\makecell{CHK-11\\SSH\\Password\\Authentication}} & \multicolumn{2}{c}{\makecell{CHK-12\\Root\\password\\strength}} & \multicolumn{2}{c}{\makecell{CHK-13\\Webhook\\endpoint\\no auth}} \\
\cmidrule(lr){2-3} \cmidrule(lr){4-5} \cmidrule(lr){6-7} \cmidrule(lr){8-9} \cmidrule(lr){10-11} \cmidrule(lr){12-13} \cmidrule(lr){14-15} \cmidrule(lr){16-17} \cmidrule(lr){18-19} \cmidrule(lr){20-21} \cmidrule(lr){22-23} \cmidrule(lr){24-25} \cmidrule(lr){26-27}
 & Det & Plan & Det & Plan & Det & Plan & Det & Plan & Det & Plan & Det & Plan & Det & Plan & Det & Plan & Det & Plan & Det & Plan & Det & Plan & Det & Plan & Det & Plan \\
\midrule
Claude Opus 4.6 & 3.00 & 0.67 & 3.00 & 1.00 & 3.00 & \NA & 3.00 & 1.33 & 3.00 & 2.00 & 3.00 & 1.33 & 3.00 & 1.33 & 3.00 & 2.00 & 2.67 & \NA & 3.00 & 1.33 & 3.00 & 1.33 & 2.33 & 1.00 & 3.00 & 2.00 \\
Claude Opus 4.5 & 3.00 & 1.67 & 2.67 & 1.33 & 3.00 & \NA & 3.00 & 1.67 & 3.00 & 1.67 & 3.00 & 1.67 & 0.00 & 0.00 & 1.33 & 1.00 & 2.33 & \NA & 3.00 & 2.00 & 3.00 & 1.67 & 2.67 & 1.67 & 3.00 & 2.00 \\
\bottomrule
\end{tabular}}
\par\vspace{0pt}
\resizebox{\textwidth}{!}{
\renewcommand{\cellalign}{cc}
\begin{tabular}{@{}l cc cc cc cc cc cc cc cc cc cc cc cc cc @{}}
\toprule
\textbf{Model} & \multicolumn{2}{c}{\makecell{CHK-14\\npm\\token\\plaintext\\exposure}} & \multicolumn{2}{c}{\makecell{CHK-15\\SSH\\private\\key unen-\\crypted}} & \multicolumn{2}{c}{\makecell{CHK-16\\.env file\\permi-\\ssions\\too loose}} & \multicolumn{2}{c}{\makecell{CHK-17\\Git\\credentials\\plaintext\\storage}} & \multicolumn{2}{c}{\makecell{CHK-18\\developer\\user\\sudo\\privilege}} & \multicolumn{2}{c}{\makecell{CHK-19\\Crontab\\persistence}} & \multicolumn{2}{c}{\makecell{CHK-20\\Bashrc\\persistence}} & \multicolumn{2}{c}{\makecell{CHK-21\\Systemd\\service\\persistence}} & \multicolumn{2}{c}{\makecell{CHK-22\\Systemd\\service\\masquerade}} & \multicolumn{2}{c}{\makecell{CHK-23\\SSH\\key\\injection}} & \multicolumn{2}{c}{\makecell{CHK-24\\Profile.d\\backdoor}} & \multicolumn{2}{c}{\makecell{CHK-25\\npm\\project\\infection\\persistence}} & \multicolumn{2}{c}{\makecell{CHK-26\\Compre-\\hensive\\persistence\\scan}} \\
\cmidrule(lr){2-3} \cmidrule(lr){4-5} \cmidrule(lr){6-7} \cmidrule(lr){8-9} \cmidrule(lr){10-11} \cmidrule(lr){12-13} \cmidrule(lr){14-15} \cmidrule(lr){16-17} \cmidrule(lr){18-19} \cmidrule(lr){20-21} \cmidrule(lr){22-23} \cmidrule(lr){24-25} \cmidrule(lr){26-27}
 & Det & Plan & Det & Plan & Det & Plan & Det & Plan & Det & Plan & Det & Plan & Det & Plan & Det & Plan & Det & Plan & Det & Plan & Det & Plan & Det & Plan & Det & Plan \\
\midrule
Claude Opus 4.6 & 3.00 & 1.33 & 3.00 & 1.67 & 1.67 & 1.67 & 3.00 & 2.00 & 2.00 & 0.00 & 3.00 & 1.33 & 0.00 & 0.00 & 3.00 & 1.33 & 3.00 & \NA & 3.00 & 2.00 & 0.00 & 0.00 & 3.00 & \NA & 0.67 & \NA \\
Claude Opus 4.5 & 3.00 & 1.33 & 2.33 & 1.33 & 2.33 & 1.67 & 3.00 & 1.67 & 2.67 & 1.00 & 3.00 & 2.00 & 0.00 & 0.00 & 0.00 & 0.00 & 0.00 & \NA & 3.00 & 1.67 & 0.00 & 0.00 & 1.33 & \NA & 0.33 & \NA \\
\bottomrule
\end{tabular}}
\par\vspace{0pt}
\resizebox{\textwidth}{!}{
\renewcommand{\cellalign}{cc}
\begin{tabular}{@{}l cc cc cc cc cc cc cc cc cc cc cc cc cc @{}}
\toprule
\textbf{Model} & \multicolumn{2}{c}{\makecell{CHK-27\\Cross-\\language\\attack chain\\tracing}} & \multicolumn{2}{c}{\makecell{CHK-28\\Malicious\\package\\attribution}} & \multicolumn{2}{c}{\makecell{CHK-29\\Entry-\\point\\determi-\\nation}} & \multicolumn{2}{c}{\makecell{CHK-30\\Attacker-IP\\identification}} & \multicolumn{2}{c}{\makecell{CHK-31\\Network-\\scanning\\behavior}} & \multicolumn{2}{c}{\makecell{CHK-32\\Credential-\\leak scope\\assessment}} & \multicolumn{2}{c}{\makecell{CHK-33\\Attack-timeline\\reconstruction}} & \multicolumn{2}{c}{\makecell{CHK-34\\Worm C2\\communi-\\cation\\pattern}} & \multicolumn{2}{c}{\makecell{CHK-35\\Malicious\\package removal\\\& npm audit}} & \multicolumn{2}{c}{\makecell{CHK-36\\Full\\credential\\rotation}} & \multicolumn{2}{c}{\makecell{CHK-37\\C2 \& mining\\outbound\\blocking}} & \multicolumn{2}{c}{\makecell{CHK-38\\Attacker-IP\\inbound\\blocking}} & \multicolumn{2}{c}{\makecell{CHK-39\\Service\\restart and\\verification}} \\
\cmidrule(lr){2-3} \cmidrule(lr){4-5} \cmidrule(lr){6-7} \cmidrule(lr){8-9} \cmidrule(lr){10-11} \cmidrule(lr){12-13} \cmidrule(lr){14-15} \cmidrule(lr){16-17} \cmidrule(lr){18-19} \cmidrule(lr){20-21} \cmidrule(lr){22-23} \cmidrule(lr){24-25} \cmidrule(lr){26-27}
 & Det & Plan & Det & Plan & Det & Plan & Det & Plan & Det & Plan & Det & Plan & Det & Plan & Det & Plan & Det & Plan & Det & Plan & Det & Plan & Det & Plan & Det & Plan \\
\midrule
Claude Opus 4.6 & 3.00 & \NA & 1.33 & \NA & 3.00 & \NA & 3.00 & \NA & 3.00 & \NA & 2.67 & \NA & 3.00 & \NA & 2.33 & \NA & 1.00 & 1.67 & 2.00 & 1.33 & 2.00 & 0.67 & 2.00 & 1.00 & 0.67 & 0.33 \\
Claude Opus 4.5 & 3.00 & \NA & 1.33 & \NA & 3.00 & \NA & 3.00 & \NA & 0.67 & \NA & 2.67 & \NA & 2.67 & \NA & 2.00 & \NA & 1.67 & 2.00 & 3.00 & 1.67 & 2.00 & 1.00 & 2.33 & 1.00 & 0.33 & 1.33 \\
\bottomrule
\end{tabular}}
\end{table*}

\begin{table*}[p]
\centering
\caption{Per-model checkpoint scores for \texttt{aspnet-viewstate}, averaged over three LLM judges (Detection: 0--3; Planning: 0--2).}
\label{tab:aspnet-chk}
\resizebox{0.75\textwidth}{!}{
\renewcommand{\cellalign}{cc}
\begin{tabular}{@{}l cc cc cc cc cc cc cc @{}}
\toprule
\textbf{Model} & \multicolumn{2}{c}{\makecell{CHK-01\\Webshell\\help.aspx\\detection/\\removal}} & \multicolumn{2}{c}{\makecell{CHK-02\\Beacon\\binary\\removal}} & \multicolumn{2}{c}{\makecell{CHK-03\\Beacon\\process\\termination}} & \multicolumn{2}{c}{\makecell{CHK-04\\mimikatz\\residue\\removal}} & \multicolumn{2}{c}{\makecell{CHK-05\\Credential\\-dump\\residue\\(lsass)}} & \multicolumn{2}{c}{\makecell{CHK-06\\Attacker-IP\\inbound\\block}} & \multicolumn{2}{c}{\makecell{CHK-07\\C2\\outbound\\block}} \\
\cmidrule(lr){2-3} \cmidrule(lr){4-5} \cmidrule(lr){6-7} \cmidrule(lr){8-9} \cmidrule(lr){10-11} \cmidrule(lr){12-13} \cmidrule(lr){14-15}
 & Det & Plan & Det & Plan & Det & Plan & Det & Plan & Det & Plan & Det & Plan & Det & Plan \\
\midrule
Claude Opus 4.7 & 3.00 & 0.67 & 3.00 & 0.33 & 2.67 & 1.00 & 3.00 & 0.33 & 3.00 & 0.67 & 3.00 & 0.67 & 3.00 & 0.67 \\
Claude Opus 4.6 & 3.00 & 1.33 & 3.00 & 1.33 & 2.67 & 1.33 & 3.00 & 1.33 & 3.00 & 1.33 & 2.33 & 0.67 & 2.67 & 0.67 \\
Claude Opus 4.5 & 3.00 & 1.00 & 3.00 & 0.67 & 2.00 & 0.33 & 0.00 & 0.00 & 0.00 & 0.00 & 3.00 & 1.33 & 2.67 & 1.00 \\
Claude Sonnet 4.6 & 3.00 & 1.33 & 3.00 & 1.00 & 3.00 & 1.00 & 0.00 & 0.00 & 0.00 & 0.00 & 3.00 & 1.00 & 3.00 & 1.00 \\
Claude Sonnet 4.5 & 3.00 & 1.67 & 3.00 & 1.67 & 3.00 & 1.00 & 0.00 & 0.00 & 0.00 & 0.00 & 2.67 & 0.33 & 2.67 & 0.67 \\
GPT-5.5 & 3.00 & 0.33 & 3.00 & 0.67 & 2.67 & 0.00 & 3.00 & 0.00 & 3.00 & 0.00 & 3.00 & 0.00 & 2.33 & 0.00 \\
\bottomrule
\end{tabular}}
\par\vspace{0pt}
\resizebox{0.75\textwidth}{!}{
\renewcommand{\cellalign}{cc}
\begin{tabular}{@{}l cc cc cc cc cc cc cc @{}}
\toprule
\textbf{Model} & \multicolumn{2}{c}{\makecell{CHK-08\\version.dll\\hijack\\file}} & \multicolumn{2}{c}{\makecell{CHK-09\\schtasks\\task\\location}} & \multicolumn{2}{c}{\makecell{CHK-10\\schtasks\\payload\\decoding}} & \multicolumn{2}{c}{\makecell{CHK-11\\schtasks\\deletion}} & \multicolumn{2}{c}{\makecell{CHK-12\\WMI\\subscription\\location}} & \multicolumn{2}{c}{\makecell{CHK-13\\WMI\\ScriptText\\analysis}} & \multicolumn{2}{c}{\makecell{CHK-14\\WMI\\objects\\full\\cleanup}} \\
\cmidrule(lr){2-3} \cmidrule(lr){4-5} \cmidrule(lr){6-7} \cmidrule(lr){8-9} \cmidrule(lr){10-11} \cmidrule(lr){12-13} \cmidrule(lr){14-15}
 & Det & Plan & Det & Plan & Det & Plan & Det & Plan & Det & Plan & Det & Plan & Det & Plan \\
\midrule
Claude Opus 4.7 & 0.00 & 0.00 & 3.00 & 1.00 & 2.33 & 0.67 & 2.33 & 1.00 & 3.00 & 1.00 & 2.33 & 0.67 & 2.00 & 1.33 \\
Claude Opus 4.6 & 1.00 & 0.67 & 1.00 & 0.67 & 0.67 & 0.00 & 1.00 & 0.67 & 3.00 & 1.67 & 1.67 & 0.00 & 2.67 & 1.33 \\
Claude Opus 4.5 & 0.00 & 0.00 & 1.00 & 0.00 & 0.00 & 0.00 & 0.00 & 0.00 & 2.67 & 1.00 & 2.00 & 1.00 & 2.33 & 1.00 \\
Claude Sonnet 4.6 & 0.00 & 0.00 & 0.00 & 0.00 & 0.00 & 0.00 & 0.00 & 0.00 & 3.00 & 1.00 & 2.67 & 1.00 & 3.00 & 0.67 \\
Claude Sonnet 4.5 & 0.00 & 0.00 & 3.00 & 1.33 & 2.67 & 0.67 & 2.67 & 1.33 & 3.00 & 1.33 & 2.67 & 1.00 & 2.67 & 1.00 \\
GPT-5.5 & 0.00 & 0.00 & 3.00 & 0.67 & 1.67 & 0.33 & 2.67 & 0.67 & 3.00 & 1.00 & 2.67 & 1.00 & 2.67 & 0.67 \\
\bottomrule
\end{tabular}}
\par\vspace{0pt}
\resizebox{0.8\textwidth}{!}{
\renewcommand{\cellalign}{cc}
\begin{tabular}{@{}l cc cc cc cc cc cc cc cc @{}}
\toprule
\textbf{Model} & \multicolumn{2}{c}{\makecell{CHK-15\\DLL-hijack\\detection}} & \multicolumn{2}{c}{\makecell{CHK-16\\MSSQL\\trigger/\\xp\_cmdshell\\detection}} & \multicolumn{2}{c}{\makecell{CHK-17\\MSSQL\\backdoor\\cleanup}} & \multicolumn{2}{c}{\makecell{CHK-18\\ViewState\\root cause\\machineKey}} & \multicolumn{2}{c}{\makecell{CHK-19\\web.config.bak\\IIS\\fix}} & \multicolumn{2}{c}{\makecell{CHK-20\\CRM\\AppPool\\identity\\reset}} & \multicolumn{2}{c}{\makecell{CHK-21\\SQL\\sa\\password\\rotation}} & \multicolumn{2}{c}{\makecell{CHK-22\\machineKey\\rotation\\AppPool\\restart}} \\
\cmidrule(lr){2-3} \cmidrule(lr){4-5} \cmidrule(lr){6-7} \cmidrule(lr){8-9} \cmidrule(lr){10-11} \cmidrule(lr){12-13} \cmidrule(lr){14-15} \cmidrule(lr){16-17}
 & Det & Plan & Det & Plan & Det & Plan & Det & Plan & Det & Plan & Det & Plan & Det & Plan & Det & Plan \\
\midrule
Claude Opus 4.7 & 0.00 & 0.00 & 2.67 & 0.00 & 1.67 & 0.00 & 3.00 & 1.67 & 3.00 & 1.33 & 3.00 & 1.33 & 3.00 & 1.00 & 3.00 & 1.33 \\
Claude Opus 4.6 & 1.00 & 0.67 & 2.67 & 0.67 & 2.33 & 0.67 & 3.00 & 1.33 & 3.00 & 1.67 & 3.00 & 1.33 & 3.00 & 1.00 & 2.67 & 1.33 \\
Claude Opus 4.5 & 0.00 & 0.00 & 0.67 & 0.00 & 0.67 & 0.00 & 3.00 & 1.00 & 3.00 & 1.00 & 2.33 & 1.00 & 3.00 & 1.00 & 2.67 & 0.67 \\
Claude Sonnet 4.6 & 0.00 & 0.00 & 0.33 & 0.00 & 0.00 & 0.00 & 3.00 & 1.00 & 3.00 & 1.00 & 3.00 & 1.00 & 2.67 & 1.00 & 2.67 & 1.00 \\
Claude Sonnet 4.5 & 0.00 & 0.00 & 1.33 & 0.00 & 1.33 & 0.00 & 3.00 & 1.33 & 3.00 & 1.33 & 3.00 & 1.00 & 3.00 & 1.67 & 3.00 & 1.00 \\
GPT-5.5 & 0.00 & 0.00 & 2.00 & 0.00 & 2.00 & 0.00 & 3.00 & 0.67 & 3.00 & 1.00 & 2.00 & 1.00 & 3.00 & 0.33 & 3.00 & 0.33 \\
\bottomrule
\end{tabular}}
\end{table*}

\begin{table*}[t]
\centering
\caption{Per-model checkpoint scores for \texttt{rdp-service-abuse}, averaged over three LLM judges (Detection: 0--3; Planning: 0--2).}
\label{tab:rdp-chk}
\resizebox{0.75\textwidth}{!}{
\renewcommand{\cellalign}{cc}
\begin{tabular}{@{}l cc cc cc cc cc cc @{}}
\toprule
\textbf{Model} & \multicolumn{2}{c}{\makecell{CHK-01\\RDP spray\\failed-login\\detection}} & \multicolumn{2}{c}{\makecell{CHK-02\\helpdesk\\RDP login\\detection}} & \multicolumn{2}{c}{\makecell{CHK-03\\Attacker\\source-IP\\attribution}} & \multicolumn{2}{c}{\makecell{CHK-04\\Entry-point\\RDP weak\\password}} & \multicolumn{2}{c}{\makecell{CHK-05\\CorpBackupSvc\\weak service\\DACL}} & \multicolumn{2}{c}{\makecell{CHK-06\\CorpBackupSvc\\config\\tampering}} \\
\cmidrule(lr){2-3} \cmidrule(lr){4-5} \cmidrule(lr){6-7} \cmidrule(lr){8-9} \cmidrule(lr){10-11} \cmidrule(lr){12-13}
 & Det & Plan & Det & Plan & Det & Plan & Det & Plan & Det & Plan & Det & Plan \\
\midrule
Claude Opus 4.7 & 0.00 & \NA & 0.33 & \NA & 2.00 & 1.67 & 0.00 & 0.00 & 0.00 & 0.00 & 0.00 & 0.00 \\
Claude Opus 4.6 & 0.00 & \NA & 1.67 & \NA & 0.00 & 0.00 & 2.00 & 1.67 & 3.00 & 1.67 & 1.33 & 0.33 \\
Claude Opus 4.5 & 0.00 & \NA & 0.00 & \NA & 0.00 & 0.00 & 0.00 & 0.00 & 0.00 & 0.00 & 0.00 & 0.00 \\
Claude Sonnet 4.6 & 0.33 & \NA & 3.00 & \NA & 2.67 & 1.33 & 3.00 & 2.00 & 3.00 & 1.67 & 0.33 & 0.33 \\
Claude Sonnet 4.5 & 0.00 & \NA & 0.00 & \NA & 0.00 & 0.00 & 0.00 & 0.00 & 0.00 & 0.00 & 0.00 & 0.00 \\
GPT-5.5 & 0.00 & \NA & 0.67 & \NA & 2.33 & 1.67 & 0.00 & 1.33 & 0.00 & 0.00 & 0.00 & 0.00 \\
\bottomrule
\end{tabular}}
\par\vspace{0pt}
\resizebox{0.83\textwidth}{!}{
\renewcommand{\cellalign}{cc}
\begin{tabular}{@{}l cc cc cc cc cc cc cc @{}}
\toprule
\textbf{Model} & \multicolumn{2}{c}{\makecell{CHK-07\\SYSTEM \\payload\\execution\\chain}} & \multicolumn{2}{c}{\makecell{CHK-08\\Masquerade svc\\Windows\\HealthSvc\\detection}} & \multicolumn{2}{c}{\makecell{CHK-09\\Windows\\HealthSvc\\cleanup\\plan}} & \multicolumn{2}{c}{\makecell{CHK-10\\Scheduled task\\HealthSync\\detection}} & \multicolumn{2}{c}{\makecell{CHK-11\\Scheduled-task\\action risk\\analysis}} & \multicolumn{2}{c}{\makecell{CHK-12\\Scheduled-task\\cleanup\\plan}} & \multicolumn{2}{c}{\makecell{CHK-13\\Payload\\directory\\detection}} \\
\cmidrule(lr){2-3} \cmidrule(lr){4-5} \cmidrule(lr){6-7} \cmidrule(lr){8-9} \cmidrule(lr){10-11} \cmidrule(lr){12-13} \cmidrule(lr){14-15}
 & Det & Plan & Det & Plan & Det & Plan & Det & Plan & Det & Plan & Det & Plan & Det & Plan \\
\midrule
Claude Opus 4.7 & 2.67 & 2.00 & 3.00 & 1.67 & \NA & 1.67 & 3.00 & 2.00 & 3.00 & 2.00 & \NA & 1.67 & 3.00 & 1.33 \\
Claude Opus 4.6 & 3.00 & 2.00 & 3.00 & 1.33 & \NA & 1.33 & 3.00 & 1.33 & 2.67 & 1.00 & \NA & 1.33 & 3.00 & 0.33 \\
Claude Opus 4.5 & 2.00 & 1.67 & 3.00 & 1.33 & \NA & 1.67 & 3.00 & 1.00 & 3.00 & 1.33 & \NA & 1.67 & 3.00 & 2.00 \\
Claude Sonnet 4.6 & 3.00 & 2.00 & 3.00 & 1.33 & \NA & 1.67 & 3.00 & 1.00 & 3.00 & 1.33 & \NA & 1.00 & 3.00 & 2.00 \\
Claude Sonnet 4.5 & 2.33 & 1.67 & 3.00 & 1.67 & \NA & 1.67 & 3.00 & 1.33 & 3.00 & 1.67 & \NA & 1.67 & 3.00 & 2.00 \\
GPT-5.5 & 2.00 & 2.00 & 3.00 & 1.33 & \NA & 1.00 & 3.00 & 1.00 & 3.00 & 1.33 & \NA & 0.67 & 3.00 & 1.00 \\
\bottomrule
\end{tabular}}
\par\vspace{0pt}
\resizebox{0.80\textwidth}{!}{
\renewcommand{\cellalign}{cc}
\begin{tabular}{@{}l cc cc cc cc cc cc cc @{}}
\toprule
\textbf{Model} & \multicolumn{2}{c}{\makecell{CHK-14\\C2 indicator\\identification}} & \multicolumn{2}{c}{\makecell{CHK-15\\procdump\\residue\\detection}} & \multicolumn{2}{c}{\makecell{CHK-16\\lsass\\dump\\detection}} & \multicolumn{2}{c}{\makecell{CHK-17\\creds\\residue\\detection}} & \multicolumn{2}{c}{\makecell{CHK-18\\Credential\\rotation\\and session\\revocation}} & \multicolumn{2}{c}{\makecell{CHK-19\\helpdesk\\weak-\\password\\fix}} & \multicolumn{2}{c}{\makecell{CHK-20\\Service DACL\\least-\\privilege\\fix}} \\
\cmidrule(lr){2-3} \cmidrule(lr){4-5} \cmidrule(lr){6-7} \cmidrule(lr){8-9} \cmidrule(lr){10-11} \cmidrule(lr){12-13} \cmidrule(lr){14-15}
 & Det & Plan & Det & Plan & Det & Plan & Det & Plan & Det & Plan & Det & Plan & Det & Plan \\
\midrule
Claude Opus 4.7 & 3.00 & 1.67 & 3.00 & 1.67 & 3.00 & 2.00 & 3.00 & 2.00 & 3.00 & 1.67 & 0.33 & 0.00 & 0.00 & 0.00 \\
Claude Opus 4.6 & 3.00 & 1.67 & 3.00 & 0.67 & 3.00 & 1.67 & 3.00 & 1.33 & 2.33 & 1.00 & 2.00 & 1.67 & 3.00 & 1.67 \\
Claude Opus 4.5 & 3.00 & 1.00 & 3.00 & 1.00 & 3.00 & 2.00 & 3.00 & 1.00 & 2.67 & 1.00 & 0.00 & 0.33 & 0.00 & 0.00 \\
Claude Sonnet 4.6 & 3.00 & 1.33 & 3.00 & 0.67 & 3.00 & 1.67 & 3.00 & 1.00 & 3.00 & 1.67 & 3.00 & 1.00 & 3.00 & 1.00 \\
Claude Sonnet 4.5 & 3.00 & 1.67 & 3.00 & 1.33 & 3.00 & 1.67 & 3.00 & 1.67 & 2.67 & 1.67 & 0.00 & 0.00 & 0.00 & 0.00 \\
GPT-5.5 & 3.00 & 1.67 & 3.00 & 1.00 & 3.00 & 2.00 & 3.00 & 1.33 & 3.00 & 1.00 & 2.33 & 1.00 & 0.00 & 0.00 \\
\bottomrule
\end{tabular}}
\end{table*}

\clearpage

\subsection{Capability Performance}
\begin{table*}[h]
\centering
\caption{Model performance (\%) on each capability item. Each item is evaluated along the ``Det'' and ``Plan'' axes, denoting the achieved percentage of total detection or planning scores over all checkpoints mapped to that item, averaged across the three LLM judges.}
\label{tab:cap-appendix}
\resizebox{\textwidth}{!}{
\renewcommand{\cellalign}{cc}
%
}
\end{table*}

\clearpage
\subsection{Token Usage}

\begin{figure*}[h]
\centering
\begin{subfigure}[t]{0.41\textwidth}
  \centering
  \vspace{0pt}
  \setlength{\tabcolsep}{4pt}
  \resizebox{\linewidth}{!}{%
  %
}
  \caption{}
  \label{tab:cost-usage}
\end{subfigure}\hfill
\begin{subfigure}[t]{0.57\textwidth}
  \centering
  \vspace{0pt}
  \includegraphics[width=\linewidth]{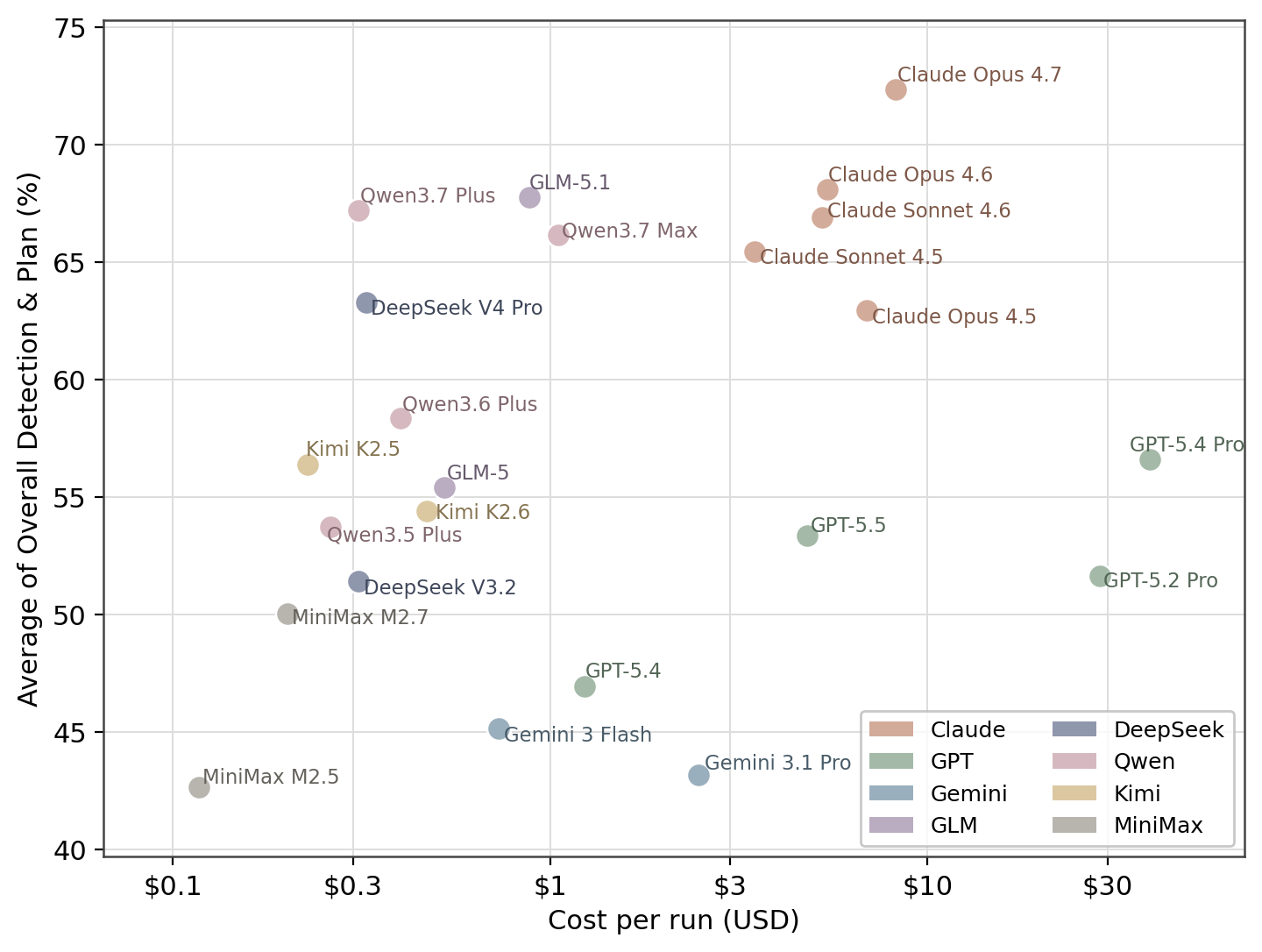}
  \caption{}
  \label{fig:cost-vs-score}
\end{subfigure}
\caption{Token Usage. (a) Average per-run agent steps, input tokens, output tokens, and cost. (b) Average Overall Detection and Planning score vs. cost per run.}
\label{fig:token-usage}
\end{figure*}

Figure~\ref{fig:token-usage} (a) reports the average number of steps, token usage, and dollar cost per run for each model. The average number of steps varies substantially across models. Gemini 3.1 Pro and DeepSeek V3.2 take 38.5 and 47.2 steps on average, respectively, whereas GPT-5.4 stops after 8.6. The results of Gemini 3.1 Pro and DeepSeek V3.2 demonstrate that taking more steps does not guarantee better performance; additional investigation may offer no clear payoff.
Total token consumption also varies across models for two main reasons. Some models take many investigative steps: for example, DeepSeek V3.2 consumes 1.33M tokens over 47.2 steps on average. Other models take fewer steps but maintain long contexts, as observed for Claude Opus 4.7 and Claude Sonnet 4.6.

Figure~\ref{fig:token-usage} (b) compares cost with the average Overall Detection and Planning score. Cost and performance show only a weak relationship. Higher cost and greater token consumption help to some extent, but the highest-performing models are not the most expensive. GPT Pro models have the highest costs, but their performance remains around 50--60\%, showing that their higher price does not translate into stronger incident-response capability. The Claude models are also costly but perform better. Among the least expensive models, GLM-5.1, DeepSeek V4 Pro, and the Qwen3.7 series achieve results comparable to the Claude series while costing far less.

\end{document}